\documentclass{jfm}

\usepackage{amstext}
\usepackage{amsmath,mathrsfs}
\usepackage{hyperref}

\usepackage{graphicx}

\usepackage{natbib}

\usepackage{subfigure}
\usepackage{sidecap}
\usepackage{units}

\usepackage{color}

\ifCUPmtlplainloaded \else
  \checkfont{eurm10}
  \iffontfound
    \IfFileExists{upmath.sty}
      {\typeout{^^JFound AMS Euler Roman fonts on the system,
                   using the 'upmath' package.^^J}%
       \usepackage{upmath}}
      {\typeout{^^JFound AMS Euler Roman fonts on the system, but you
                   dont seem to have the}%
       \typeout{'upmath' package installed. JFM.cls can take advantage
                 of these fonts,^^Jif you use 'upmath' package.^^J}%
      }
  \else
  \fi
\fi

\ifCUPmtlplainloaded \else
  \checkfont{msam10}
  \iffontfound
    \IfFileExists{amssymb.sty}
      {\typeout{^^JFound AMS Symbol fonts on the system, using the
                'amssymb' package.^^J}%
       \usepackage{amssymb}%
         \let\leq=\leqslant
         \let\geq=\geqslant
      }{}
  \fi
\fi

\ifCUPmtlplainloaded \else
  \IfFileExists{amsbsy.sty}
    {\typeout{^^JFound the 'amsbsy' package on the system, using it.^^J}%
     \usepackage{amsbsy}}
    {\providecommand\boldsymbol[1]{\mbox{\boldmath $##1$}}}
\fi

\newcommand\Real{\mbox{Re}} 
\newcommand\Imag{\mbox{Im}} 
\newcommand\Rey{\mbox{\textit{Re}}}  
\newcommand{\D}{\text{d}}
\newcommand{\I}{\text{i}}

\newsavebox{\astrutbox}
\sbox{\astrutbox}{\rule[-5pt]{0pt}{20pt}}

\newcommand\eg{e.g.\ }

\usepackage{framed}
\usepackage[framed]{ntheorem}
 
\newframedtheorem{result}{Result}

\title[Transient growth in Taylor--Couette flows]{Transient growth in linearly stable Taylor--Couette flows}

\author[S. Maretzke, B. Hof and M. Avila]%
{S\ls I\ls M\ls O\ls N\ns M\ls A\ls R\ls E\ls T\ls Z\ls K\ls E$^1$%
  \thanks{Email address for correspondence: simon.maretzke@googlemail.com},
  B\ls J\ls \"O\ls R\ls N\ns H\ls O\ls F$^2$\break
\and M\ls A\ls R\ls C\ns A\ls V\ls I\ls L\ls A$^3$}

\affiliation{$^1$Faculty of Physics, University of G\"ottingen, 37073 G\"ottingen, Germany\\
[\affilskip]$^2$Max Planck Institute for Dynamics and Self-Organization (MPIDS), 37077 G\"ottingen, Germany\\
[\affilskip]$^3$Institute of Fluid Mechanics, Friedrich-Alexander-Universit\"at Erlangen-N\"urnberg, 91058 Erlangen, Germany}

\begin{document}

\maketitle

\begin{abstract}
Non-normal transient growth of disturbances is considered as an essential prerequisite for subcritical transition in shear flows, i.e. transition to turbulence despite linear stability of the laminar flow. In this work we present numerical and analytical computations of linear transient growth covering all linearly stable regimes of Taylor--Couette flow. Our numerical experiments reveal comparable energy amplifications in the different regimes. For high shear Reynolds numbers $\Rey$ the optimal transient energy growth always follows a $\Rey^{2/3}$ scaling, which allows for large
amplifications even in regimes where the presence of turbulence remains debated. In co-rotating Rayleigh-stable flows the optimal perturbations become increasingly columnar in their structure, as the optimal axial wavenumber goes to zero. In this limit of axially invariant perturbations we show that linear stability and transient growth are independent of the cylinder rotation ratio and we derive a universal $\Rey^{2/3}$ scaling of optimal energy growth using Wentzel–Kramers–Brillouin theory. Based on this, a semi-empirical formula for the estimation of linear transient growth valid in all regimes is obtained.
\end{abstract}

\begin{keywords}
instability, transition to turbulence
\end{keywords}

%
%
%
%
%
%
%
%
\section{Introduction}
\begin{figure}
  \centering
  \begin{tabular}{cc}
    (a) & (b)\\
    \includegraphics[width=0.33\textwidth, clip=true, trim = -1.5cm -0.5cm -1.5cm -0.5cm]{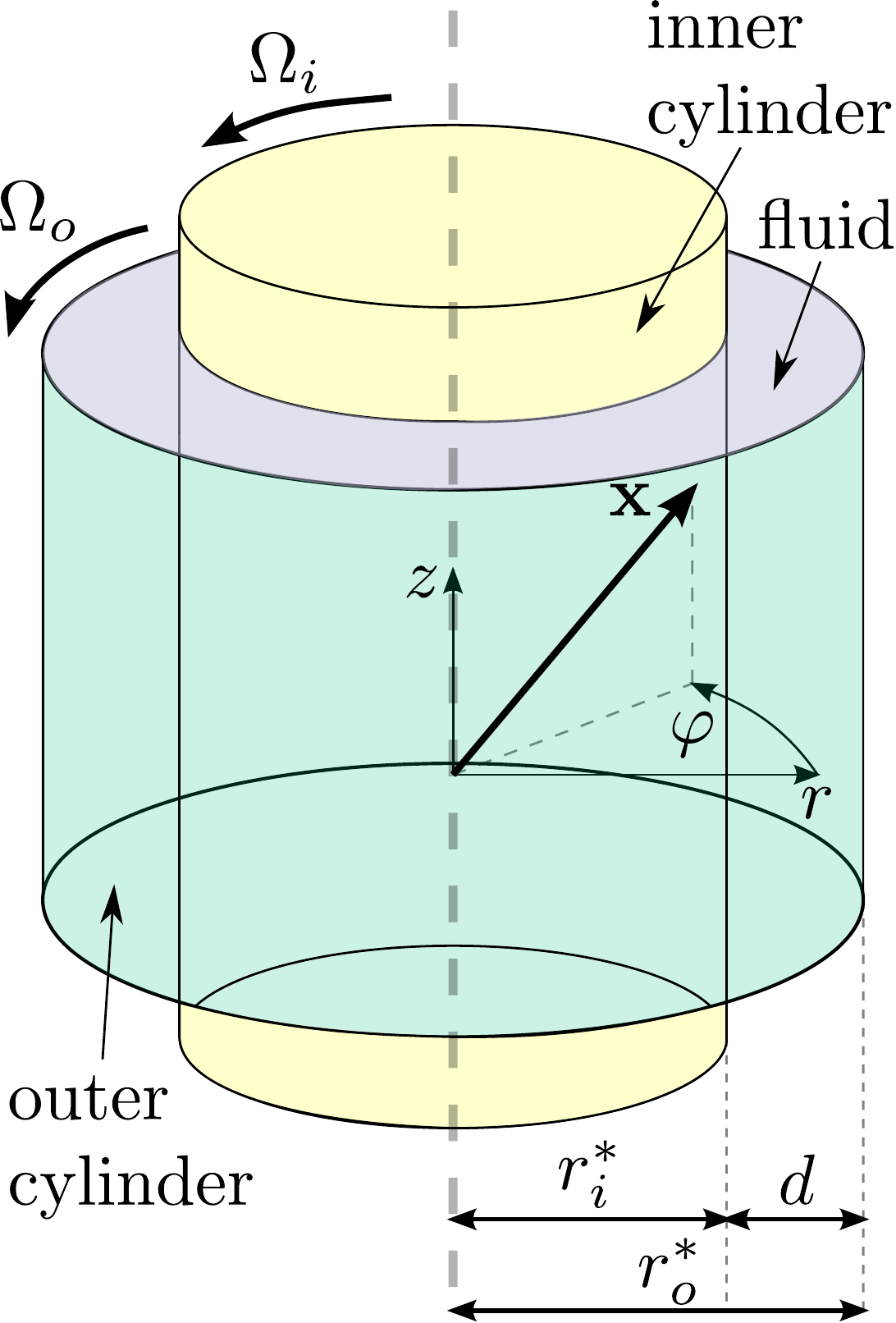} &
    \includegraphics[width=0.63\textwidth, clip=true, trim = 0cm 0.0cm 2cm 0.0cm]{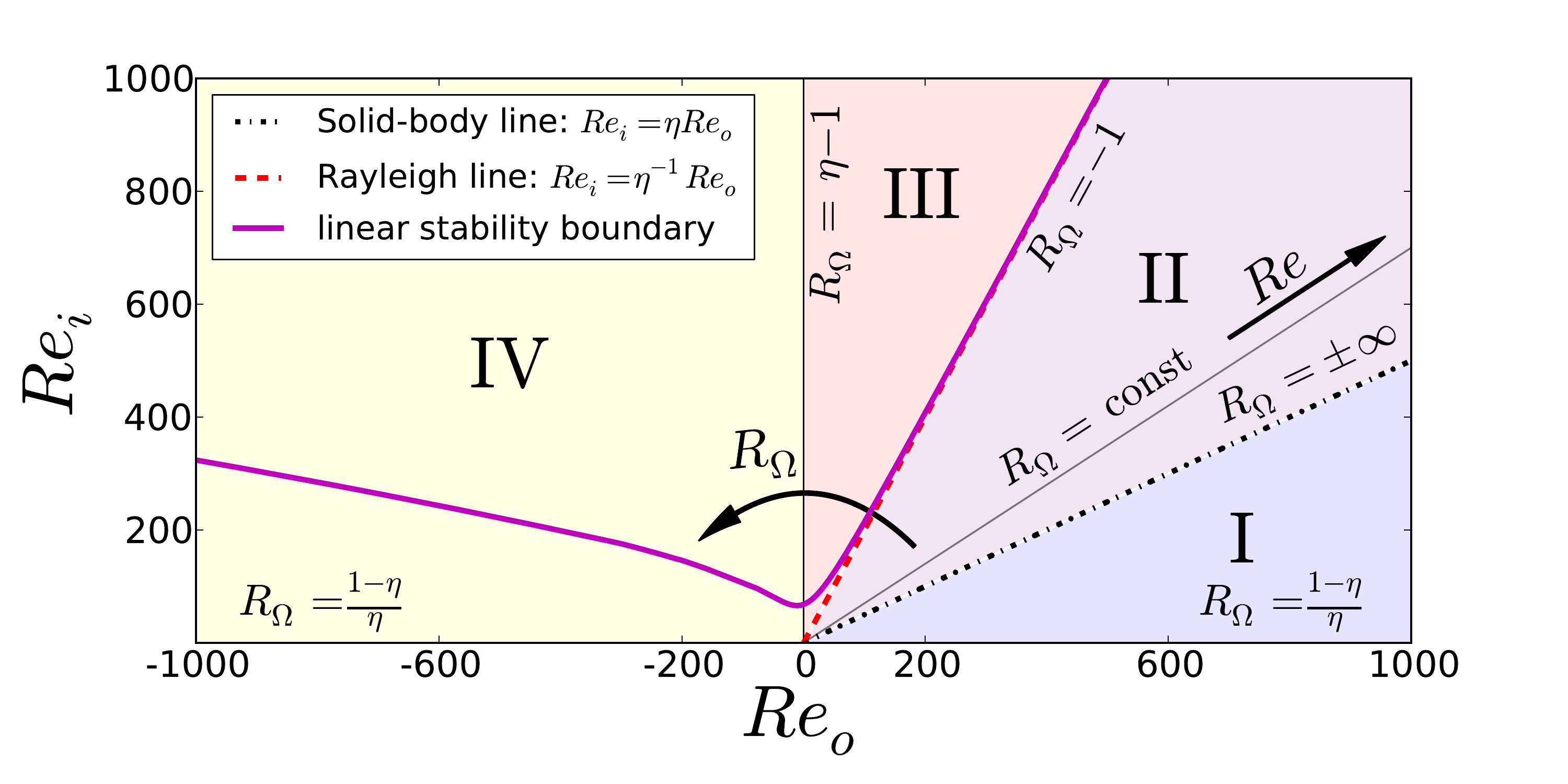}
    \end{tabular}
  \caption{{(a) Sketch of the studied Taylor--Couette geometry: a viscous, incompressible fluid is confined between two coaxial independently rotating cylinders; the system is assumed to be infinite in axial direction. (b) Taylor--Couette flow regimes in the $\Rey_i$-$\Rey_o$ plane following the parametrization of \citet{DubrulleEtAl2005} ($\eta = 0.5$). The rotation number $R_\Omega$ uniquely determines the regimes I to IV, whereas the shear Reynolds number gives the magnitudes of $\Rey_i$ and $\Rey_o$, as visualized in the plot (annotated $R_\Omega$ values correspond to the regime boundary lines; see table \ref{tab 1.1} for details). The laminar flo is linearly stable/unstable below/above the stability boundary.}}
    \label{fig sketch}
\end{figure}

The flow of viscous fluid between two coaxial independently and uniformly rotating cylinders, Taylor--Couette flow, is a paradigmatic system to study the stability and dynamics of rotating shear flows. For simplicity, we assume here that the system is infinite in the axial direction so that the annular geometry is uniquely determined by the dimensionless radius ratio $\eta$ of the inner and outer cylinders. A sketch of the Taylor--Couette system is shown in figure \ref{fig sketch}a.

The laminar Couette flow is determined by the inner and outer Reynolds numbers $\Rey_i$ and $\Rey_o$, which are proportional to the rotation frequencies of the cylinders, $\Omega_i$ and $\Omega_o$, respectively (see figure \ref{fig sketch}a). It is well known that the stability of Couette flow not only depends on the magnitudes of $\Rey_i$ and $\Rey_o$, but also changes qualitatively with their ratio. In particular, Couette flow is stable to infinitesimal inviscid disturbances if and only if the fluid particles' angular momentum increases in the radial direction. This result is known
as Rayleigh's criterion \citep{Rayleigh1917}. Consequently, inviscid instabilities solely depend on the ratio $\Rey_i / \Rey_o$. Throughout this work, the term \emph{Rayleigh (un)stable} is used to refer to the stability of Couette flow to inviscid disturbances. For viscous disturbances, there is a complex interplay of shear and centrifugal mechanisms determining the stability of the laminar flow (solid curve in figure~\ref{fig sketch}b). Herein we use the expression \emph{linearly (un-) stable} to refer to the \emph{viscous} case.
{In an attempt to separate the different effects that govern viscous stability, we adopt the parametrization introduced by \citet{DubrulleEtAl2005}, using shear Reynolds number $\Rey$ and rotation number $R_\Omega$ to parametrize the $\Rey_i$-$\Rey_o$ plane (see figure \ref{fig sketch}b). As the name suggests, $\Rey \sim \Omega_o- \Omega_i$ is a measure for the (absolute) shear in the flow, whereas $R_\Omega$ depends solely on the ratio $\Omega_o/\Omega_i$.}

In the Rayleigh-stable regimes I and II in figure \ref{fig sketch}b the flow is also linearly stable. The remaining regimes III and IV are Rayleigh-unstable. Here viscosity has a stabilizing effect and the laminar flow first develops linear instabilities at finite non-zero Reynolds numbers. These already appear at moderate $\Rey = O(10^2 - 10^3 )$, except when approaching the boundaries of regimes I and II \citep{Taylor1923}. {Indeed, the viscous linear stability boundary in figure \ref{fig sketch}b, determined from our numerical eigenvalue computations, shows that regime IV contains a relatively large range of moderate Reynolds numbers $(\Rey_o, \Rey_i)$ just above the $\Rey_o$ axis, in which the viscous laminar flow is linearly stable. Note that $\Rey_i =0$ defines the boundary to regime I, as the sign of $\Rey_o$ is irrelevant here. In contrast, the region of linear stablity in III turns out to be negligibly small. Therefore, this regime is not studied in the 
present work, as the focus lies on linearly stable flows.}


We subdivide the Rayleigh-stable regime according to the angular velocity profile $\Omega^B$ of the base flow: The \emph{quasi-Keplerian} regime II is characterized by $\partial_r \Omega^B <0$, i.e. radially decreasing angular velocity, whereas regime I is defined by a positive gradient $\partial_r \Omega^B > 0$. In figure \ref{fig sketch}b these domains are separated by the \emph{solid-body line} given by $\Rey_i = \eta \Rey_o$. Because of the absence of shear, for these configurations $\Omega^B$ is constant, corresponding to a solid-body rotation flow profile. The transition from regime II to III defines the \emph{Rayleigh line} where Rayleigh's stability criterion ceases to be fulfilled and a centrifugal (linear) instability of the laminar flow emerges. In experiments, this results in the formation
of a new stationary flow, characterized by the famous toroidal Taylor vortices \citep{Taylor1923}. Similar instabilities and associated patterns occur in the \emph{counter-rotating regime} IV {above the linear stability boundary plotted in figure \ref{fig sketch}b. For moderate $\Rey$ very good agreement between this theoretical curve and experimentally observed instabilities has been achieved.}

However, similarly to plane Couette and Poiseuille flow \citep[cf.][]{Romanov1973, Davey1973, DrazinReid1981}, certain Taylor--Couette flows may undergo \emph{subcritical} transition to turbulence in the absence of unstable eigenvalues. This phenomenon has been observed both by \citet{Coles1965} in the Rayleigh-unstable counter-rotating regime IV as well as by \citet{Wendt1933} and \citet{Taylor1936} for a stationary inner cylinder (i.e. at the lower boundary of the Rayleigh-stable regime I: see figure \ref{fig sketch}b). Recent studies by \citet{Borrero2010} have confirmed the rapid lifetime increase of turbulent spots with the Reynolds number in the latter setting. Hence, we may infer the existence of subcritical turbulence within regime I in spite of the lack of experimental and numerical data for such flows.

On the other hand, the existence of turbulence remains controversial in the equally Rayleigh-stable quasi-Keplerian regime II \citep{Yecko2004,JiBurin2006,PaolettiLathrop2011,Balbus2011}. As the name suggests, these flows are of great importance in modelling astrophysical objects with Keplerian velocity profiles, such as accretion disks \citep[for details, see][]{Pringle1981}. However, endcap effects render this regime difficult to explore experimentally. In fact, \citet{Avila2012} has shown state-of-the-art Taylor--Couette apparatus to be possibly unsuited to adequately produce the respective flow fields at the required Reynolds numbers. 
Based on $\Rey$ bounds derived from a variational formulation of the stability problem, \citet{Busse2007} conjectured that turbulence cannot exist in the quasi-Keplerian regime. Yet, this result is predicated on the hypothesis that the extremizing vector fields are independent of the streamwise coordinate. To the best of our knowledge there is no general proof ruling out the existence of turbulence in the literature.

Whether linear or nonlinear, stability analysis boils down to the evolution of initial perturbations to a stationary state. For stationary flows, the development of the perturbation energy is given by the Reynolds-Orr equation, which is valid for both fully nonlinear and linearized dynamics \citep{SchmidHenningson2001}. Remarkably, this implies that nonlinear instabilities may exist only if the linearized Navier-Stokes equations have solutions that grow in energy, i.e. transition requires linear growth.

At first glance, this theory seems contradictory to subcritical transition being a manifestation of nonlinear instability \emph{despite} linear stability. However, the apparent paradox is resolved by the non-normality of the linearized Navier-Stokes operator, i.e. the non-orthogonality of its eigenmodes \citep{Kato1976}. This potentially allows for \emph{transient growth} of infinitesimal perturbations \citep{BobergBrosa1988,TrefethenWithoutEV}, i.e. \emph{temporary} energy growth even in the case of linear stability \citep[as illustrated, for example, by][]{Grossmann2000}. As in other flow geometries the non-normality of the Taylor--Couette operator grows with the shear Reynolds number $\Rey$ so that the maximum energy amplification, $G_{\max}$, may reach several orders of magnitude at sufficiently large $\Rey$ \citep{ReddyHenningson1993}. For instance, numerical simulations by \citet{Yecko2004} of the rotating plane Couette geometry showed an asymptotic scaling of $G_{\max} \sim Re^{\frac{2}{3}}$ for one 
quasi-
Keplerian 
flow configuration in the limit $\Rey \rightarrow  \infty$.

{\citet{Hristova2002} and \citet{Meseguer2002} were the first to study transient growth in the Taylor--Couette system. Both studies investigate counter-rotating flows. The former focuses on the growth behaviour of a single axisymmetric mode, whereas the latter computes optimal linear energy amplifications at the subcritical stability boundary $\Rey_T(R_\Omega)$ measured by \citet{Coles1965}.} Most prominently, \citet{Meseguer2002} partly observes a strong correlation and finds a sharp threshold value $G_{\max, T} = 71.58 \pm 0.16$ for relaminarization in the experiments. These results reinforce the potential significance of non-normal growth in subcritical transition.

This article is concerned with transient growth in all regimes of linearly stable Taylor--Couette flows, identifying universal properties, especially in the limit of high Reynolds numbers. After briefly presenting the governing equations of the Taylor--Couette problem and our numerical formulation in $\S\S$ \ref{S2} and \ref{S3}, we discuss some tests of the method and numerical issues of transient growth computations in $\S$\ref{S4}. In $\S$\ref{S5} the main numerical results for the asymptotic scaling $G_{\max} \sim \Rey^{\alpha}$ of optimal transient growth and the corresponding optimal perturbations are presented. Furthermore, a semi-empirical formula for the estimation of $G_{\max}$ by $\Rey$ and the cylinder radius ratio $\eta$ is obtained. The latter is revealed to be
universal by the analytical results for axially independent perturbations derived in $\S$\ref{S6}. For such disturbances we further verify the characteristic scaling $G_{\max} \sim \Rey^{2/3}$ via a Wentzel–Kramers–Brillouin (WKB) approximation to the 
linearized evolution equations in $\S$\ref{S7}. In the final section $\S$\ref{S8} we discuss our results and draw some conclusions concerning subcritical instability.
%
%
%
%
%
%
%
%
\section{The linearized Taylor--Couette problem}\label{S2}

\subsection{Principal equations}\label{SS2.1}

We consider an incompressible Newtonian fluid with kinematic viscosity $\nu$ confined between two coaxial independently rotating cylinders with radii $r'_i < r'_o$ that are infinite in the axial direction. The annular geometry and its governing parameters are visualized in figure \ref{fig sketch}a. Non-dimensionalized with the gap width $d:= r'_o - r'_i$ as length scale, viscous time $\nu^{-1}d^2$ and the pressure scale $\nu^{-2}d^2$, the system is governed by the dimensionless incompressible Navier--Stokes equations and continuity equation
\begin{subequations}\label{Eq 2.1.1}
\begin{equation}
\partial_{t} \boldsymbol{v}=-(\boldsymbol{v} \cdot \boldsymbol{\nabla})\boldsymbol{v}-\boldsymbol{\nabla}\tilde{p}+ \Delta \boldsymbol{v} \label{Eq 2.1.1a}
\end{equation}
\begin{equation}
\boldsymbol{\nabla} \cdot \boldsymbol{v} = 0 \label{Eq 2.1.1b}
\end{equation}
\end{subequations}
where $\tilde{p}$ is the reduced pressure and $\boldsymbol{v}$ the velocity field of the fluid.

The independent variables are the viscous time $t$ and the spatial vector $\boldsymbol{x}$ parametrized in cylindrical coordinates $\boldsymbol{x} =: (r, \varphi, z)^\intercal$ {(see figure \ref{fig sketch}a)}. The dimensionless geometry parameters are given by $r_i:=r'_id^{-1}$, $r_o:= r'_od^{-1}$ and the radius ratio $\eta := r_ir_o^{-1}$. Let $\Omega_i$ and $\Omega_o$ be the (constant) angular velocities of the inner and outer cylinder, respectively. Defining the {\it inner} and {\it outer Reynolds numbers} $\Rey_i := \frac{d}{\nu}r'_i\Omega_i$ and $\Rey_o := \frac{d}{\nu}r'_o\Omega_o$ the {no-slip} boundary condition at the inner and outer cylinder walls read
\begin{equation}
\boldsymbol{v}_{|r = r_i} = \Rey_{i}\boldsymbol{e}_\varphi  \,\,\,\,\,\, \text{and} \,\,\,\,\,\, \boldsymbol{v}_{|r = r_o}=\Rey_{o}\boldsymbol{e}_\varphi \label{Eq 2.1.2}
\end{equation}
where $\boldsymbol{e}_r =: (1,0,0)^{\intercal}$, $\boldsymbol{e}_\varphi =: (0,1,0)^{\intercal}$ and $\boldsymbol{e}_z =: (0,0,1)^{\intercal}$ denote the orthonormal radial, azimuthal and axial unit vectors. {The unusual appearance of the Reynolds number in the boundary conditions is due to the non-dimensionalization with the viscous timescale $d^2/\nu$.}

A well-known solution of the boundary value problem \eqref{Eq 2.1.1} and \eqref{Eq 2.1.2} is laminar Couette flow $(\boldsymbol{v}^B, \tilde{p}^B)$, given by
\begin{subequations}\label{Eq 2.1.3}
\begin{equation}
\boldsymbol{v}^B = \left( Ar + \frac{B}{r} \right) \boldsymbol{e}_\varphi \,\,\,\,\,\, \text{and} \,\,\,\,\,\, \tilde{p}_B = \frac{1}{2}A^2r^2 +2AB \ln (r) - \frac{B^2}{2r^2} \label{Eq 2.1.3a} \\
\end{equation}
\begin{equation}
\,A:=\frac{\Rey_o- \eta \Rey_i}{1+ \eta}\,\,\,\,\,\, \text{and} \,\,\,\,\,\,B:=\frac{\eta(\Rey_i- \eta \Rey_o)}{(1- \eta)(1- \eta^2)} \label{Eq 2.1.3b}.
\end{equation}
\end{subequations}

In order to investigate its stability, equations \eqref{Eq 2.1.1} are linearized about $(\boldsymbol{v}^B, \tilde{p}^B)$ yielding the linearized Navier--Stokesequations for the evolution of an infinitesimally small perturbation $(\boldsymbol{\tilde{u}}, \tilde{q})$:
\begin{subequations}\label{Eq 2.1.4}
\begin{equation}
\partial_t \boldsymbol{\tilde{u}} = -(\boldsymbol{v}^B \cdot \boldsymbol{\nabla})\boldsymbol{\tilde{u}} -(\boldsymbol{\tilde{u}} \cdot \boldsymbol{\nabla})\boldsymbol{v}^B  - \boldsymbol{\nabla}\tilde{q} + \Delta \boldsymbol{\tilde{u}} \label{Eq 2.1.4a}
\end{equation}
\begin{equation}
\boldsymbol{\nabla} \cdot \boldsymbol{\tilde{u}} = 0 \label{Eq 2.1.4b}
\end{equation}
\begin{equation}
\boldsymbol{\tilde{u}}_{|r=r_i} = \boldsymbol{\tilde{u}}_{|r=r_o} = 0 \label{Eq 2.1.4c}
\end{equation}
\end{subequations}
By a Fourier ansatz in the azimuthal and axial coordinates $\boldsymbol{\tilde{u}}(r,\varphi,z) := \boldsymbol{u}(r) e^{\I (n \varphi + kz)}$, $\tilde{q}(r,\varphi,z) := q(r) e^{\I (n \varphi + kz)}$ for $k \in \mathbb{R}$, $n \in \mathbb{Z}$ the evolution equation can be written as
\begin{equation}
\partial_t \boldsymbol{u} = \mathcal{L} \boldsymbol{u} - \boldsymbol{\nabla}_{\rm c} q. \label{Eq 2.1.5}
\end{equation}
Herein a subscript $\rm c$ for an operator $\mathcal{T}$ denotes the conjugate with $ e^{\I (n \varphi + kz)}$, i.e. $\mathcal{T}_{\rm c} :=  e^{-\I (n \varphi + kz)} \mathcal{T} e^{\I (n \varphi + kz)}$. The operator $\mathcal{L}$ is given by \citep{Meseguer2002}
\begin{subequations}\label{Eq 2.1.6}
\begin{equation}
\mathcal{L} \boldsymbol{u} = -(\boldsymbol{v}^B \cdot \boldsymbol{\nabla})_{\rm c}\boldsymbol{u} -(\boldsymbol{u} \cdot \boldsymbol{\nabla})\boldsymbol{v}^B   + \Delta_{\rm c} \boldsymbol{u} =: \begin{pmatrix} \mathcal{L}_{rr} &\mathcal{L}_{r\varphi} &0 \\ \mathcal{L}_{\varphi r} &\mathcal{L}_{\varphi \varphi} &0 \\ 0 &0 &\mathcal{L}_{zz} \end{pmatrix} \begin{pmatrix} u_r \\ u_\varphi \\ u_z \end{pmatrix} \label{Eq 2.1.6a}
\end{equation}
\begin{eqnarray}
\mathcal{L}_{rr} = \mathcal{L}_{\varphi \varphi} = \mathcal{L}_{zz} - \frac{1}{r^2} &=& \mathcal{D}_+\mathcal{D} -\frac{n^2+1}{r^2} - k^2 - \frac{in}{r} v_\varphi^B \nonumber \\
\mathcal{L}_{r\varphi} &=& \frac{2}{r} v_\varphi^B -\frac{2in}{r^2} \nonumber \\
\mathcal{L}_{\varphi r} &=& \frac{2in}{r^2} - \mathcal{D}_+v_\varphi^B  \label{Eq 2.1.6b}
 \end{eqnarray}
 \end{subequations}
with the abbreviations $\mathcal{D} := \partial_r$ and $\mathcal{D}_+ := \partial_r + 1/r$. The domain of admissible velocity fields $\boldsymbol{u} = (u_r, u_\varphi, u_z)^\intercal$ in equation \eqref{Eq 2.1.5} is the twice continuously differentiable subspace
\begin{equation}
 \mathbb{V} := \left\lbrace \boldsymbol{v} \in \mathbb{H}^3 \cap \mathscr{C}^2((r_i; r_o)): \,\, \boldsymbol{v}(r_i) = \boldsymbol{v}(r_o)=0,\,\, \boldsymbol{\nabla}_{\text{c}} \cdot \boldsymbol{v} = 0 \right\rbrace \label{Eq 2.1.65}
\end{equation}
 of the Hilbert space $\mathbb{H}^3$. Here we define $\mathbb{H}:=\mathbb{L}^2((r_i; r_o))$ with the inner product
\begin{equation}
 \left\langle\cdot , \cdot \right\rangle  : \mathbb{H} \times \mathbb{H} \rightarrow \mathbb{C}; \,\, (q_1, q_2) \mapsto \int_{r_i}^{r_o} q_1^\ast q_2 \, r \D r \label{Eq 2.1.7}.
\end{equation}
where the superscript $^\ast$ denotes the conjugate transpose of a scalar, vector or matrix. For simplicity, we likewise denote the canonical inner product in $\mathbb{H}^3$, $(\boldsymbol{u}_1, \boldsymbol{u}_2) \mapsto \left\langle u_{1,r}, u_{2,r} \right\rangle + \left\langle u_{1,\varphi}, u_{2,\varphi}  \right\rangle +\left\langle u_{1,z}, u_{2,z}  \right\rangle$, by $\left\langle \cdot , \cdot \right\rangle$. The induced norm squared $\| \boldsymbol{u}\|^2 :=\left\langle \boldsymbol{u} , \boldsymbol{u} \right\rangle$ is proportional to the total kinetic energy of a perturbation $\boldsymbol{u}$ and is therefore denoted as the \emph{energy norm}.

A modal ansatz in the time coordinate $t$, i.e. $\boldsymbol{u} := \boldsymbol{u}_\lambda e^{\lambda t}$ and $q := q_\lambda e^{\lambda t}$ for $\lambda \in \mathbb{C}$ yields the eigenvalue problem
\begin{equation}
\lambda \boldsymbol{u}_\lambda = \mathcal{L} \boldsymbol{u}_\lambda - \boldsymbol{\nabla}_{\rm c} q_\lambda, \,\,\,\,\,\, (\boldsymbol{u}_\lambda, q_\lambda) \in  \mathbb{V} \times  \mathbb{H}. \label{Eq 2.1.8}
\end{equation}
For the axisymmetric case $n=0$, \citet{DiPrimaHabetler1969} have shown the discreteness of the eigenvalues $\left\lbrace \lambda \right\rbrace$ and completeness of the corresponding generalized eigenfunctions in $\mathbb{V}$. If we assume that this remains true for $n\neq 0$, then the laminar Couette flow \eqref{Eq 2.1.3} is linearly stable if and only if all eigenvalues of \eqref{Eq 2.1.8} have negative real parts.
\newline
\subsection{The parameter space for transient growth}\label{SS2.2}
In addition to the experimental parameters $\Rey_i$, $\Rey_o$ and $\eta$, the evolution problem \eqref{Eq 2.1.5} depends on the azimuthal and axial wavenumbers $n$ and $k$. Owing to the cylindrical  symmetry of the Taylor--Couette geometry the parametric analysis may be confined to $\Rey_i,n,k \geq 0$ (for details, see \citet{MeseguerMarques2000}). The parameter $\eta \in (0;\,1)$ determines the curvature of the system and thus the rotational influence. The limit $\eta \rightarrow 1$ corresponds to plane Couette flow as demonstrated with respect to transient growth by \citet{Hristova2002}, whereas $\eta \rightarrow 0$ implies infinite curvature at the inner cylinder wall.

For reasons discussed in $\S$1 we introduce the shear Reynolds number $\Rey$ and the rotation number $R_\Omega$. Assuming $\Rey_i \geq 0$ and $\Rey_i \neq \eta \Rey_o$, the mapping $(\Rey_i, \Rey_o) \mapsto (\Rey, R_\Omega)$ is one-to-one so that the flow parameters $A$ and $B$ can be expressed via $\Rey$ and $R_\Omega$:
\begin{subequations}\label{Eq 2.2.1}
\begin{equation}
 Re := \frac{2|\eta Re_o - Re_i|}{1+\eta} \,\,\,\,\,\, \text{and} \,\,\,\,\,\, R_\Omega := \frac{(1-\eta)(Re_i+ Re_o)}{\eta Re_o - Re_i}\label{Eq 2.2.1a}
\end{equation}
\begin{equation}
A = \frac{\text{sgn}(R_\Omega)Re}{2}(R_\Omega +1) \,\,\,\,\,\, \text{and} \,\,\,\,\,\, B = -\frac{\text{sgn}(R_\Omega)\eta Re}{2(1-\eta)^2}.\label{Eq 2.2.1b}
\end{equation}
\end{subequations}
{The parametrization of the different Taylor--Couette flow regimes in figure \ref{fig sketch}b by $R_\Omega$ is summarized in table \ref{tab 1.1}. As can be seen from \eqref{Eq 2.2.1b}, only the parameter $A$, which governs the solid-body rotation part $\propto Ar$ of the base flow (see \eqref{Eq 2.1.3a}) depends on the \emph{rotation} number $R_\Omega$. The shear term $\propto \frac{B}{r}$ is independent of $R_\Omega$ modulo sign, which will be essential for the results of $\S\S$ \ref{S6} and \ref{S7}. On the other hand, we have $A,B \propto \Rey$, so that the shear Reynolds number determines the overall magnitude of the flow.}
 \begin{table}
\def~{\hphantom{0}}
\centering
\begin{tabular}{cccccccc} 
&& Regime I & Regime II & Regime III & Regime IV & Solid-body line & Rayleigh line \parbox[0pt][2em][c]{0cm}{}  \\
$R_\Omega \in $ && $\left(\frac{1-\eta}{\eta}; \, \infty \right)$ & $ \left(-\infty, \,-1\right)$ & $ \left(-1 ; \, \eta -1\right)$ & $\left(\eta -1; \,\frac{1-\eta}{\eta}\right)$   &  $\left\lbrace \pm \infty \right\rbrace $ & $ \left\lbrace -1 \right\rbrace $ \parbox[0pt][2em][c]{0cm}{}  \\
\end{tabular} \caption{Parametrization of the Taylor--Couette flow regimes by the rotation number $R_\Omega$ as visualized in figure \ref{fig sketch}b; lines of constant $R_\Omega$ are axes meeting in the origin \label{tab 1.1} }
\end{table}

Computing the commutator of the operator $\mathcal{L}$ given by \eqref{Eq 2.1.6} with its adjoint $\mathcal{L}^\ast$, $[\mathcal{L}^\ast, \mathcal{L}] = O(\Rey^2)$, reveals its non-normality, scaling with the shear Reynolds number. The eigenspaces are therefore non-orthogonal to one another \citep{Kato1976}, which potentially allows for significant transient growth at large $\Rey$. Detailed discussions of this mechanism can be found in \citet{Grossmann2000} and \citet[pp. 99-101]{SchmidHenningson2001}.

As a consequence, initial perturbations $\boldsymbol{u}(0)$ may temporarily grow in energy before they ultimately decay -- even if $\mathcal{L}$ has only stable eigenvalues $\lambda \in \mathbb{C}$ with $\Real ( \lambda) < 0 $. The maximum transient growth at time $t\geq 0$ is given by $G(t) := \sup_{\|\boldsymbol{u}(0)\| = 1}  \|\boldsymbol{u}(t)\|^2$. {The evolution of $\boldsymbol{u}$ may be written as a linear equation of the form $\partial_t \boldsymbol{u} =\tilde{\mathcal{L}}\boldsymbol{u}$ (where, strictly speaking, $\tilde{\mathcal{L}} \neq \mathcal{L}$ due to the remaining pressure dependence in \eqref{Eq 2.1.5}).} Thus $G$ can be expressed using the operator norm \citep{TrefethenWithoutEV}:
\begin{equation}
G(t) := \sup_{\|\boldsymbol{u}(0)\| = 1}  \|\boldsymbol{u}(t)\|^2 = \sup_{\|\boldsymbol{u}(0)\| = 1}  \|\exp( \tilde{\mathcal{L}} t)\boldsymbol{u}(0)\|^2 = \|\exp( \tilde{\mathcal{L}} t)\|^2 \label{Eq 2.2.2}
\end{equation}
If $\| \cdot \|$ denotes the energy norm, $G(t)$ is equal to the greatest kinetic energy amplification that an initial perturbation $\boldsymbol{u}(0) \in \mathbb{V}$ can attain {at time $t\geq 0$}.
\vspace{1cm}

For a Taylor--Couette flow configuration given by the parameters $\Rey_i$, $\Rey_o$ and $\eta$, the {\it optimal transient growth} is defined by $G_{\max} := \sup_{t,n,k} G(t)$. A perturbation $\boldsymbol{u}$ with $\|\boldsymbol{u}(0)\| = 1$ is called {\it optimal} if $\|\boldsymbol{u}(t)\|^2 = G_{\max}$ for some $t \geq 0$. Note that $G_{\max}$ is finite if and only if all eigenvalues of $\mathcal{L}$ are stable.
\newline
%
%
%
%
%
%
%
%
\section{Numerical formulation and implementation} \label{S3}

\subsection{The Galerkin method}\label{SS3.1}

The eigenvalue problem \eqref{Eq 2.1.8} is numerically solved using a Galerkin method. The implementation is similar to the Petrov-Galerkin method described by \citet{MeseguerMarques2000} and \citet{Avila2007}, but based on Legendre rather than Chebyshev polynomials so that trial and projection basis are identical.

The basis choice is $U := \lbrace \boldsymbol{u}^j_m\rbrace_{m\in \mathbb{N}_0}^{j = 1,2}$, where $\boldsymbol{u}^1_m$ and $\boldsymbol{u}^2_m$ are defined according to table \ref{tab 4.1} for different wavenumbers $n$ and $k$. The functions $h_m$ and $g_m$ are given by
\begin{equation}
 h_m (r):=r(1-x^2)L_m(x) \,\,\,\,\,\, \text{and} \,\,\,\,\,\, g_m (r):=r(1-x^2)^2L_m(x) \,\,\,\,\,\, \text{for} \,\,\,\,\,\, r \in [r_i; r_o] \label{Eq 4.1.1}.
 \end{equation}
where $L_m$ is the Legendre polynomial of degree $m$ and $x := 2r - (1+ \eta)(1- \eta)^{-1}$. Then every $\boldsymbol{u}^j_m$ satisfies both the continuity condition $ \boldsymbol{\nabla}_{\rm c} \cdot \boldsymbol{u}^j_m = 0$ and the boundary conditions by definition, since we have by construction
\begin{equation}
 h_m (r_i) = h_m (r_o) = g_m (r_i) = g_m (r_o) = g_m' (r_i) = g_m' (r_o) = 0.
\end{equation}

The problem is discretized by truncating $U$ at the {\it polynomial resolution} $N\in \mathbb{N}$, i.e. defining $U_N :=  \lbrace \boldsymbol{u}^j_m\rbrace_{m<N}^{j = 1,2}$, and expanding possible solutions $\boldsymbol{u}_\lambda$ to the eigenvalue problem \eqref{Eq 2.1.8} in terms of $U_N$, $\boldsymbol{u}_\lambda := \sum_{m<N , \, j=1,2} a_m^j \boldsymbol{u}_m^j$. Plugging this ansatz into equation \eqref{Eq 2.1.8} and projecting on some $\boldsymbol{u}_l^i$ yields
\begin{equation}
\lambda\sum_{\substack{m<N \\ j=1,2}}  \left\langle \boldsymbol{u}_l^i , \boldsymbol{u}_m^j \right\rangle a_m^j = \sum_{\substack{m<N \\ j=1,2}}  \left\langle\boldsymbol{u}_l^i , \mathcal{L} \boldsymbol{u}_m^j\right\rangle  a_m^j- \underbrace{\left\langle \boldsymbol{u}_l^i , \boldsymbol{\nabla}_{\rm c} q \right\rangle}_{=0} \label{Eq 4.1.2}.
 \end{equation}
The pressure terms vanish due to the boundary and divergence conditions.

Thus, equations \eqref{Eq 4.1.2} for all $l<N$, $i=1,2$, can be written in {the} form of a $2N \times 2N$ generalized eigenvalue problem
 \begin{equation}
\lambda \mathsfbi{G} \boldsymbol{a} =  \mathsfbi{H} \boldsymbol{a} \,\,\,\,\,\, \text{with} \,\,\,\,\,\, \mathsfbi{G} := \left(\left\langle \boldsymbol{u}_l^i , \boldsymbol{u}_m^j \right\rangle\right), \,\,\,\,\,\,  \mathsfbi{H} := \left(\left\langle \boldsymbol{u}_l^i , \mathcal{L} \boldsymbol{u}_m^j \right\rangle\right)\label{Eq 4.1.3}
 \end{equation}
 for the coefficient vector $\boldsymbol{a}:=(a^1_0, \ldots, a_{N-1}^1, a^2_0 , \ldots , a_{N-1}^2)^{\intercal}$ where $\mathsfbi{G}$ and $\mathsfbi{H}$ are $2N \times 2N$-matrices $\mathsfbi{G}$ being Hermitian positive definite \citep{MeseguerMarques2000}.
 \begin{table}
\centering
\begin{tabular}{ccccc} 
& $n=0$, $k=0$ & $n=0$, $k\neq 0$ & $n\neq0$, $k= 0$ & $n\neq0$, $k \neq 0$ \\ 
\hline
\hline
 $\boldsymbol{u}_m^1 :=$ & $\begin{pmatrix} 0 \\ h_m \\ 0  \end{pmatrix}$  & $\begin{pmatrix} 0 \\ h_m \\ 0  \end{pmatrix}$ & $\begin{pmatrix} -\I ng_m \\ \mathcal{D}(rg_m) \\ 0  \end{pmatrix}$ &  $\begin{pmatrix} -\I ng_m \\ \mathcal{D}(rg_m) \\ 0  \end{pmatrix}$ \\ 
 \hline
 $\boldsymbol{u}_m^2 :=$ & $\begin{pmatrix} 0 \\ 0 \\ h_m  \end{pmatrix}$  & $\begin{pmatrix} -\I krg_m \\ 0 \\ \mathcal{D}_+(rg_m)  \end{pmatrix}$ & $\begin{pmatrix} 0 \\ 0 \\ h_m  \end{pmatrix}$ &  $\begin{pmatrix} 0 \\ -\I krh_m \\ \I nh_m  \end{pmatrix}$  \\ 
\end{tabular} \caption{Spectral basis functions for $m \in \mathbb{N}$ used for the discretization of the eigenvalue problem \eqref{Eq 2.1.8} via equations \eqref{Eq 4.1.1} and \eqref{Eq 4.1.2} according to \citet{Avila2007} \label{tab 4.1} }
\end{table}
\newline
\subsection{Computation of transient growth}\label{SS3.2}

Now let $Q:=\lbrace \boldsymbol{q}_1, \ldots \boldsymbol{q}_{2N} \rbrace$ be the eigenfunctions corresponding to the eigenvalues $\boldsymbol{\lambda}:=\lbrace\lambda_1, \ldots \lambda_{2N}\rbrace$ and eigen(-coefficient-)vectors $\lbrace \boldsymbol{a}_1, \ldots \boldsymbol{a}_{2N}\rbrace$ solving the generalized eigenvalue problem \eqref{Eq 4.1.3}. Consider some perturbation expanded in $Q$, i.e. $\boldsymbol{u} = \sum_{i=1}^{2N} b_i \boldsymbol{q}_i$ where $\boldsymbol{b} = (b_1, \ldots, b_{2N})^\intercal$ denotes the time-dependent coefficient vector. Since the $\boldsymbol{q}_i$ are (approximate) solutions to the eigenvalue problem \eqref{Eq 2.1.8}, it follows that
 \begin{equation}
\boldsymbol{b}(t) = \exp\left(\text{diag}(\boldsymbol{\lambda} ) t \right) \boldsymbol{b}(0)\label{Eq 4.1.35}
 \end{equation}
where $\text{diag}(\boldsymbol{\lambda} )$ denotes the diagonal matrix constructed from $\boldsymbol{\lambda}$ and exp is the matrix exponential. Thus the evolution of the perturbations kinetic energy reads
\begin{equation}
\| \boldsymbol{u} \|^2 = \left\langle \boldsymbol{u} , \boldsymbol{u} \right\rangle = \sum_{i,j=1}^{2N} b_i^\ast b_j \left\langle \boldsymbol{q}_i , \boldsymbol{q}_j \right\rangle = \boldsymbol{b}^\ast \mathsfbi{M} \boldsymbol{b} = \| \mathsfbi{F}\boldsymbol{b} \|_2^2 = \| \mathsfbi{F} \exp\left(\text{diag}(\boldsymbol{\lambda} ) t \right)  \boldsymbol{b}(0) \|_2^2 \label{Eq 4.1.4}
 \end{equation}
Here $\mathsfbi{M}$ is the Hermitian positive definite Gramian matrix $ \mathsfbi{M} := \left(\left\langle \boldsymbol{q}_i , \boldsymbol{q}_j \right\rangle\right)$, $\mathsfbi{M} = \mathsfbi{F}^\ast \mathsfbi{F}$ a Cholesky decomposition and $\| \cdot \|_2$ denotes the standard 2-norm on $\mathbb{C}^{2N}$. Hence, the maximum transient growth at time $t\geq 0$ is given by \citep[see][]{Meseguer2002}
\begin{eqnarray}
G(t) &=& \sup_{\|\boldsymbol{u}(0)\|=1} \| \boldsymbol{u}(t) \|^2 = \sup_{\|\mathsfbi{F}\boldsymbol{b}\|_2=1} \| \mathsfbi{F}\exp\left(\text{diag}(\boldsymbol{\lambda} )t \right) \boldsymbol{b} \|_2^2 \nonumber \\ & \stackrel{\boldsymbol{v}  = \mathsfbi{F}\boldsymbol{b}}{=} & \sup_{\|\boldsymbol{v}\|_2=1} \| \mathsfbi{F}\exp\left(\text{diag}(\boldsymbol{\lambda} )t \right) \mathsfbi{F}^{-1} \boldsymbol{v} \|_2^2 = \|\mathsfbi{F} \exp\left( \text{diag}(\boldsymbol{\lambda} ) t\right) \mathsfbi{F}^{-1} \|^2_2. \label{Eq 4.1.5}
 \end{eqnarray}
 
So $G(t)$ is equal to the squared maximum singular value $\sigma_0^2$ of $\mathsfbi{F} \exp\left( \text{diag}(\boldsymbol{\lambda} t) \right) \mathsfbi{F}^{-1}$. Moreover, if $\boldsymbol{v}_0$ denotes the corresponding right-singular vector, $\mathsfbi{F} \boldsymbol{v}_0$ is the initial $Q$-coefficient vector of a perturbation that attains optimal transient growth at time $t$. By means of singular value decomposition, we thus obtain both maximum transient growth $G(t)$ and corresponding perturbations in the finite-dimensional subspace spanned by $Q$. This yields a lower bound to the maximum attainable by arbitrary initial conditions in $\mathbb{V}$. As discussed in $\S$\ref{SS4.2} we find convergence of this estimate to the total maximum.
\newline
\subsection{Outline of the code}\label{SS3.3}
By definition of $U_N$ only integrals over polynomial functions have to be evaluated in order to calculate $\mathsfbi{G}$ and $\mathsfbi{H}$. Hence, these are computed exactly using Gauss--Legendre quadrature with Gauss-Lobatto collocation points of degree $M$, where $M \geq N + 6$ \citep[see][pp. 69 ff.]{SpectralMethods2006}. Moreover, the derivatives in the operator $\mathcal{L}$ are implemented by means of the corresponding differentiation matrices given in \citet[p. 76]{SpectralMethods2006}.
%

The code used in this work is based on the scientific computing package \texttt{Scipy} for the interactive language \texttt{Python}. The linear algebra algorithms are provided by the package \texttt{Scipy.Linalg} based on the standard \texttt{ATLAS}, \texttt{LAPACK} and \texttt{BLAS} implementations.

The optimization of $G$ in the time coordinate $t\in[0; t_{\text{cut}}]$ and in the continuous wavenumber $k\in[0; k_{\text{cut}}]$ are performed via the \texttt{Scipy.Optimize} implementation of Brent's method \citep[for details see][sec. 9.3]{NumRecipes}. With respect to the discrete wavenumber $n\in \left\lbrace 0, 1, \ldots , n_{\text{cut}} \right\rbrace$, $G$ is optimized by brute force. If the optimal transient growth is found at the upper boundary of the considered domains, i.e. for $t = t_{\text{cut}}$, $k = k_{\text{cut}}$ or $n = n_{\text{cut}}$ the respective intervals are enlarged in subsequent steps until a local maximum is located in their interior.
%
%
%
%
%
%
%
%
\section{Numerical issues} \label{S4}
In this section, the performance of the numerical implementation presented in section \ref{S3} is tested by comparison to results in the literature. Furthermore, eigenvalue and transient growth convergence are studied for test cases in order to justify the choice of polynomial resolution $N$ used to obtain the numerical results in $\S$\ref{S5}. We find that the optimal transient growth may converge without the Y-shaped spectrum being properly resolved. This observation suggests a minor significance of the spectrum in transient growth computations, disagreeing with the conclusions drawn by \citet{ReddyHenningson1993} for channel flows.
\newline
\subsection{Eigenvalue decomposition}\label{SS4.1}

Our discretization of the eigenvalue problem \eqref{Eq 2.1.8} has been tested against the results on eigenvalue-critical Reynolds numbers presented in \citet[table 2]{Krueger1966} as well as by replication of the plotted spectra given by \citet[fig. 3a-d]{GebhardtGrossmann1993}. Agreement within the respective accuracies has been found. Additionally, we have compared our Galerkin method to the Petrov-Galerkin scheme of \citet{Avila2007}. No significant deviations are found between the converged spectra.

\begin{figure}
    \centering
    \subfigure[$\Rey=8000$]
    {\includegraphics[width=0.495\textwidth]{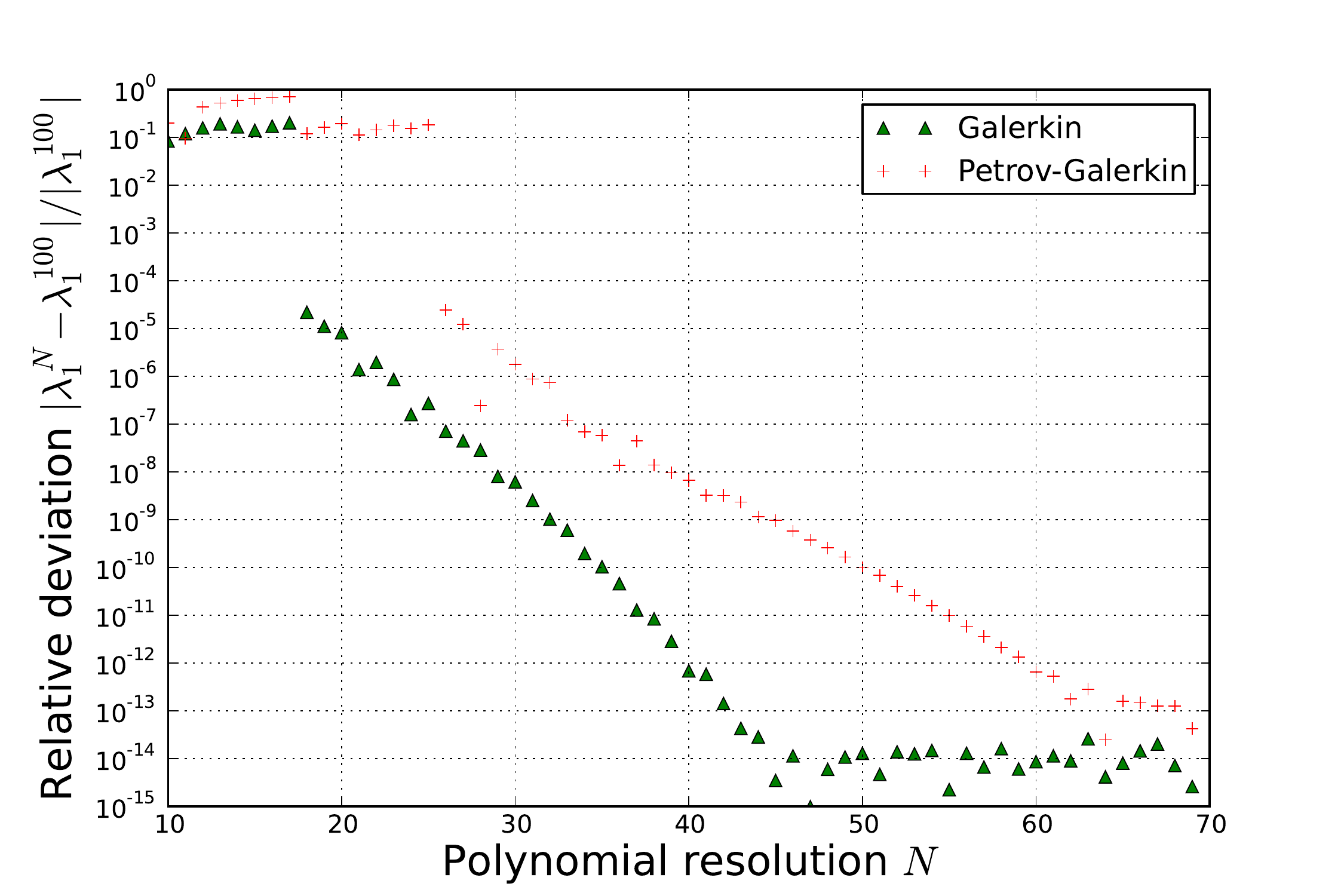}\label{fig 5.1.1a}}
    \hfill
    \subfigure[$\Rey=128000$]
    {\includegraphics[width=0.495\textwidth]{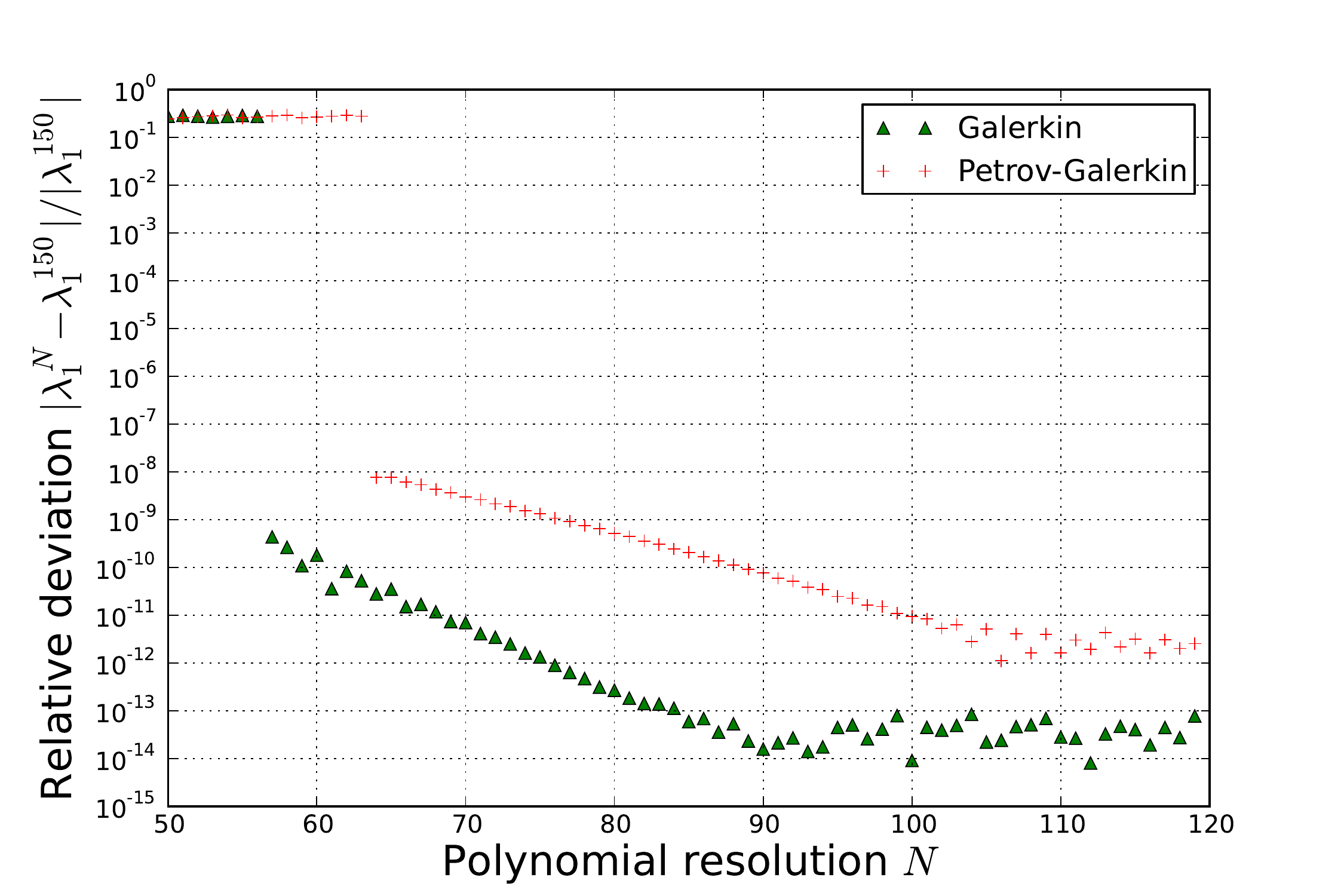}\label{fig 5.1.1b}}
    \caption{Convergence of the least stable eigenvalue $\lambda_1$ of $\mathcal{L}$ for $R_\Omega = -2$, $\eta = 0.5$, $n=5$, $k=1$ and $\Rey = 8000$ (a)), $\Rey = 128000$ (b)) computed using our Galerkin method (triangles) and the Petrov-Galerkin scheme of \citet{Avila2007} (crosses); $\lambda_1^N$ denotes the approximation to $\lambda_1$ computed using $N$ Legendre {or} Chebyshev polynomials; $|\lambda_1^{N}-\lambda_1^{N_\text{ref}}| /|\lambda_1^{N_\text{ref}}|$ is the relative deviation of $\lambda_1^N$ from the converged result $\lambda_1^{N_\text{ref}}$}
    \label{fig 5.1.1}
 \end{figure}

For these methods we study the convergence of the approximated least stable eigenvalue $\lambda_1^{N}$ as the number $N$ of Legendre {or} Chebyshev polynomials is increased. In figure \ref{fig 5.1.1} the relative errors $|\lambda_1^{N}-\lambda_1^{N_\text{ref}}| |\lambda_1^{N_\text{ref}}|^{-1} $ compared to (converged) reference values $\lambda_1^{N_\text{ref}}$ are plotted against $N$. The test parameters are $R_\Omega = -2$, $\eta = 0.5$, $n=5$ and $k=1$ at shear Reynolds numbers $\Rey = 8000$ for figure \ref{fig 5.1.1a} and $\Rey = 128000$ for figure \ref{fig 5.1.1b}. 

The plots in figure \ref{fig 5.1.1} show plateaus of non-convergence for low $N$, which are due to the difficulty of identifying the respective eigenvalue in a non-converged spectrum. For moderate {resolutions ($N \in \left[20; 45  \right]$ in figure \ref{fig 5.1.1a} and $N \in \left[  45; 90  \right]$ in \ref{fig 5.1.1b})} spectral accuracy, i.e. exponential convergence rates, is observed for both methods.
Notably however, the convergence turns out to be significantly quicker {for the Legendre-polynomial-based Galerkin method} presented in this work: spectral accuracy is attained using significantly fewer polynomials and the limiting machine precision is reached already for $N = 43$ ($\Rey = 8000$) and $N = 83$ ($\Rey = 128000$) compared to $N = 62$ and $N = 104${, respectively,} {in the case of the} Petrov-Galerkin scheme (see figure \ref{fig 5.1.1}).

The required resolution $N$ for convergence grows with the shear Reynolds number $\Rey$ and -- much more significantly -- as soon as subsequent, more stable eigenvalues are considered. In fact, it turns out to be numerically impossible to resolve significant parts of the eigenvalue spectrum for $\Rey \geq O(10^5)$. This also affects the computation of transient growth discussed in the next subsection.

\subsection{Computation of Transient Growth}\label{SS4.2}
In table \ref{tab 5.1} our results concerning the optimal transient growth $G_{\max}:=\sup_{n,k,t}G(t)$ for $\eta = 0.881$ and the corresponding optimal wavenumbers $n_{\max}$ are compared to the numerical data of \citet[table 1]{Meseguer2002}. The values of $k_{\max}$ and $G_{\max}$ differ by less than $\unit[0.3]{\%}$.
\begin{table}
\centering
\begin{tabular}{cccccccccc}
 & && \multicolumn{3}{c}{\citet{Meseguer2002}} && \multicolumn{3}{c}{Present work ($N=50$)} \\
  \hline
$\Rey_i$ & $\Rey_o$ && $n_{\max}$ & $k_{\max}$ & $G_{\max}$  && $n_{\max}$ & $k_{\max}$ & $G_{\max}$ \\ 
\hline
\hline
$591$ & $-2588$ && $10$ & $1.994$ & $71.36$ && $10$ & $1.997$ & $71.58$  \\
$523$ & $-2975$ && $11$ & $1.996$ & $71.58$ && $11$ & $1.998$ & $71.81$ \\
$473$ & $-3213$ && $11$ & $1.920$ & $71.64$ && $11$ & $1.922$ & $71.87$ \\
$405$ & $-3510$ && $11$ & $1.839$ & $71.75$ && $11$ & $1.841$ & $71.99$ \\ 
\end{tabular} \caption{Optimal transient growth $G_{\max}:=\sup_{n,k,t}G(t)$ according to \citet[Table 1]{Meseguer2002} and {present results}; parameters are $\eta = 0.881$ and $N=50$; $n_{\max}$ and $k_{\max}$ denote the azimuthal and axial wavenumbers which attain optimal transient growth $G_{\max}$ \label{tab 5.1}}
\end{table}

The convergence of the maximum transient growth $G$ shows remarkable characteristics which partly {contradict} the significance of the linearized Navier-Stokes operator's spectrum for such computations claimed, for example, by \citet{ReddyHenningson1993}. 

\begin{figure}
    \centering
    \subfigure[$N=5$]
    {\includegraphics[width=0.325\textwidth]{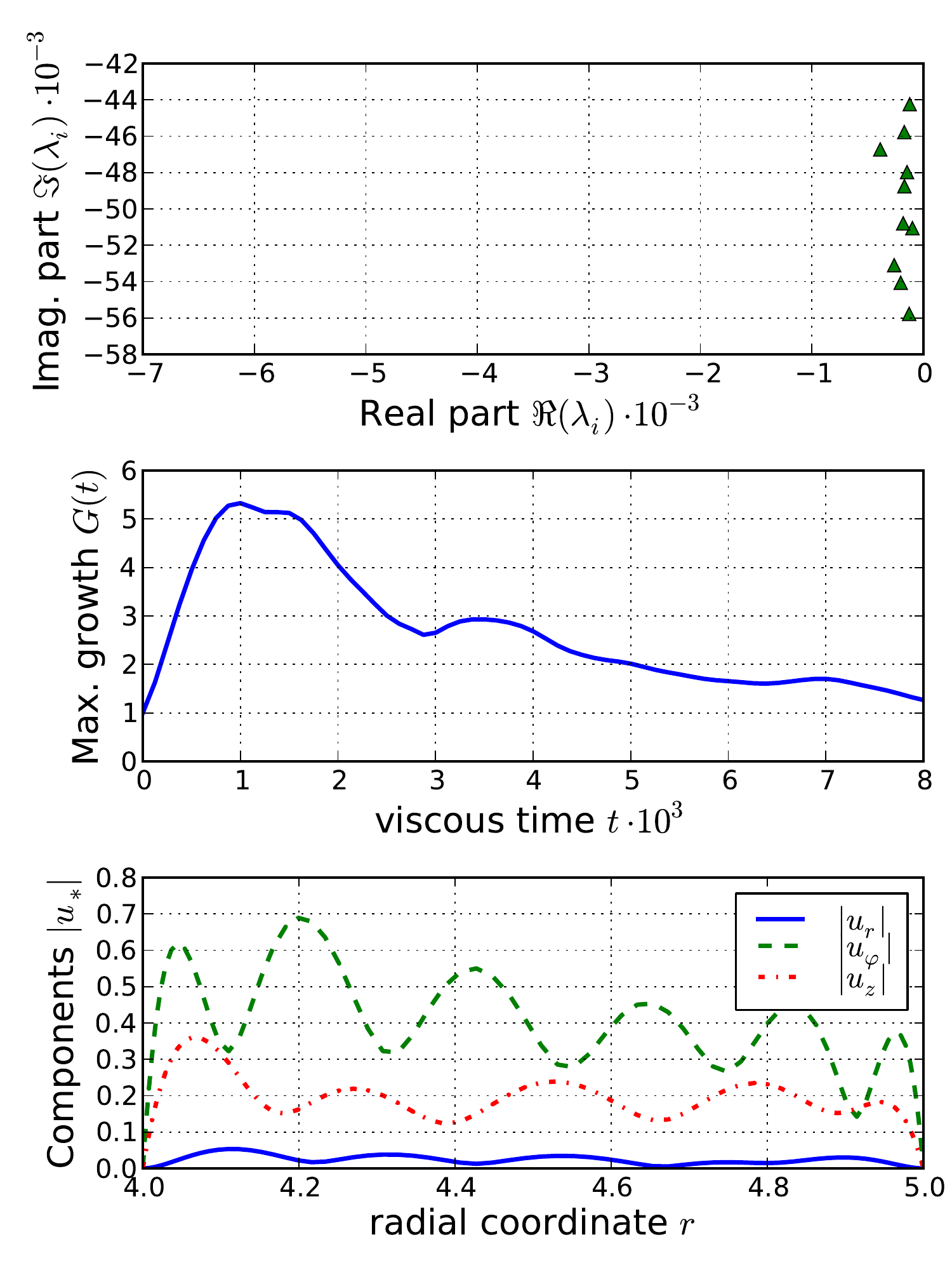}\label{fig 5.2.1a}}
    \hfill
   \subfigure[$N=15$]
    {\includegraphics[width=0.325\textwidth]{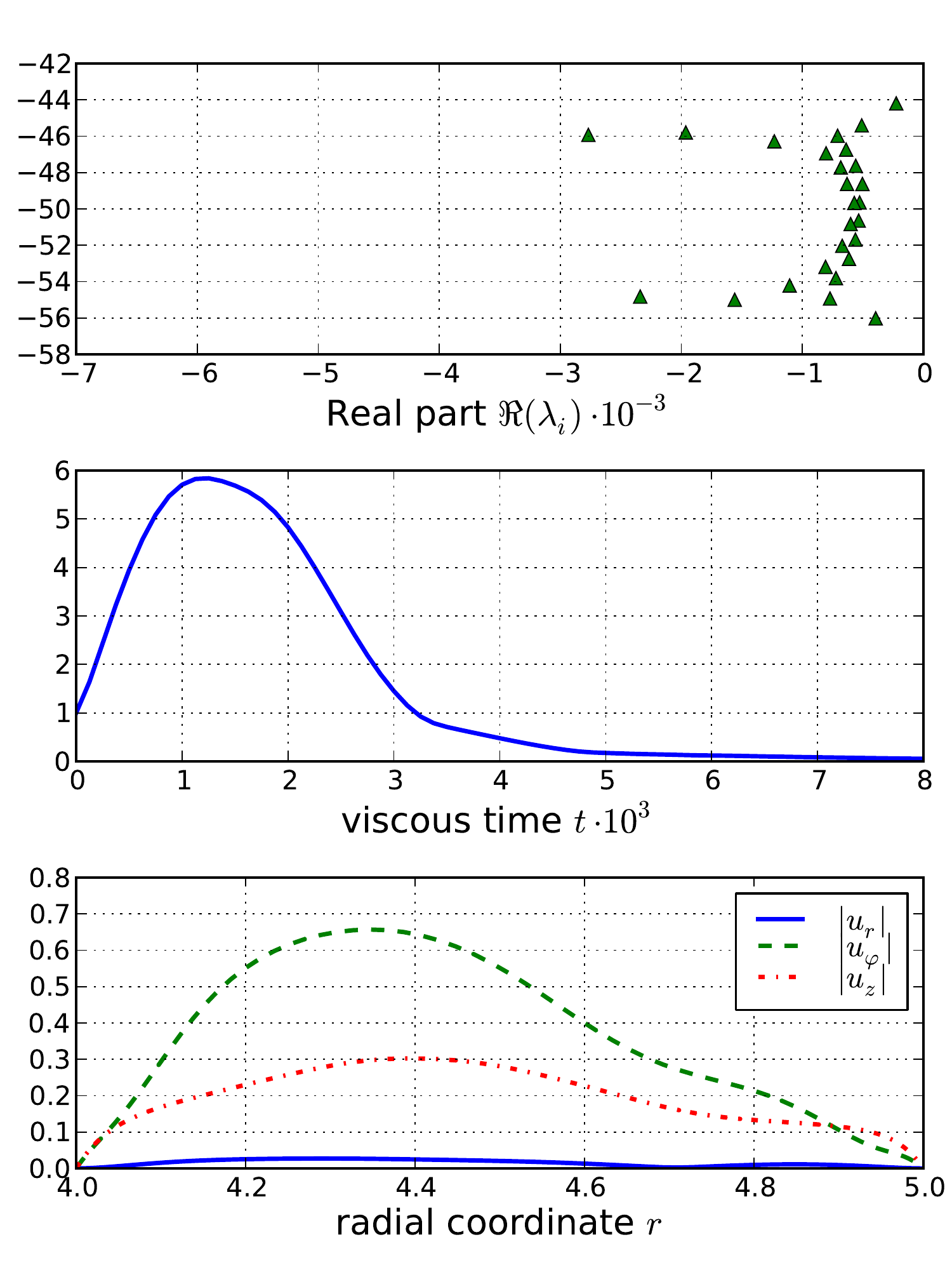}\label{fig 5.2.1b}}
    \hfill
   \subfigure[$N=50$]
    {\includegraphics[width=0.325\textwidth]{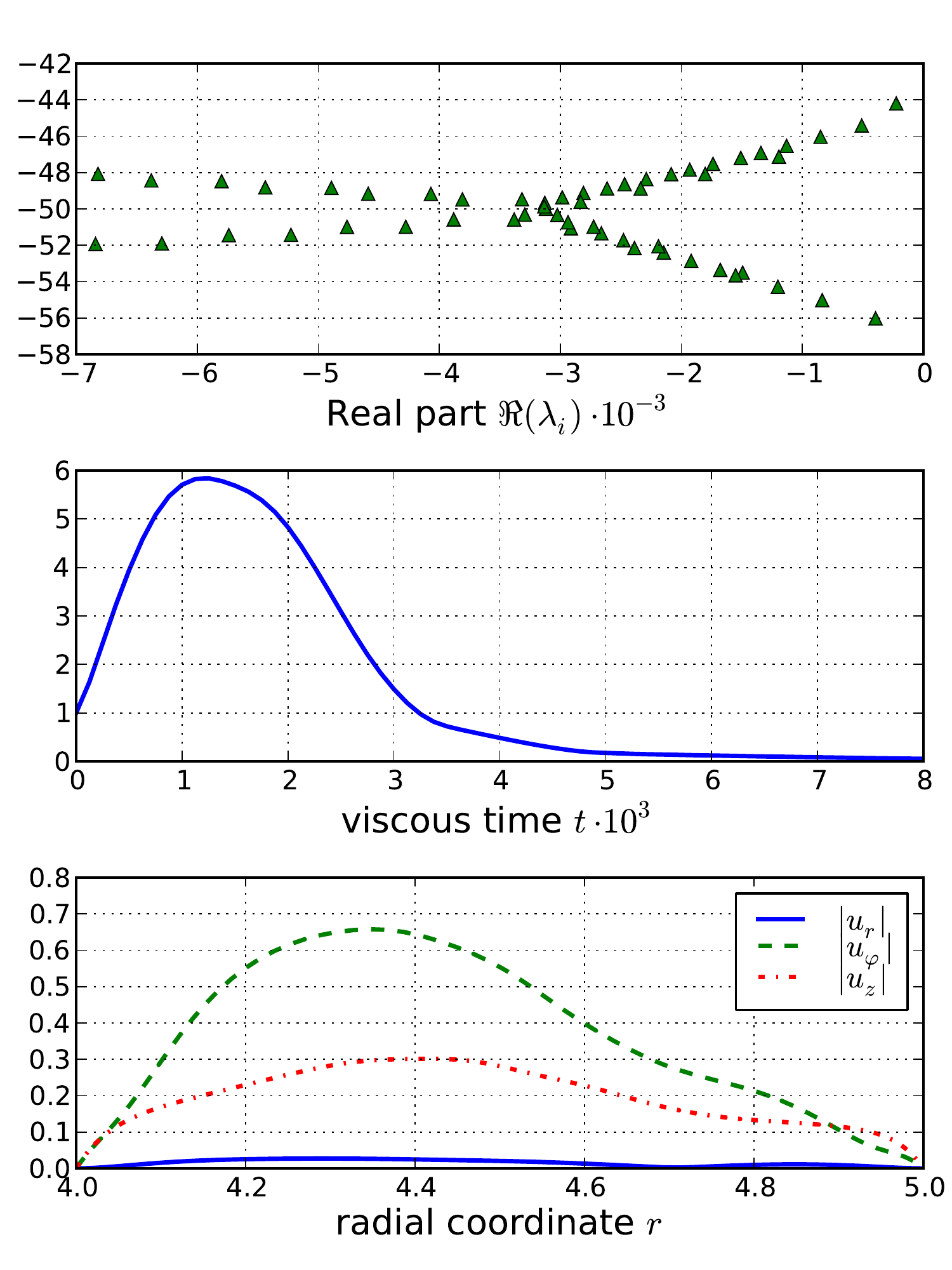}\label{fig 5.2.1c}}
    \caption{Eigenvalues $\lambda_i$ (top), time dependent maximum transient growth $G(t)$ (middle) and modulus $|u_r(r)|$, $|u_\varphi(r)|$ and $|u_z(r)|$ of the radial, azimuthal and axial components of the perturbation $\boldsymbol{u}$ attaining optimal growth $ \sup_{t \geq0} G(t)$ (bottom) approximated by different resolutions $N$; parameters: $\Rey=10000$, $R_\Omega = -2$, $\eta = 0.8$, $n=5$ and $k=1$}
    \label{fig 5.2.1}
 \end{figure}

These features are discussed with reference to the example displayed in figure \ref{fig 5.2.1}: for three different resolutions $N\in \left\lbrace 5, 15, 50 \right\rbrace$ (corresponding to figures \ref{fig 5.2.1a}, \ref{fig 5.2.1b} and \ref{fig 5.2.1c}), the eigenvalues (top), the evolution of the maximum transient growth $G(t)$ (middle) and the moduli of the components $|u_r|$, $|u_\varphi|$ and $|u_z|$ of the corresponding optimal perturbation $\boldsymbol{u}(0)$ are plotted for comparison. The example parameters are $\Rey = 10000$, $R_\Omega = -2.0$, $\eta = 0.8$, $n = 5$ and $k =1$.

A few aspects are noteworthy. Around its maximum $G$ is already surprisingly well approximated by only $N = 5$ Legendre polynomials, whereas the optimal perturbation is far from its actual shape (see figure \ref{fig 5.2.1a}). For $N = 15$ (figure \ref{fig 5.2.1b}) the curve $\left\lbrace (t, G(t)) \right\rbrace $ is converged within an error $\leq 1 \%$ while its maximum is even approximated up to $\approx 0.01 \, \%$. Likewise, the optimal perturbation $\boldsymbol{u}(0)$ is practically converged. At the same time the characteristic Y-like structure of the eigenvalue spectrum \citep[cf.][]
{GebhardtGrossmann1993} is by no means well resolved for $N=15$ not to mention $N=5$ (top). In fact, it takes as many as $N = 50$ polynomials for convergence of the two meeting branches (see figure \ref{fig 5.2.1c}). However, this does not seem to affect the transient growth quantities -- even though the converged spectrum in figure \ref{fig 5.2.1c} (top) is even much more stable as a whole than its approximation for $N=15$ (figure \ref{fig 5.2.1b}).

In contrast to these observations \citet{ReddyHenningson1993} stress the significance of the two eigenvalue branches and especially their meeting point for transient growth in channel flows. As for Taylor--Couette flow, this is only confirmed if the Y structure is resolved in the first place. This turns out not to be necessary, which is a lucky circumstance in two respects. On the one hand, the two branches consist of $O(\Rey^{\alpha})$ discrete eigenvalues for $\alpha \approx \frac{1}{2}$, rendering their convergence numerically infeasible for $\Rey\geq O(10^5)$. This agglomeration of eigenvalues can be explained by the dominance of the $O(\Rey)$ convective multiplicative terms in the linearized Taylor--Couette operator $\mathcal{L}$ over the viscous differential contributions in
the limit $\Rey \rightarrow \infty$ (see equations \eqref{Eq 2.1.6a} and \eqref{Eq 2.1.6b}). The asymptotic degeneracy into a pure multiplication operator, i.e. a mapping $\boldsymbol u \mapsto   \mathsfbi{A} \boldsymbol u$ with a continuous, matrix-valued function $\mathsfbi{A}: [r_i;\,r_o] \to \mathbb{R}^{3\times3}$, corresponds to a transition from discrete eigenvalues to a continuous spectrum.


On the other hand the standard Cholesky decomposition of the matrix $\mathsfbi{M}$ (see $\S$\ref{SS3.2}) tends to fail at large $\Rey$ if the eigenvalue spectrum is over-resolved. In the example shown in figure \ref{fig 5.2.1} this happens for $N \geq 51$ -- just as the crucial meeting point is resolved. Accordingly, one might expect to miss a sudden jump in the maximum transient growth $G$ if the method breaks down precisely at this point. Note, however, that no such {discontinuity} is observed in those cases where the intersection can still be resolved, i.e. for smaller $\Rey$.

We may thus conclude that the transient growth of the linearized Taylor--Couette operator $\mathcal{L}$ is already converged while its approximated spectrum is still far from its natural shape. Startling at first glance, this is  yet another manifestation of transient growth's non-modal nature: the non-eigendirections are those of significance.

Nevertheless, numerical artifacts in {the} form of spurious unstable eigenvalues have to be avoided by choosing sufficiently high resolutions $N$. However, $N$ must not be too large either in order to keep the Cholesky decomposition stable (although preconditioning or more stable algorithms such as the one presented by \citet{OgitaOishi2012} might be another alternative). For a given set of parameters $\eta$, $\Rey$, $R_\Omega$ it turns out that resolving the transient growth peak for optimal wavenumbers $n= n_{\max}$, $k= k_{\max}$ tends to require the highest resolutions. Moreover, the necessary $N$ are {mostly} independent of $R_\Omega$ and at least of the same magnitude for different $\eta$. Here greater curvature, i.e. $\eta \rightarrow 0$, results in slower convergence. Consequently, for
practical computations, suitable resolutions $N_{Re}$ are determined for different ranges of $\Rey$ by the convergence of (computationally challenging) test cases, more precisely less than $0.3\%$ deviation in the optimal transient growth for $\eta = 0.2$ and $N\in [N_{Re}-3; N_{Re}]$.
\begin{table}
\centering
\begin{tabular}{ccccccccccc}
Maximum $\Rey$ && $8000$ & $16000$ & $32000$ & $64000$ & $128000$ & $256000$ & $512000$ & $1024000$ & $2048000$ \\
 \hline
Resolution $N_{Re} $ && $31$ & $38$ & $47$ &  $58$  & $71$ & $88$  & $107$  & $131$ & $159$ \\ 
\end{tabular} \caption{Canonical resolutions $N_{Re}$ for the computation of optimal transient growth $G_{\max}$ for $\Rey$ below the given upper bounds; determined by the convergence of $G_{\max}$ for $\eta = 0.2$ \label{tab 5.2.2.1} }
\end{table}

For greater $\eta$ lower resolutions $N$ may be sufficient and greater Reynolds numbers than $\Rey = 2048000$ might be resolvable. However, universal convergence for any parameters $R_\Omega$, $\eta$, $n$ and $k$ within about $1 \%$ may be assumed if $N$ is chosen according to the resulting canonical resolutions $N_{Re}$ given in table \ref{tab 5.2.2.1}. They are found to approximately follow a power law of the form $N_{Re} = N_0 \Rey^{\alpha}$ with $N_0 = 2.28 \pm 0.06$ and $\alpha = 0.293 \pm 0.002$.

Starting from these, $N$ is temporarily reduced in subsequent steps whenever the Cholesky decomposition fails and temporarily increased if unstable eigenvalues occur in order to identify possible numerical artifacts. In the case of converged unstable eigenvalues the computation of the matrix $\mathsfbi{M}$ and thus of the transient growth is confined to the stable eigenmodes in $Q$ in agreement with the analysis of \citet{Meseguer2002}.

These computation guidelines have been applied to obtain the numerical results presented in $\S$\ref{S5}.

%
%
%
%
%
%
%
%
\section{Numerical results} \label{S5}

In this section the numerical results concerning stability and transient growth in Taylor--Couette flows are presented.
\newline
\subsection{Optimal transient growth in various regimes}\label{SS5.2}
According to the numerical strategy discussed in $\S \S$ \ref{S3} and \ref{SS4.2}, the optimal transient growth $G_{\max} = \sup_{n,k,t} G(t)$ is computed for logarithmically equidistant shear Reynolds numbers $250 \leq \Rey \leq 2 \cdot 10^6$ and $\eta \in \left\lbrace 0.2, 0.5, 0.8 \right\rbrace$.

By studying test cases we find that $t\in [0; \tau_0]$ with $\tau_0 = \frac{2\pi}{Re^{\alpha}(1-\eta)}$ and $\alpha = 0.85$ is {a} suitable choice to determine the transient growth maximum in time for all considered parameter regimes. Optimization in the wavenumbers is carried out by default in the range $n \in \left\lbrace 0,1 \ldots, 8\right\rbrace$ and {$ 0 \leq k \leq 5 $}. Additionally, as discussed in $\S$\ref{SS3.3} the ranges for $t$, $n$ and $k$ are enlarged whenever the optimization terminates near one of the upper bounds.

\begin{figure}
    \centering
	\includegraphics[width=0.65\textwidth]{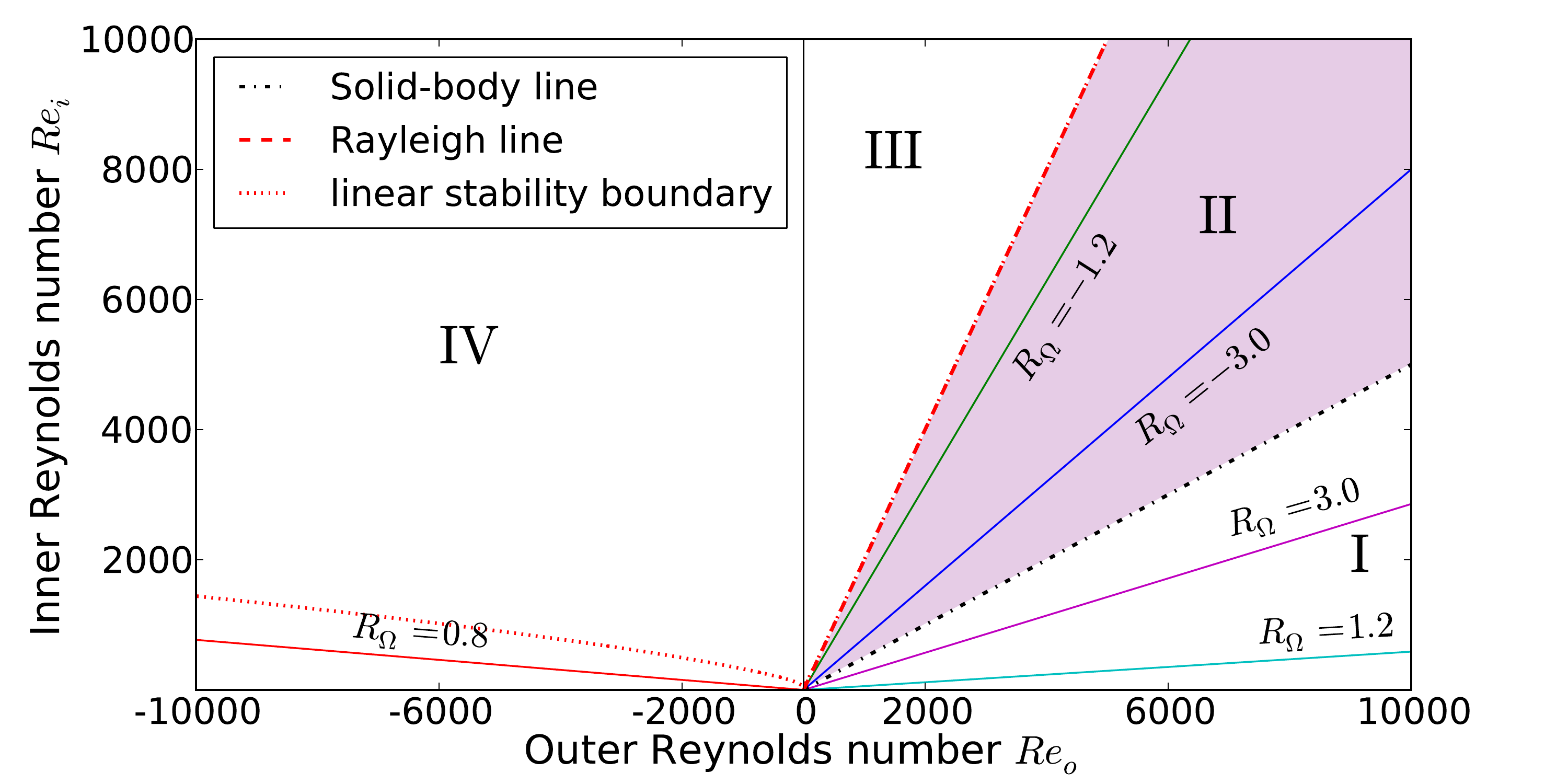}
	\caption{Representative lines in the $\Rey_i$-$\Rey_o$ plane in the case $\eta = 0.5$ for which the optimal transient growth $G_{\max}$, and corresponding optimal axial wavenumbers $k_{\max}$, are plotted in figures \ref{fig 6.2.2a} and \ref{fig 6.2.2b}, respectively. The quasi-Keplerian regime II is shaded for orientation.} \label{fig 6.2.1}
 \end{figure}

\begin{figure}
   \centering
   \subfigure[Optimal transient growth $G_{\max}$]
   {\includegraphics[width=0.48\textwidth]{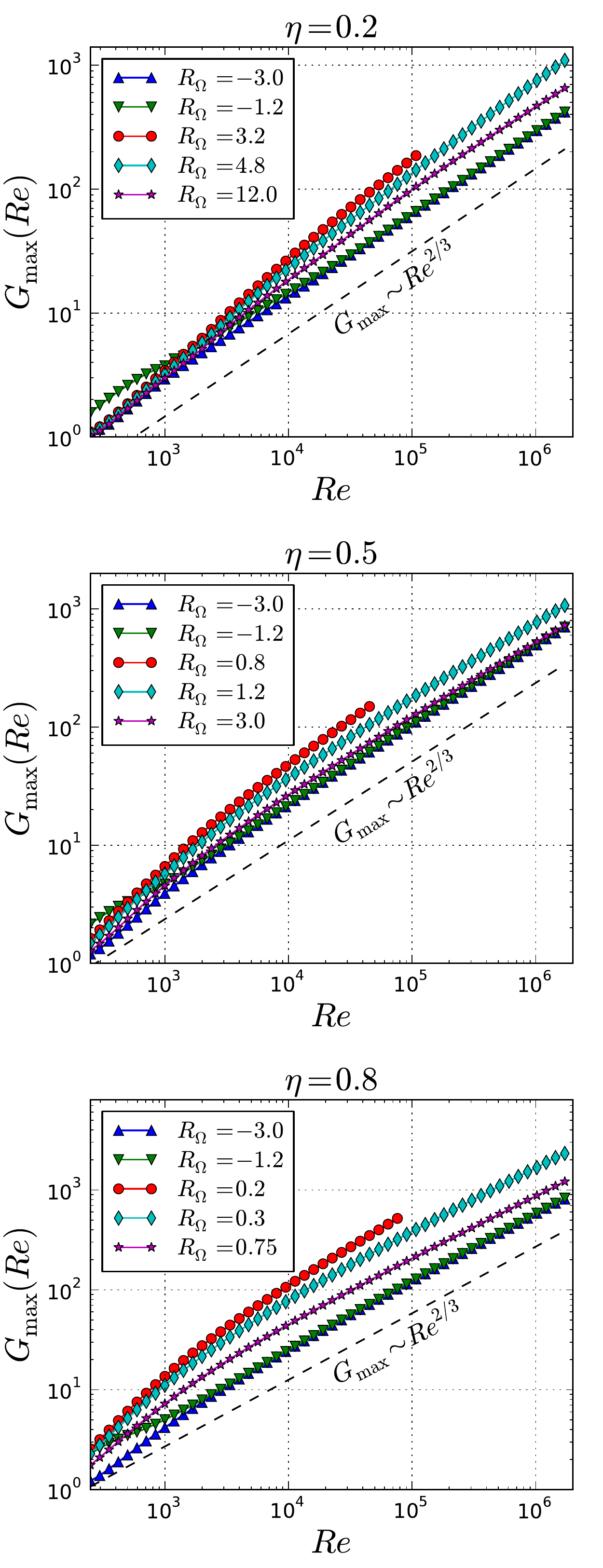}\label{fig 6.2.2a}}
   \hfill
   \subfigure[Optimal axial wavenumber $k_{\max}$]
   {\includegraphics[width=0.48\textwidth]{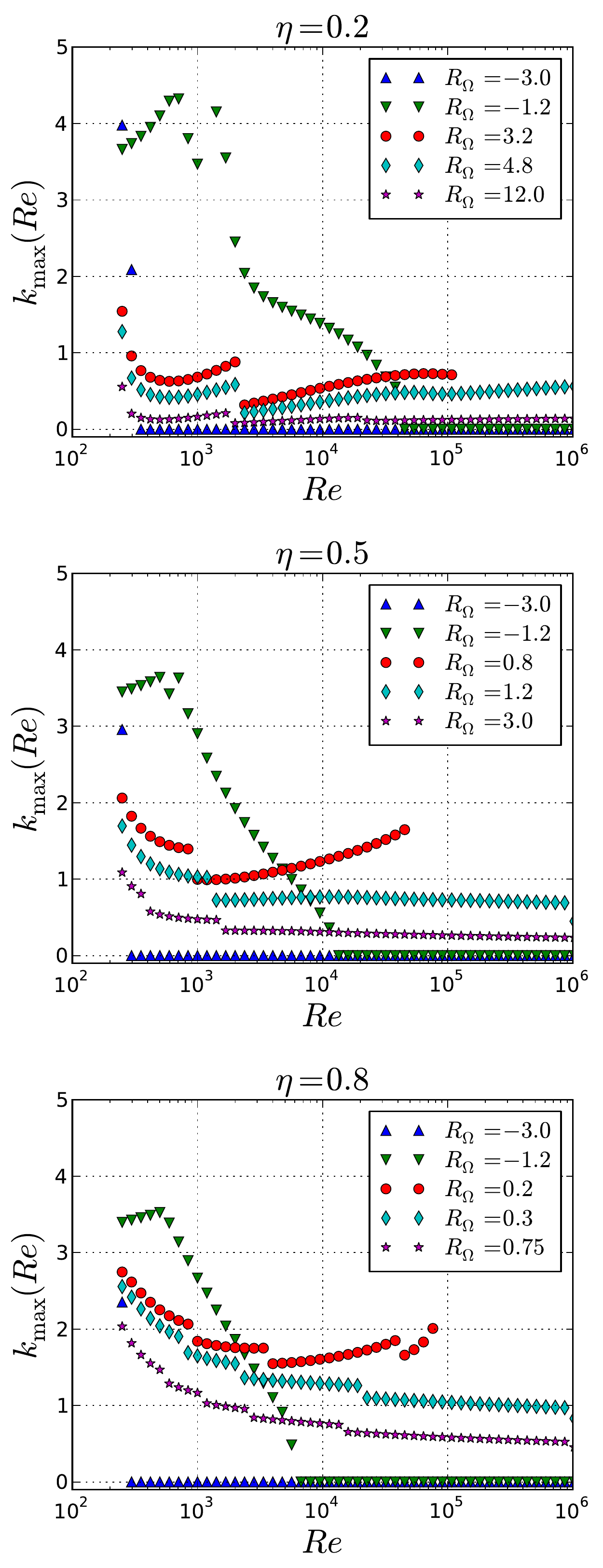}\label{fig 6.2.2b}}
   \caption{Numerical results concerning (a) optimal transient growth $G_{\max}$ and (b) {respective} optimal axial wavenumbers $k_{\max}$ against the shear Reynolds number $\Rey$ for different $\eta$ and $R_\Omega \in \lbrace -3, -1.2, 0.8\frac{1-\eta}{\eta}, 1.2\frac{1-\eta}{\eta}, 3\frac{1-\eta}{\eta} \rbrace$ corresponding to the lines in figure \ref{fig 6.2.1} in regimes I, II and IV; discontinuities in (b) are due to changes in the discrete optimal azimuthal wavenumber $n_{\max}$; the asymptotic slopes in (a) show a common scaling of $G_{\max} \sim \Rey^\alpha$ for $\alpha \approx \frac{2}{3}$ for high Reynolds numbers $\Rey \rightarrow \infty$ (dashed line)}
   \label{fig 6.2.2}
\end{figure}

By the choice of rotation numbers $R_\Omega$ the linearly stable regimes I and II are parametrized considering  $R_\Omega \in [\frac{1- \eta}{\eta} ; 10\frac{1- \eta}{\eta}]$ (I) and $R_\Omega \in [-10 ; -1.1]$ (II) (see table \ref{tab 1.1}). Furthermore, transient growth is studied in the counter-rotating regime IV {near} the $\Rey_i = 0$ line in the $\Rey_i$-$\Rey_o$ {left quadrant} by choosing $R_\Omega \in [0.1 \frac{1- \eta}{\eta} ; 0.9  \frac{1- \eta}{\eta}]$. For a global overview the results for $R_\Omega \in \lbrace -3, -1.2, 0.8\frac{1- \eta}{\eta}, 1.2\frac{1- \eta}{\eta}, 3\frac{1- \eta}{\eta} \rbrace$ are presented. The lines in the $\Rey_i$-$\Rey_o$ plane defined by this choice for $\eta = 0.5$ are visualized in figure \ref{fig 6.2.1} for orientation, along with the numerically computed (viscous) linear stability boundary. Figure \ref{fig 6.2.2} shows the optimized transient growth $G_{\max}$ {and} the corresponding optimal axial wavenumber $k_{\max}${, respectively,} 
against $\Rey$ for the considered parameter sets. 

The most prominent feature in the double-logarithmic plots of figure \ref{fig 6.2.2a} are the nearly identical asymptotic slopes of the lines in the linearly stable regimes for $R_\Omega \in \left\lbrace -3, -1.2, 1.2, 3 \right\rbrace$ showing a characteristic power law $G_{\max} \sim \Rey^\alpha$ with $\alpha \approx \frac{2}{3} \pm \unit[7]{ \%}$ (compare dashed line in figure \ref{fig 6.2.2a}). Notably, even in the Rayleigh-unstable counter-rotating regime IV (circles in figure \ref{fig 6.2.2a}), $G_{\max}$ seems to approach this scaling as long as the computation is not destabilized by dominant linear instability. In fact, 
for constant $\Rey$ the energy amplifications $G_{\max}(\Rey)$ in the different regimes differ only by $O(1)$ factors and not -- as possibly expected -- by orders of magnitude. Within the linearly stable regimes I and II these deviations are most distinct in the vicinity of the Rayleigh line {and} the boundary to regime IV where larger amplifications occur.

Hence, the numerical results suggest that optimal transient growth in linearly stable Taylor--Couette flows roughly follows a common scaling $G_{\max} \sim \Rey^{\frac{2}{3}}$ for $\Rey \rightarrow \infty$. Note that this scaling result is in perfect agreement with those by \cite{Yecko2004} obtained for Keplerian flows at fixed $R_\Omega = 1.5$ in rotating plane Couette geometry.
\newline
\subsection{Optimal axial wavenumber}\label{SS5.3}

Beyond the magnitude of transient growth studied in $\S$\ref{SS5.2} the spatial structure of the optimal perturbations $\boldsymbol{u}_{\max}$ is of great interest. The latter is determined by the optimal axial and azimuthal wavenumbers $k_{\max}$ and $n_{\max}$ which attain the optimal transient growth $G_{\max}$ shown in figure \ref{fig 6.2.2a}. The $k_{\max}$ are plotted in figure \ref{fig 6.2.2b} with logarithmic horizontal axes. Note the discontinuities of the curves whenever the (discrete) optimal azimuthal wavenumbers $n_{\max}$ changes.

The plots reveal a characteristic {quasi-two-dimensional, \emph{columnar} structure} of the optimal perturbations in regime II (also observed by \cite{Yecko2004}): for $\Rey > \Rey_0$ the optimal transient growth $G_{\max}(\Rey)$ is consistently attained by axially independent perturbations, i.e. $k_{\max} = 0$ (compare $R_\Omega = -3$ and $R_\Omega = -1.2$ in figure \ref{fig 6.2.2b}). The transition to $k_{\max} = 0$ typically occurs already for Reynolds numbers as small as $\Rey_0 = O(10^3)$. Only near the Rayleigh line -- that is, for $-1.2 \leq R_\Omega < -1 $ -- is a sharp divergence of $\Rey_0$ for $R_\Omega \rightarrow -1$ is observed. Here $k_{\max} \approx 1$ holds up to the greatest studied shear Reynolds numbers $\Rey = O(10^6)$.

 \begin{figure}
     \begin{center}
 	\includegraphics[width=\textwidth]{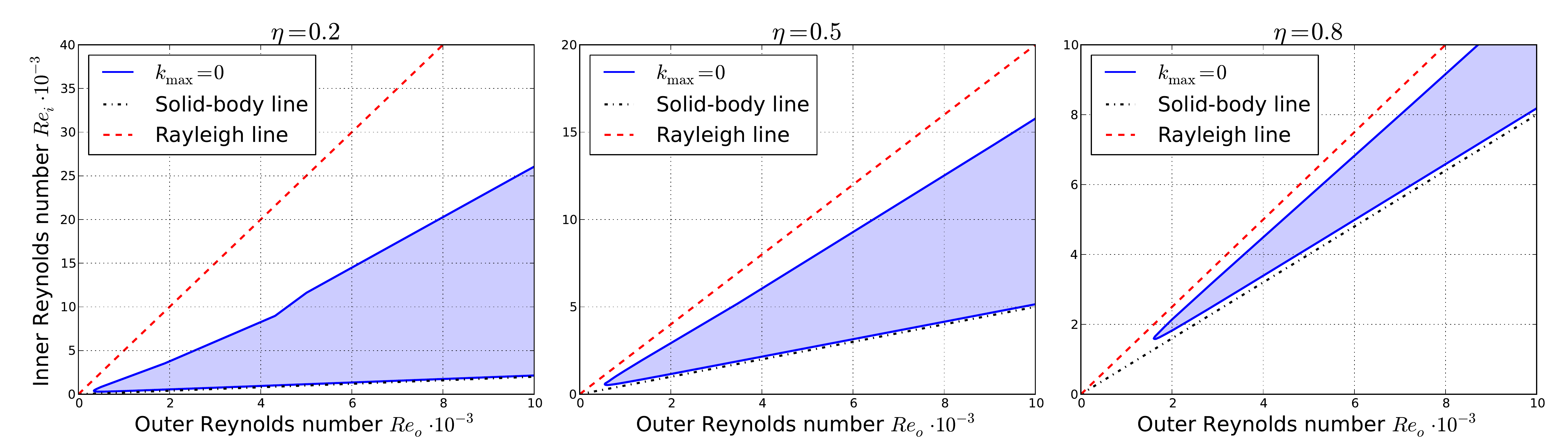}
 	\caption{Domain of the quasi-Keplerian regime (II) where the optimal perturbation is axially independent (blue shading), i.e. $k_{\max}=0$, for $\eta = 0.2, 0.5, 0.8$; the boundary line (blue solid line) has been determined by a bisecting algorithm with relative accuracy $\epsilon = 10^{-2}$; in the white regions between Rayleigh line (red) and solid-body line (black) $k_{\max}\neq0$} \label{fig 6.3.2}
     \end{center}
 \end{figure}

While $k_{\max} = 0$ is only obtained in the quasi-Keplerian regime (II), in regime I (represented by $R_\Omega = 3, 1.2$) $k_{\max}$ seems to decay (slowly) to zero for $\Rey \rightarrow \infty$. At least weak axial dependence $k_{\max} \lesssim 1$  is observed for $\Rey\geq  O(10^4)$ in these flows. Once again, the asymptotic decay $k_{\max} \rightarrow 0$ is most distinct near the solid-body line $R_\Omega \rightarrow \infty$ and is lost near the transition to counter-rotation at $\Rey_i = 0$. Here an almost constant optimal wavenumber $k_{\max} = O(1)$ is  observed.

\begin{figure}
    \centering
	\includegraphics[width=0.7\textwidth]{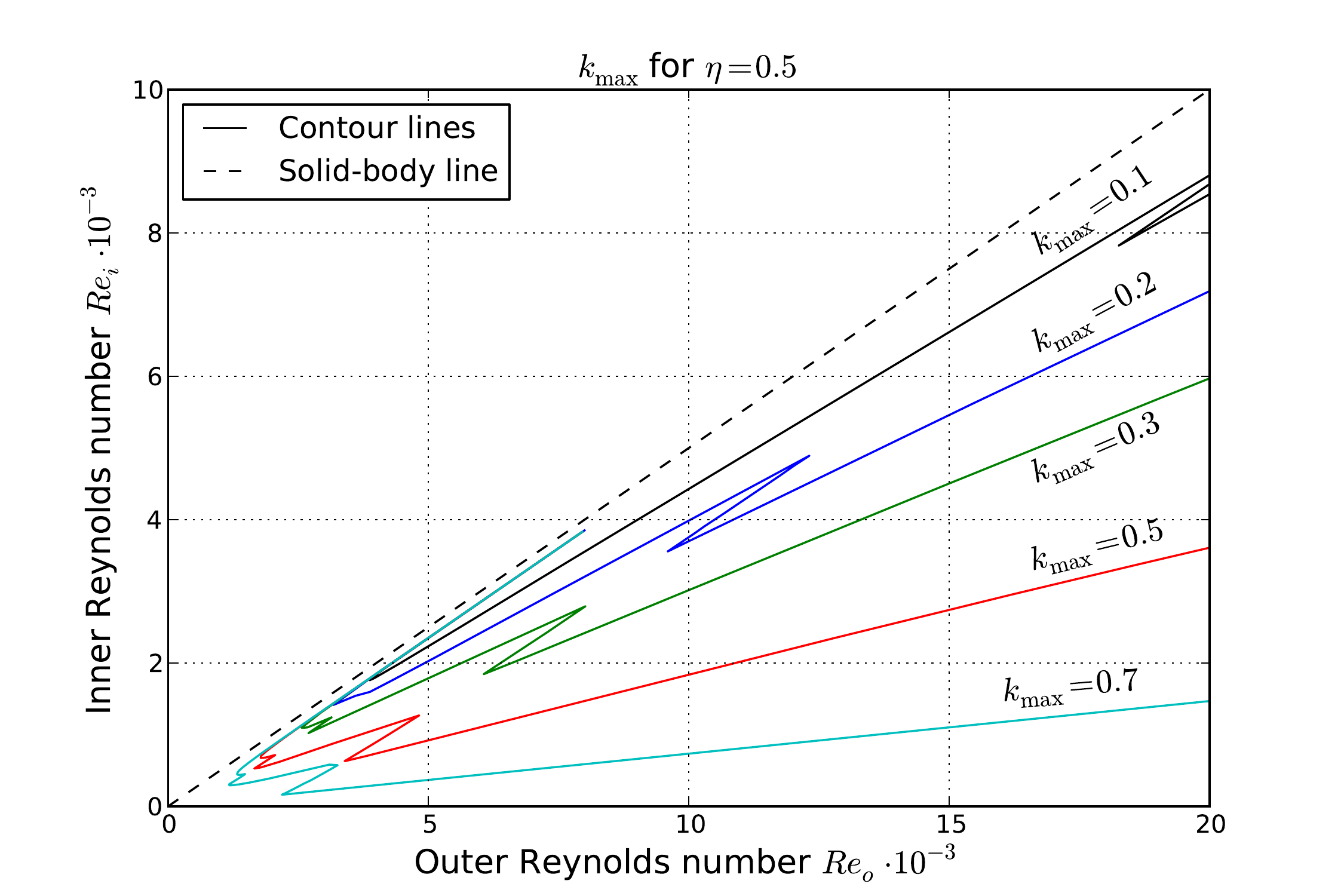}
	\caption{Contour plot of the optimal axial wavenumber $k_{\max}$ attaining optimal transient growth $G_{\max}$ within the regime I of the $\Rey_i$-$\Rey_o$ parameter space; lines determined by bisection at $\epsilon = 10^{-2}$; discontinuities are due to optimization in the discrete azimuthal wavenumber $n$} \label{fig 6.3.3}
 \end{figure}

For further illustration figures \ref{fig 6.3.2} and \ref{fig 6.3.3} show contour plots of $k_{\max}$ in {the regimes II and I, respectively}. The boundary lines have been computed by a bisecting algorithm with relative accuracy $\epsilon = 10^{-2}$. The extent of the shaded regions in figure \ref{fig 6.3.2} emphasizes the dominance of axially independent{, columnar} perturbations for quasi-Keplerian flows.
\sidecaptionvpos{figure}{c}

In the counter-rotating regime (IV) we observe a growing $k_{\max}$ with $\Rey$. This difference might be explained by emerging linear instabilities which first appear for $k > 1$ in this regime and thus render fully three-dimensional perturbations less dissipative.

The behaviour of the optimal \emph{azimuthal} wavenumber $n_{\max}$ is not discussed in detail here. Notably however, axisymmetric perturbations (corresponding to $n = 0$) never attain significant energy growth $G>O(1)$ up to high Reynolds numbers $\Rey = O (10^6)$ except for a small neighbourhood of $R_\Omega = -1$ where the dominant Taylor-vortex-related instability of regime III emerges. {On the other hand, usually transient growth of the same order is attained for different $n\neq 0$}. Numerical results indeed suggest that for sufficiently large shear Reynolds numbers, $n_{\max}$ depends more on the geometrical parameter $\eta$ rather than on $\Rey$ or $R_\Omega$ which parametrize the base flow. In general, an azimuthal wavenumber $n$ seems to be optimal if the associated wavelength is $\lambda \approx 2 \pi \eta /(n(1-\eta)) \approx 2$, i.e. twice the gap width, leading to vortices, that are of about the same radial and streamwise dimension (see e.g. figure \ref{fig 6.4.2a}, centre right).

In contrast, the dominant \emph{axial} wavenumbers $k< 1$ in regime I correspond to wavelengths of $O(10)$ rather than $O(1)$ gap widths. The axial dependence of the optimal perturbations is thus indeed weak compared to azimuthal (and radial) variations. {For comparison, recall that one observes axial symmetry and order-one axial wavelengths for the usual Taylor vortices corresponding to $n=0$ and $k= \pi$.} Moreover{, we observe that}, the stronger the rotational influence on the fluid's stability expressed by smaller $\eta$ and/or larger $|R_\Omega|$, the smaller are the $k_{\max}$ attained for $\Rey \rightarrow \infty$ (figures \ref{fig 6.2.2b}). The observed {columnwise preference of the optimal perturbations} is thus in good agreement with the Taylor--Proudman theorem,
stating that a rapidly rotating inviscid fluid is (preferably) uniform along its rotational axis. {On the other hand, this preference does not seem to be manifested in the dominant least stable eigenmodes observed in quasi-Keplerian flows: numerical optimization of the principal eigenvalue's real part over $n$ and $k$ in the test cases $\eta = 0.5$, $R_\Omega = -2.0$ and $Re = 1000, 2000, \ldots 128\,000$ (data not plotted) indeed shows significantly non-columnar modes with $k \sim 5$ to be least dissipative for $n \geq 1$. The principal zero mode $n= k =0$, on the other hand, is 
found to decay about one order of magnitude more slowly than the optimal non-axisymmetric ones in the considered parameter range. Note, furthermore, that eigenvalues corresponding to perturbations with predominantly streamwise (i.e. azimuthal) or spanwise (axial) flow, respectively, alternate along the real axis in the least stable parts of all studied spectra, where the spanwise modes even turn out to be slightly less stable. The study thus demonstrates that the structure of optimal non-modal perturbations may be entirely different from that of the dominant eigenmodes.} 

Changing $\eta$ does not seem to have any further qualitative effects on transient growth according to the results in figure \ref{fig 6.2.2}, as long as none of the limits $\eta \rightarrow \lbrace 0, 1 \rbrace$ is considered. A further study of this parameter is therefore omitted in the following.
\newline
\subsection{Evolution of optimal perturbations}\label{SS5.4}

In the sequel, three different optimal perturbations $\boldsymbol{u}_{\max,1}$, $\boldsymbol{u}_{\max,2}$ and $\boldsymbol{u}_{\max,3}$ are considered at a constant shear Reynolds number $\Rey = 8000$ and $\eta = 0.5$. The rotation numbers are given by $R_{\Omega,1} = -2.0$, $R_{\Omega,2} = 2.0$ and $R_{\Omega,3} = 0.8$ corresponding to regimes II, I and IV. The optimal wavenumbers are given by $k_{\max,1} = 0$, $k_{\max,2} \approx 0.464$ and $k_{\max,3} \approx 1.200$ and $n_{\max,1} = n_{\max,2} = n_{\max,3} = 3$. The time evolution of these modes is computed by eigenmode decomposition at a polynomial resolution $N=50$.
\begin{figure}
   \centering
   \subfigure[$\boldsymbol{u}_{\max,1}$, $R_{\Omega,1} = -2.0$ (quasi-Keplerian regime II)]
   {\includegraphics[width=\textwidth]{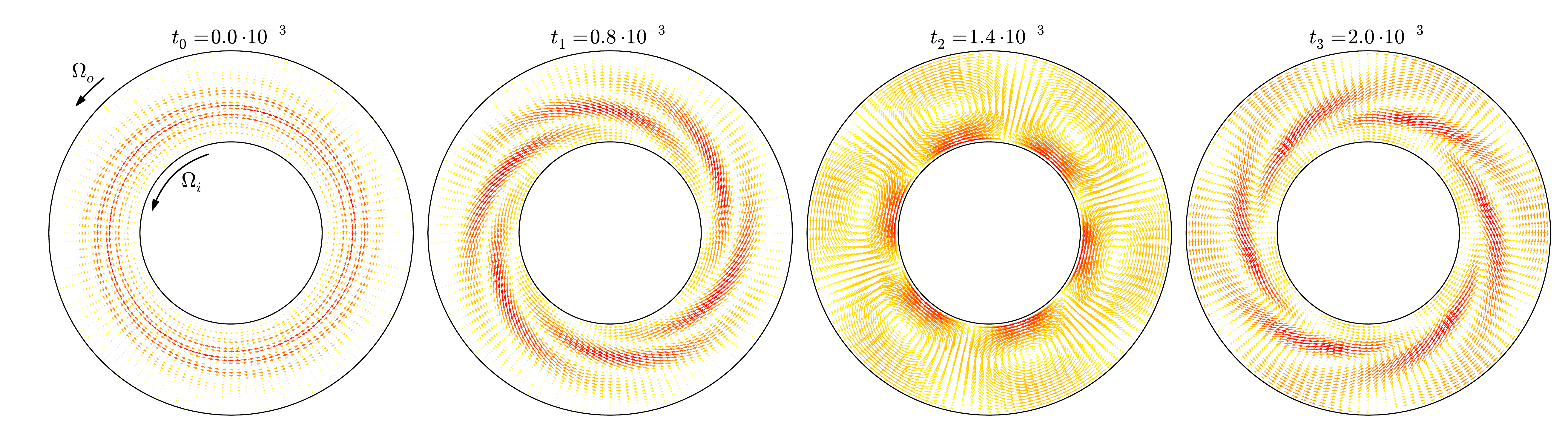}\label{fig 6.4.2a}}
   \\
   \subfigure[$\boldsymbol{u}_{\max,2}$, $R_{\Omega,2} = 2.0$ (regime I)]
   {\includegraphics[width=\textwidth]{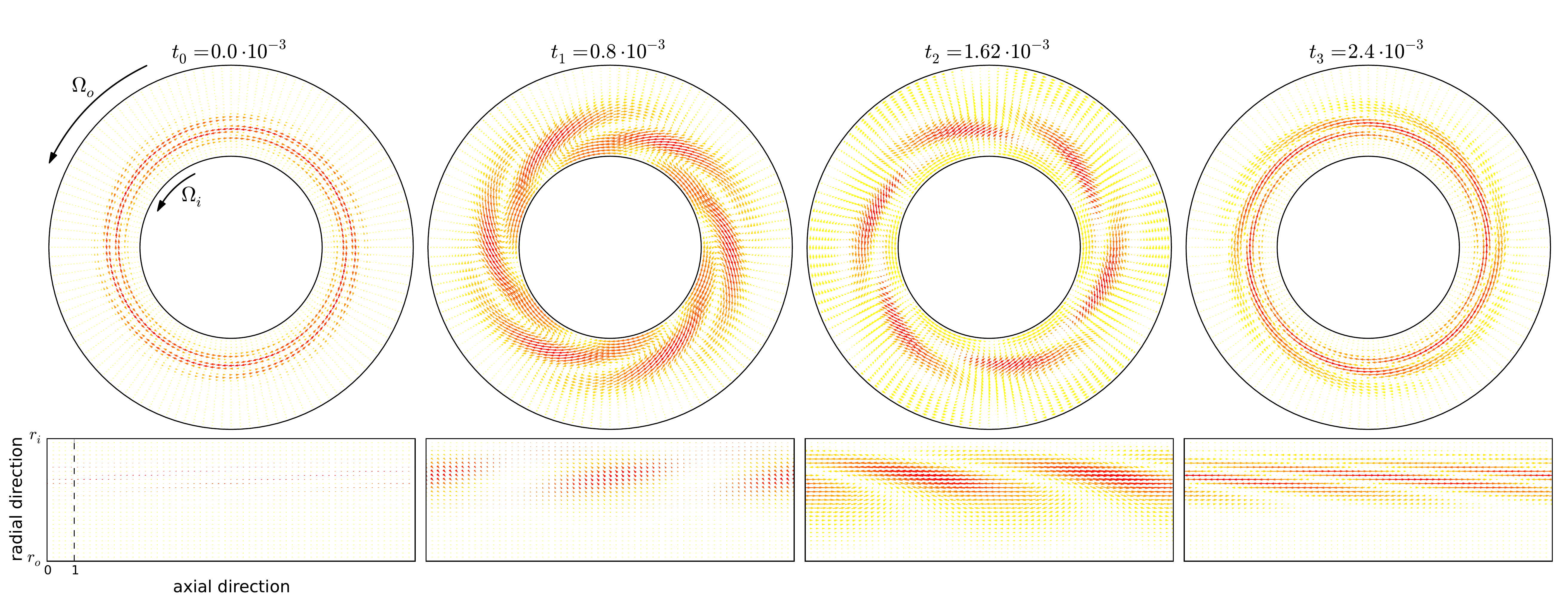}\label{fig 6.4.2b}}
   \\
   \subfigure[$\boldsymbol{u}_{\max,3}$, $R_{\Omega,3} = 0.8$ (counter-rotating regime IV)]
   {\includegraphics[width=\textwidth]{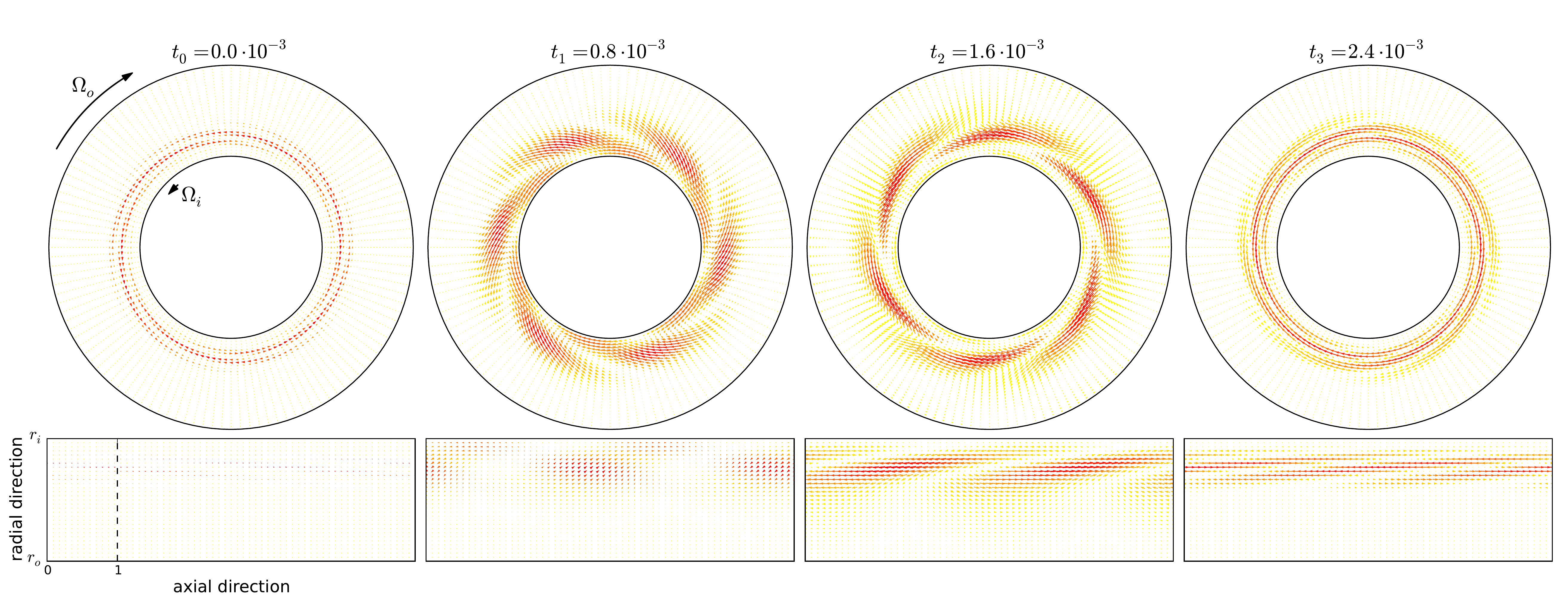}\label{fig 6.4.2c}}
   \caption{{Evolution of the optimal perturbations for $\Rey = 8000$ and $\eta = 0.5$ ($N=50$) in the regimes II ((a), quasi-Keplerian), I (b) and IV ((c), counter-rotating). Radial--azimuthal and radial--axial projections are shown. In the latter the plots are scaled to show exactly one axial wavelength {along the horizontal axis, and, to aid visualization, a unit length ($d$ = gap width) in the axial direction is indicated by the dashed line}. The subfigures each show subsequent {snapshots} at times $t=t_j$ during the transient growth evolution; the $t_j$ are also marked in the energy
   evolution curves plotted in figure \ref{fig 6.4.1}. Arrow lengths are scaled  with the flow velocities whereas their shading from lighter to stronger colours (yellow to red) reflects energy densities $|\boldsymbol{u}_{\max,i}|^2$. The relative rotation of the inner and outer cylinder in the different settings is indicated by arrows visualizing the frequencies $\Omega_o$ and $\Omega_i$, respectively. The corresponding optimal axial wavenumbers are $k_{\max,1} = 0$, $k_{\max,2} \approx 0.464$ and $k_{\max,3} \approx 1.200$}}
   \label{fig 6.4.2}
\end{figure}

In figure \ref{fig 6.4.2} the resulting real parts of $\boldsymbol{u}_{\max,1}$, $\boldsymbol{u}_{\max,2}$ and $\boldsymbol{u}_{\max,3}$ are shown at a sequence of snapshots $t_j$ throughout the transient growth evolution. The flow fields are plotted in radial--azimuthal projection (top) and radial-axial projection (bottom) with $z$ on the horizontal axis except for $\boldsymbol{u}_{\max,1}$ where the latter is omitted due to the axial independence. The radial-axial plots have been rescaled so that exactly one axial wavelength is displayed. Arrow lengths scale with the absolute flow velocities although different scalings are applied in figures \ref{fig 6.4.2a}, \ref{fig 6.4.2b} and \ref{fig 6.4.2c}. The colour map from yellow to red marks regions with relatively {low or high energy densities} $|\boldsymbol{u}_{\max,i}|^2$ in the current fields.

\begin{figure}
    \centering
	\includegraphics[width=0.64\textwidth]{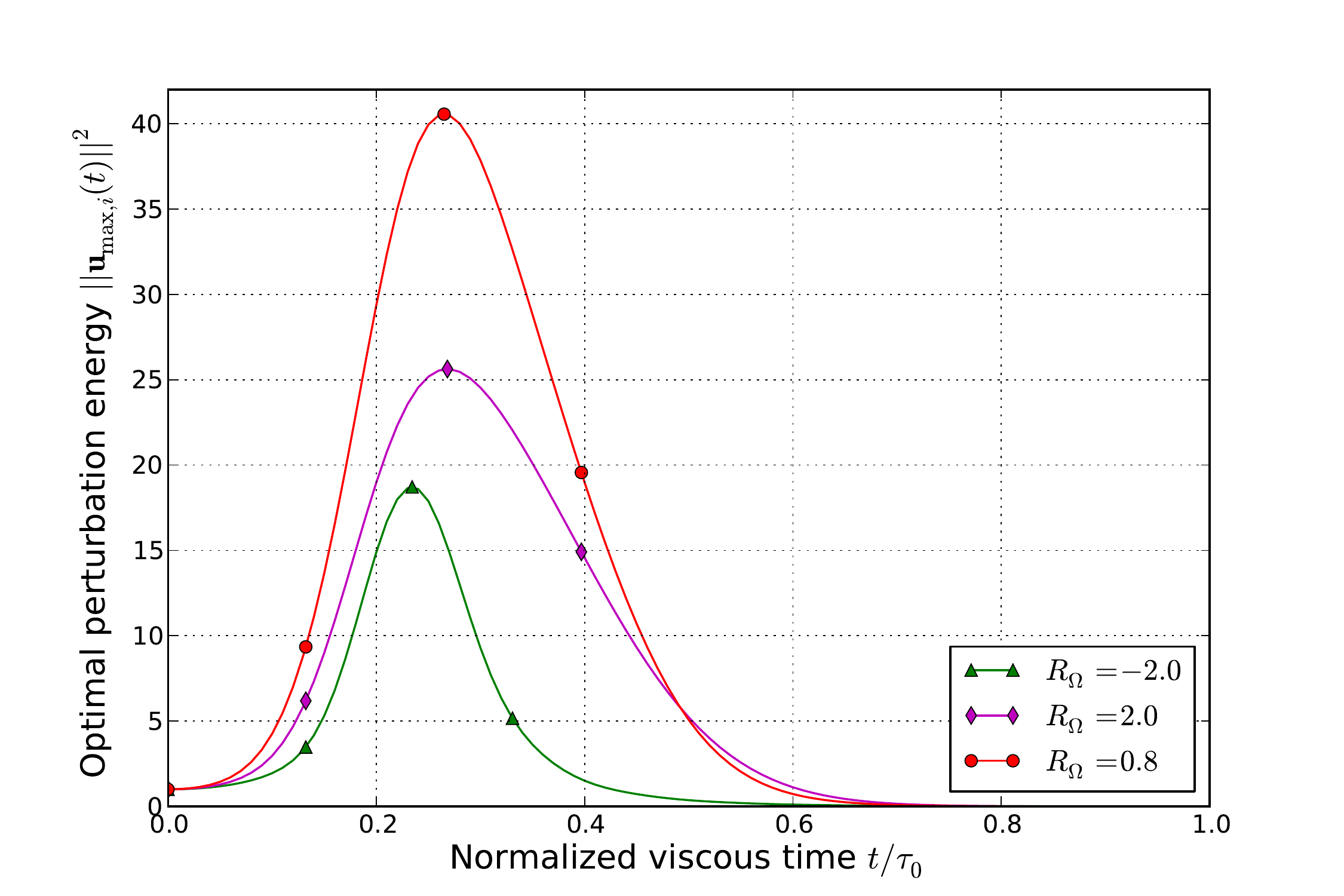}
	\caption{Evolution of the kinetic energy of the optimal perturbations $\|\boldsymbol{u}_{\max,i}\|^2$ throughout the transient growth dynamics for $\Rey = 8000$, $\eta = 0.5$ ($N=50$) and $R_{\Omega,1} = -2.0$ (quasi-Keplerian regime II), $R_{\Omega,2} = 2.0$ (regime I) {and} $R_{\Omega,3} = 0.8$ (counter-rotating regime IV). The time axis is normalized with $\tau_0 = \frac{2\pi}{Re^{0.85}(1-\eta)}$. {Snapshots} of the velocity fields at times $t_j$, indicated by markers, are shown in figure \ref{fig 6.4.2}. \label{fig 6.4.1} }
 \end{figure}

The perturbations' total kinetic energy evolution $\|\boldsymbol{u}_{\max,i}(t)\|^2$ in relation to the transient growth maxima $G_{\max,i}$ are plotted in figure \ref{fig 6.4.1} with time scale renormalized by $\tau_0 = \frac{2\pi}{Re^{0.85}(1-\eta)}$. The $t_j$ considered in figure \ref{fig 6.4.2} are identified by markers.
\sidecaptionvpos{figure}{c}

 The radial--azimuthal projections in figure \ref{fig 6.4.2} reveal essentially similar transient growth mechanisms of the considered modes: the optimal initial perturbations have a spiral-like structure of $2n$ streamwise elongated vortices. Recalling the different angular velocities $\Omega_i$ and $\Omega_o$ of the driving inner and outer cylinders(i.e. $\Omega_i > \Omega_o > 0 $ for $R_{\Omega} = -2.0$, $\Omega_o > \Omega_i > 0 $ for $R_{\Omega} = 2.0$ {and} $\Omega_i > 0 > \Omega_o$ in the counter-rotating case $R_{\Omega} = 0.8$, respectively), we find that the initial spiral orientations are always \emph{misfit} to the base flow. This ``misfit'' character is a manifestation of the perturbations' non-modal nature and thus typical of transient growth as emphasized
 by \citet{Grossmann2000}. The spiral velocity fields are tilted by the base flow and thereby gain energy (compare figures \ref{fig 6.4.2} centre-left and figure \ref{fig 6.4.1}). As for the axially independent perturbation in \ref{fig 6.4.2a} the energy maximum then occurs exactly at the turning point of the spiral orientation whereas in cases 2 and 3 it is attained shortly after this point (centre-right in figure \ref{fig 6.4.2}). Subsequently, the perturbation is further deformed into a ``fit'' flow direction, i.e. an eigendirection, and meanwhile decays.
 
{This shear-induced detilting dynamics of perturbations, with initial vorticity leaning against the background shear profile, essentially represents a Taylor--Couette analogue of the so-called Orr mechanism. The latter has been identified, e.g. in the early two-dimensional studies of optimal transient growth by \citet{farrell1988}, as an important ingredient of linear non-modal growth in plane channel flows, providing a potential explanation for the emergence of finite-amplitude disturbances required for nonlinear instabilities. Notably, in the cases studied here, this mechanism leads to transient spiral structures that resemble those of the linearly unstable, axially independent eigenmodes reported by \cite{gallet2010} -- compare our figure \ref{fig 6.4.2a} center-left with 4(d) in \citet{gallet2010}. The latter arise in the case of an additionally imposed radial inflow through the outer cylinder, which seemingly stabilizes the misfit tilt of the vortices, rendering the transient energy growth, observed in 
the present work, sustained.}
 
Especially for the {columnar axially independent} perturbation $\boldsymbol{u}_{\max,1}$, the energy growth and decay is rather sudden, leading to the sharp peak depicted in figure \ref{fig 6.4.1}. {This phenomenon is similar to the dynamics observed in plane channel flows} and is possibly due to the rapid flow near the inner cylinder wall (see figure \ref{fig 6.4.2a}, centre-right), which leads to high dissipation around the energy maximum. On the other hand, the optimal perturbations $\boldsymbol{u}_{\max,2}$ and $\boldsymbol{u}_{\max,3}$ in the regimes I and IV seem to be stabilized in this respect by their axial dependence, leading to $\unit[40]{\%}$ {and} $\unit[115]{\%}$ larger growth than that attained for $R_\Omega = -2.0$ and slower decay in figure \ref{fig 6.4.1}. This interpretation is {supported} by the fact that, in spite of the small wavenumber $k_{\max,2} \approx 0.464$ of $\boldsymbol{u}_{\max,2}$, up to $\unit[86]{\%}$ of the kinetic energy is transferred into the axial component 
around the transient growth maximum. {These three-dimensional effects go beyond the classical Orr mechanism.}

The characteristic structure of deforming elongated vortices is also reflected in the radial-axial projections in figures \ref{fig 6.4.2b} and \ref{fig 6.4.2c}. A unique feature of the counter-rotating flow ($R_{\Omega,3} = 0.8$) is the localization of the optimal perturbation near the inner cylinder walls, where the base flow is locally Rayleigh-unstable. {This localization has also been observed in the spiral eigenvectors and in the saturated spiral instability \citep[e.g.][]{Langford1988}.} Hence, although the flow remains eigenvalue stable for the chosen parameters, emerging instabilities already seem to interact with non-modal growth mechanisms. This possibly explains the greater energy amplifications in regime IV.

\subsection{Transient growth scaling for $k=0$}\label{SS5.5}

The previous numerical results, especially those for the quasi-Keplerian regime II, motivate the transient growth analysis of axially independent perturbations with $k=0$. Moreover, it will be shown in $\S$\ref{SS6.2} that $G_{\max}^{k=0}$, i.e. the optimal transient growth of $k=0$ perturbations, is indeed independent of the rotation number $R_\Omega$.

\begin{figure}
    \centering
    \subfigure[log-log-plot of $G_{\max}^{k=0}$ against $\Rey$]
    {\includegraphics[width=0.495\textwidth]{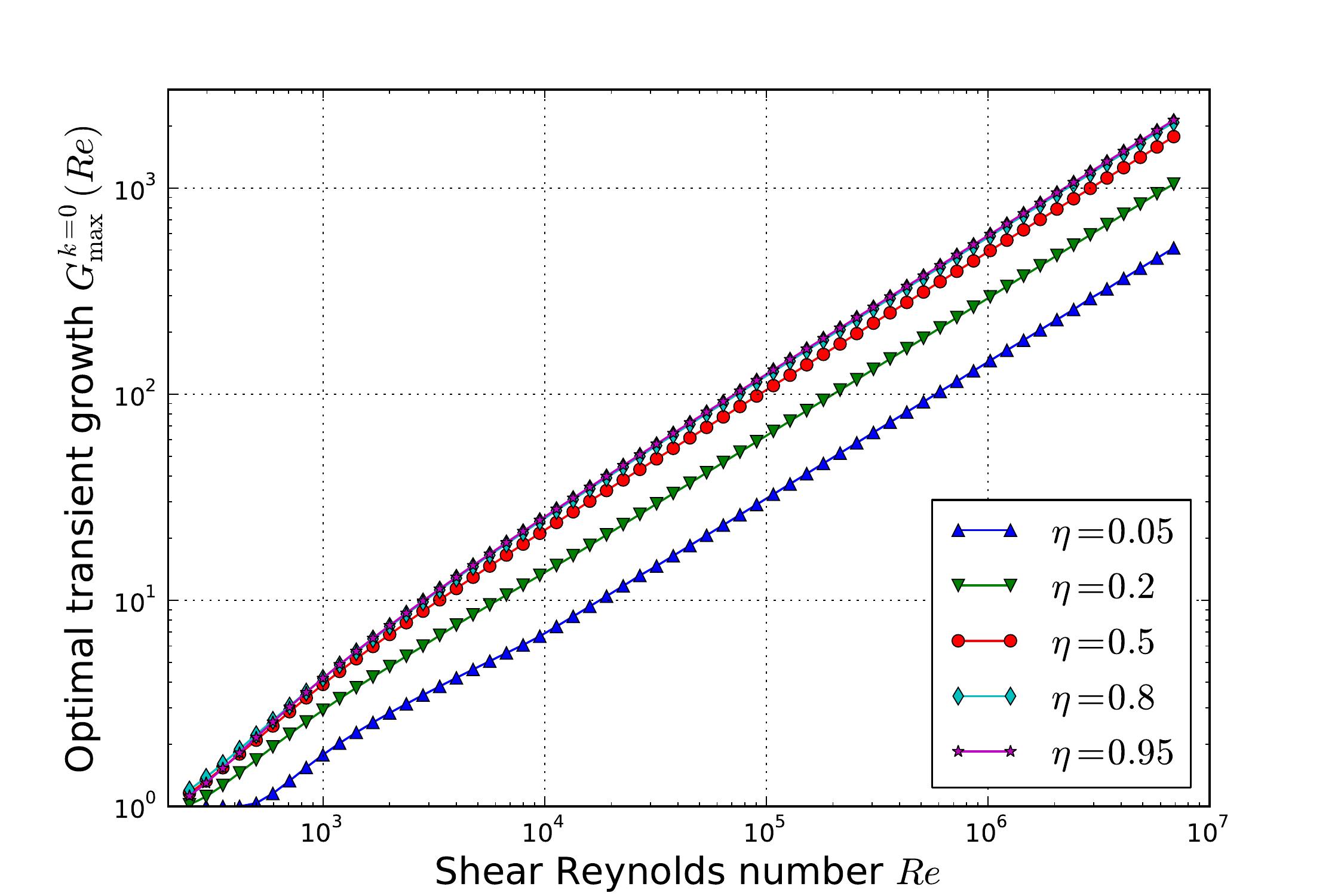}\label{fig 6.5.1a}}
    \hfill
    \subfigure[Fitted $a(\eta)$ such that $G_{\max}^{k=0} = a(\eta) \Rey^{\frac{2}{3}}$]
    {\includegraphics[width=0.495\textwidth]{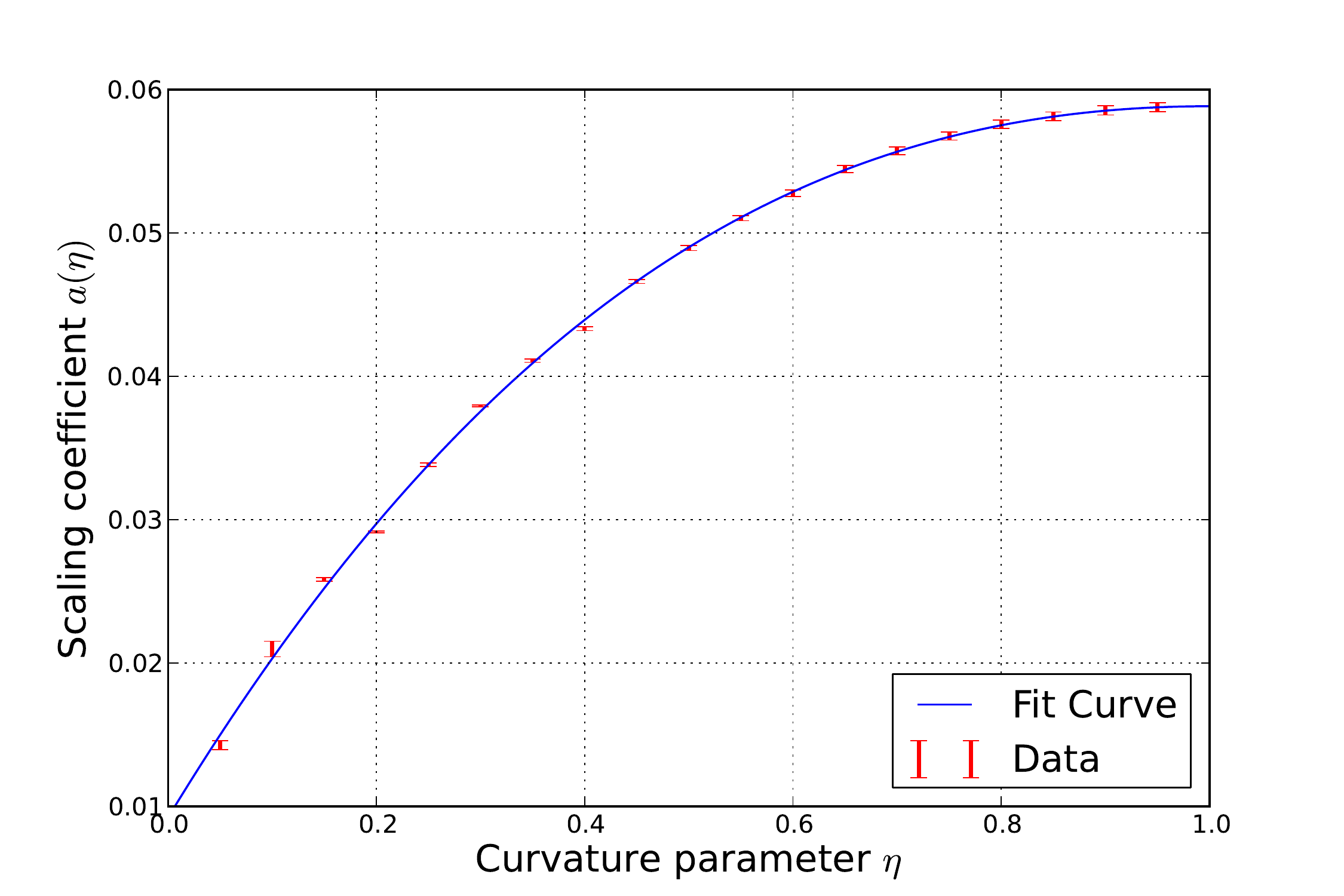}\label{fig 6.5.1b}}
    \caption{(a): Numerically computed optimal transient growth $G_{\max}^{k=0}$ for axially independent perturbations plotted against the shear Reynolds number $\Rey$; the curves are independent of $R_\Omega$ and parallel for $\Rey \geq O(10^4)$ corresponding to a common scaling $G_{\max}^{k=0} = a(\eta) \Rey^{\frac{2}{3}}$; (b): Fitted scaling coefficients $a(\eta)$ for the respective $\eta$ and high shear Reynolds numbers $\Rey \geq O(10^4)$; red bars: data determined by fitting of numerical results for $G_{\max}^{k=0}(\Rey)$ with errors by mean square deviation; blue line: third degree polynomial fit according to equations \eqref{Eq 6.5.1} and \eqref{Eq 6.5.2} \label{fig 6.5.1}}
 \end{figure}

In figure \ref{fig 6.5.1a} numerically computed optimal transient growth $G_{\max}^{k=0}$ is depicted in a log--log plot in the range $\Rey \in [250 ; 8\cdot 10^6]$ for different $\eta \in \left\lbrace 0.05, 0.2, 0.5, 0.8, 0.95 \right\rbrace $.

The parallel slopes for high Reynolds numbers $\Rey \geq O(10^4)$ show a common scaling $G_{\max}^{k=0}\sim \Rey^{\gamma}$, where the proportionality factor may depend {only} on $\eta$. In order to estimate this, $G_{\max}^{k=0}(\Rey)$ is computed for logarithmically equidistant $\Rey \in [10^5; 4\cdot 10^6]$ for $\eta \in \left\lbrace 0.05, 0.1, 0.15, \ldots, 0.95 \right\rbrace$. Fits of the form $G_{\max}^{k=0}(\Rey) = a(\eta) (\Rey)^{\gamma(\eta)}$ for each $\eta$ yield exponents $\gamma(\eta) \approx \frac{2}{3}$ within an error $\leq \unit[0.5]{ \%}$ except for $\gamma(\eta = 0.2)\approx 0.657$. Hence, a common exponent $\gamma = \frac{2}{3}$ is assumed to be universal and the factor $a(\eta)$ is independently determined by another fit. The results are plotted in figure \ref{fig 6.5.1b}, where the error bars have been determined by the mean square deviation from the data.

In order to obtain an analytical formula for $G_{\max}^{k=0}(\Rey)$ a third-degree polynomial
\begin{align}
 a(\eta) = a_0 +a_1 \eta \left(1- \frac{1}{3}\eta^2\right) + a_2 \eta^2 \left(1-\frac{2}{3} \eta\right) \label{Eq 6.5.1}
\end{align}
is fitted to the data in figure \ref{fig 6.5.1b} taking into account the extremum of $a$ at $\eta = 1$ which is due to the system's symmetry with respect to exchanging of $r_i$ and $r_o$. The result is
\begin{align}
 a_0 \approx 9.218 \cdot 10^{-3}, \,\,\,\,\,\, a_1 \approx 0.1198, \,\,\,\,\,\, \text{and} \,\,\,\,\,\, a_2 \approx -9.072 \cdot 10^{-2} \label{Eq 6.5.2}
\end{align}
and the corresponding curve is also shown in figure \ref{fig 6.5.1b}. Good agreement between fit and data is found especially for $\eta \geq 0.5$, possibly due to the lesser impact of the azimuthal wavenumber's discreteness on the attainable optimal transient growth compared to $\eta < 0.5$. For arbitrary $\eta$ test cases give less than $\unit[7]{\%}$ error if the analytical formula is applied for $\Rey \in [10^4; 8\cdot 10^6]$ and less than $\unit[5]{\%}$ in the interval $[10^5; 2\cdot 10^6]$.

The maximum amplification of axially independent perturbations $G_{\max}^{k=0} = a(\eta) \Rey^{\frac{2}{3}}$ defines a lower bound for the total ($k\neq 0$) transient growth $G_{\max}(\Rey)$ in every flow regime. Moreover, the estimate can be expected to hold within a factor of $O(1)$ and is exact in the shaded regions of the quasi-Keplerian regime II in figure \ref{fig 6.3.2}.
%
%
%
%
%
%
%
%
\section{Analytical results for axially independent perturbations} \label{S6}

The prominent role played by columnar, axially independent perturbations together with their geometrical simplicity motivates an analytical study of their properties, which is pursued in this section.
We begin by applying the conjugated curl operator
\begin{equation}
(\boldsymbol{\nabla} \times)_{\rm c} :=   e^{-\I(n\varphi +kz)}(\boldsymbol{\nabla} \times) e^{\I(n\varphi +kz)} = \begin{pmatrix} 0 & -\I k & \frac{\I n}{r} \\ \I k & 0 & -\mathcal{D} \\ -\frac{\I n}{r} & \mathcal{D}_+ & 0 \end{pmatrix} \label{Eq 7.1.2}
\end{equation}
to the linearized Navier--Stokes equation \eqref{Eq 2.1.5}. This eliminates the pressure gradient terms, yielding
\begin{equation}
\begin{pmatrix} 0 & -\I k & \frac{\I n}{r} \\ \I k & 0 & -\mathcal{D} \\ -\frac{\I n}{r} & \mathcal{D}_+ & 0 \end{pmatrix}  \begin{pmatrix} \partial_t u_r \\ \partial_t u_\varphi \\ \partial_t u_z \end{pmatrix}= \begin{pmatrix} 0 & -\I k & \frac{\I n}{r} \\ \I k & 0 & -\mathcal{D} \\ -\frac{\I n}{r} & \mathcal{D}_+ & 0 \end{pmatrix} \cdot \begin{pmatrix} \mathcal{L}_{rr} &\mathcal{L}_{r\varphi} &0 \\ \mathcal{L}_{\varphi r} &\mathcal{L}_{\varphi \varphi} &0 \\ 0 &0 &\mathcal{L}_{zz} \end{pmatrix} \begin{pmatrix} u_r \\ u_\varphi \\ u_z \end{pmatrix}.  \label{Eq 7.1.3}
\end{equation}
For axially independent perturbations ($k=0$) the azimuthal velocity
$u_\varphi$ is determined from $u_r$ via the divergence condition
\begin{equation}
0 = \boldsymbol{\nabla}_{\rm c} \cdot \boldsymbol{u} = \mathcal{D}_+ u_r + \frac{\I n}{r} u_\varphi + \underbrace{\I k u_z}_{=0}  \,\,\,\,\,\, \Longrightarrow \,\,\,\,\,\, u_\varphi = \frac{\I r}{n} \mathcal{D}_+ u_r,  \label{Eq 7.1.1}
\end{equation}
and the evolution equations for $u_r $ and $u_z$ decouple \citep{GebhardtGrossmann1993}. Using $\mathcal{L}_{rr} = \mathcal{L}_{\varphi \varphi}$ the resulting equations read
\begin{equation}
\begin{pmatrix} \frac{\I n}{r}\partial_t u_z \\ -\mathcal{D} \partial_t u_z \\ (-\frac{\I n}{r} +\mathcal{D}_+ \frac{\I r}{n}\mathcal{D}_+)\partial_t u_r  \end{pmatrix} = \begin{pmatrix} \frac{\I n}{r} \mathcal{L}_{zz} u_z \\ -\mathcal{D} \mathcal{L}_{zz}  u_z \\( -\frac{\I n}{r} \mathcal{L}_{rr} + \mathcal{D}_+ \mathcal{L}_{rr} \frac{\I r}{n} \mathcal{D}_+ + \mathcal{D}_+ \mathcal{L}_{\varphi r} + \mathcal{L}_{r\varphi}\mathcal{D}_+) u_r  \end{pmatrix} \label{Eq 7.1.4}.
\end{equation}
The first and second equations, which are equivalent, determine the evolution of $u_z$:
\begin{eqnarray}
\partial_t u_z &=& \mathcal{L}_{zz}  u_z =\left( \mathcal{D}_+\mathcal{D} -\frac{n^2}{r^2}  - \frac{\I n}{r} v_\varphi^B \right)u_z \nonumber \\
 &=& \left(\partial_r^2 + \frac{1}{r} \partial_r -\frac{n^2}{r^2}  -\I n  \left( A+\frac{B}{r^2}\right) \right) u_z. \label{Eq 7.1.5}
\end{eqnarray}
Using the results $\mathcal{D}_+ \mathcal{L}_{\varphi r} + \mathcal{L}_{r\varphi}\mathcal{D}_+ = \frac{2B}{r^2}\mathcal{D}_+ - \frac{4\I n}{r^3}$ and $[\mathcal{L}_{rr}, r\partial_r]= 2\mathcal{L}_{rr} + 2\I nA=: 2\mathcal{L}_{rr}^0$ obtained in \S\ref{SSA1}, the evolution equation for $u_r$ becomes
\begin{multline}
\partial_t \left( r\mathcal{D}_+ r\mathcal{D}_+ -n^2 \right) u_r = \mathcal{L}_{rr} \left(r\mathcal{D}_+ r\mathcal{D}_+  -n^2 \right) u_r \\ - \left(\underbrace{\left[ \frac{\I r}{n}\mathcal{D}_+ , \mathcal{L}_{rr}\right] \I rn\mathcal{D}_+}_{=\left[ \mathcal{L}_{rr} , r\partial_r \right]r\mathcal{D}_+ = 2\mathcal{L}^0_{rr}r\mathcal{D}_+ }  +\;\; \I rn \left[   \frac{2B}{r^2}\mathcal{D}_+ - \frac{4\I n}{r^3} \right] \right) u_r\label{Eq 7.1.6}
\end{multline}
Further using $\partial_r\frac{2}{r} \left(r\mathcal{D}_+ r\mathcal{D}_+ -n^2  \right) = 2\mathcal{L}_{rr}^{0}r\mathcal{D}_+ + \I rn \left(   \frac{2B}{r^2}\mathcal{D}_+ - \frac{4\I n}{r^3} \right)$ (see \S\ref{SSA1}) yields
\begin{multline}
\partial_t \left( r\mathcal{D}_+ r\mathcal{D}_+ -n^2 \right) u_r = \left( \mathcal{L}_{rr} -2 \partial_r \frac{1}{r}\right)  \left(r\mathcal{D}_+ r\mathcal{D}_+  -n^2 \right) u_r  \\
=\left( \partial_r^2 -\frac{1}{r} \partial_r - \frac{n^2+1}{r^2}  -\I n  \left( A+\frac{B}{r^2}\right)  \right)  \left(r\mathcal{D}_+ r\mathcal{D}_+  -n^2 \right) u_r \label{Eq 7.1.7}. 
\end{multline}
This fourth-order partial differential equation is supplemented with the boundary conditions
$u_r(r_i)= u_r(r_o) = \partial_r u_r(r_i) = \partial_r u_r(r_o)=0$,
which correspond to the {no-slip} boundary conditions at the cylinders
$u_r(r_i)= u_r(r_o)=u_\varphi(r_i)= u_\varphi(r_o)=0$.


\subsection{Advection of perturbations by the basic flow and universal stability properties}\label{SS6.1}

A remarkable property of the equations \eqref{Eq 7.1.5} and \eqref{Eq 7.1.7} is revealed by considering the transformation $ \tilde{u}_r := e^{\I nAt} u_r$ and  $\tilde{u}_z := e^{\I nAt} u_z$. The derivatives then read $\partial_r \tilde{u}_\ast = e^{\I nAt}\partial_r u_\ast$ and $\partial_t \tilde{u}_\ast = e^{\I nAt}(\partial_t + \I nA) u_\ast$ so substituting into \eqref{Eq 7.1.5} and \eqref{Eq 7.1.7} yields 
\begin{eqnarray}
\partial_t \tilde{u}_z   &=& e^{\I nAt}(\partial_t + \I nA) u_z  = \left(\partial_r^2 + \frac{1}{r} \partial_r -\frac{n^2}{r^2} -\frac{\I nB}{r^2} \right) \tilde{u}_z \label{Eq 7.2.1a}  \\
\partial_t \tilde{f}_r &=& e^{\I nAt}(\partial_t +\I nA) f_r =\left( \partial_r^2 -\frac{1}{r} \partial_r - \frac{n^2+1}{r^2} -\frac{\I nB}{r^2} \right) \tilde{f}_r, \label{Eq 7.2.1b} 
\end{eqnarray}
where $\tilde{f}_r := \left( r\mathcal{D}_+ r\mathcal{D}_+ -n^2 \right) \tilde{u}_r$ and $f_r := \left( r\mathcal{D}_+ r\mathcal{D}_+ -n^2 \right) u_r$. As $\tilde{u}_z$ and $\tilde{u}_r$ satisfy equations \eqref{Eq 7.1.5} and \eqref{Eq 7.1.7} with $A = 0$, the $A$ dependence of the perturbation's evolution $\boldsymbol{u}$ is entirely described by the factor $e^{-\I nAt}$. This factor corresponds to a pure advection of the perturbation with the shear-free, uniformly rotating part of the basic flow $\boldsymbol{v}^B$ and thus it is locally and globally energy-conserving ($|u_r|^2 = e^{\I nAt} e^{-\I nAt}|\tilde{u}_r|^2 = |\tilde{u}_r|^2$). Although these conclusions might seem obvious at first glance note that they are \emph{not} true in the general three-dimensional case $k \neq 0$.

The minor importance of the parameter $A$ has crucial consequences. Without loss of generality, $A=0$ can be assumed when analysing the stability of Taylor--Couette flow to axially independent perturbations. The remaining parameter $B$ characterizing the base flow $\boldsymbol{v}^B$ depends only on the shear Reynolds number $\Rey$ and \emph{not} on the rotation number $R_\Omega$ (see \eqref{Eq 2.2.1b}), which parametrizes the flow regime. Hence the linear stability of Taylor--Couette flow to axially independent perturbations is independent of $R_\Omega$ and thus is identical in all regimes. {Furthermore}, the optimal transient growth $G_{\max}^{k=0}$ for $k=0$ provides a lower bound for the absolute maximum $G_{\max}$ which is universal in the sense that it depends only on $\eta$ and $\Rey$. We note that these results can be expected to apply approximately also for weakly axially dependent perturbations in the vicinity of $k=0$.

\subsection{Global analysis of the evolution equations}\label{SS6.2}

First, consider the evolution of $u_z$ described by equation \eqref{Eq 7.1.5}. The operator $\mathcal{L}_{zz}$ is the sum of a self-adjoint negative definite operator and a skew hermitian one. As these do not commute $\mathcal{L}_{zz}$ is an example of a non-normal operator, which nonetheless does \emph{not} allow for transient growth (see appendix \ref{SSA2} for a proof). The evolution equation \eqref{Eq 7.1.7} for the radial component $u_r$ may be split into two independent problems,
\begin{equation}
\partial_t f_r = \left( \mathcal{L}_{rr} -2 \partial_r \frac{1}{r}\right) f_r \,\,\,\,\,\, \text{and} \,\,\,\,\,\, \left(r\mathcal{D}_+ r\mathcal{D}_+  -n^2 \right) u_r = f_r \label{Eq 7.2.2}
\end{equation}
where the second is of Sturm-Liouville type (the solution is given in appendix \ref{SSA5}) and the first resembles equation \eqref{Eq 7.1.5}.
{Using this factorization, it might be possible to construct an exact analytical solution of the evolution problem \eqref{Eq 7.1.7} by incorporating the boundary conditions via an influence matrix method. However, the outer problem in \eqref{Eq 7.2.2} remains cumbersome to solve, and, even if one were to write down an expression for an exact solution of \eqref{Eq 7.1.7}, this would most likely turn out to be too involved to interpret the underlying physics.}
In the following, the analysis of the evolution equation \eqref{Eq 7.1.7} is therefore confined to the limit of asymptotically large Reynolds numbers $\Rey \rightarrow\infty$ and is studied by means of scale analysis.

In order to identify and motivate the scales to be studied quantitatively in the WKB Analysis {of} $\S$\ref{S7}, we consider the energy evolution of a perturbation $\boldsymbol{u} = u_r \boldsymbol{e}_r+ u_\varphi \boldsymbol{e}_\varphi$
\begin{equation}
\partial_t \|\boldsymbol{u}\|^2 = 2 \Real \left\langle \boldsymbol{u}, \mathcal{L} \boldsymbol{u} \right\rangle =   -2 \Real \left\langle \boldsymbol{u}, (\boldsymbol{u} \cdot \boldsymbol{\nabla})\boldsymbol{v}^B \right\rangle   + 2 \Real \left\langle \boldsymbol{u}, \Delta_r \boldsymbol{u} \right\rangle \label{Eq 7.3.1}
\end{equation}
where the pressure and convective terms drop out as in the derivation of the Reynolds-Orr equation. Using $u_\varphi = \frac{\I r}{n} \mathcal{D}_+ u_r$ the non-normal term in \eqref{Eq 7.3.1} becomes
\begin{equation}
N(\boldsymbol{u}) := -2 \Real \left\langle \boldsymbol{u}, (\boldsymbol{u} \cdot \boldsymbol{\nabla})\boldsymbol{v}^B \right\rangle = - \frac{4B}{n} \Imag \left\langle u_r, \partial_r u_r \right\rangle  \label{Eq 7.3.2}
\end{equation} 
while the self-adjoint, dissipative summand reads
\begin{eqnarray}
 D(\boldsymbol{u}) := 2 \Real \left\langle \boldsymbol{u}, \Delta_r \boldsymbol{u} \right\rangle =-2 (&n^{-2}&( \| \mathcal{D} r \mathcal{D}_+ u_r\| ^2 + (n^2+1) \|\mathcal{D}_+ u_r\|^2) + \|\mathcal{D} u_r\|^2\nonumber \\  &-& 4 \Real \left\langle u_r, r^{-1} \mathcal{D}_+u_r  \right\rangle +  (n^2+1) \|r^{-1} u_r\|^2 ). \label{Eq 7.3.3}
\end{eqnarray}

Assume that $u_r$ varies on a typical length scale of order $O((n\Rey)^{-\alpha})$ with $\alpha > 0$. In the limit $\Rey \rightarrow \infty$, the highest-order $r$-derivative dominates in each term of \eqref{Eq 7.3.1}. As $B \sim \Rey$ and $\|\boldsymbol{u}\|^2 = \|u_r\|^2 + n^{-2} \| r \mathcal{D}_+ u_r\|^2$ we obtain from equations \eqref{Eq 7.3.2} and \eqref{Eq 7.3.3}
\begin{eqnarray}
N(\boldsymbol{u}) =& n^{-2} O((n\Rey)^{1+ \alpha} \|u_r\|^2) &=  O((n\Rey)^{1 -  \alpha}) \|\boldsymbol{u}\|^2 \nonumber \\
D(\boldsymbol{u}) =& n^{-2} O((n\Rey)^{4\alpha} \|u_r\|^2) &=  O((n\Rey)^{2 \alpha}) \|\boldsymbol{u}\|^2 \label{Eq 7.3.4}
\end{eqnarray}
According to \eqref{Eq 7.3.3}, $D(\boldsymbol{u})$ is strictly negative, so by virtue of \eqref{Eq 7.3.4} dissipation always dominates for $\alpha > \frac{1}{3}$. On the other hand, the non-normal term $N(\boldsymbol{u})$ may be positive, so that, for $\alpha \leq \frac{1}{3}$, growth rates $\partial_t \ln \|\boldsymbol{u}\|^2 = O((n\Rey)^{1 -  \alpha})$ are possible.

The question remains how long such growth may last. Let us consider a Fourier-type ansatz $u_r \sim e^{\I m r}$ with wavenumber $m = O((n\Rey)^{\alpha})$. Note that locally this is valid because in the limit $\Rey \rightarrow \infty$ boundary effects are confined to thin layers near the cylinder walls. Then $N(\boldsymbol{u})$ is of optimal order in \eqref{Eq 7.3.4} and $N(\boldsymbol{u})>0$ if and only if $n^{-1}B m < 0$ by virtue of \eqref{Eq 7.3.2}.

The total velocity field is $\boldsymbol{\tilde{u}} =  e^{\I (n \varphi + kz )} \boldsymbol{u} \sim e^{\I (n \varphi +  m r)}$, so the curves of constant phase (characteristics) are (locally) given by $\varphi(r) = \varphi(r_i) - n^{-1}m (r-r_i)$. Starting at the inner cylinder the set of these lines form streamwise elongated 
spiral structures like the vortices in figure \ref{fig 6.4.2}. To attain growth they have to be oriented according to the sign
\begin{equation}
\text{sgn} ( \partial_r \varphi) = -\text{sgn}(n^{-1}m) = \text{sgn}(B) = - \text{sgn}(\partial_r \Omega).\label{Eq 7.3.5}
\end{equation}
Thus, the characteristics of the perturbations have to be \emph{misfit} to the base flow's angular velocity profile $\Omega^B = r^{-1} v^B_\varphi$, as observed in the numerical computations of $\S$\ref{SS5.4}. Therefore, energy amplification may only occur \emph{transiently} until the perturbation has been sheared into the ``fit'' orientation by advection,  {analogously to the perturbation dynamics associated with the Orr mechanism in channel flows \citep[see e.g][]{farrell1988}.} Within the advective time scale $T = O(\Rey^{-1})$, i.e. a cylinder rotation period, the shear uniformly distorts the flow profile between the inner and outer cylinders by a length of order $O(1)$. Consequently, as the initial streamwise elongation of the characteristics is $O(n^{-1}m)$ and $m = O((n\Rey)^{\alpha})$, the time $t_0$ for the perturbation to be tilted into {the} fit direction is 
\begin{equation}
t_{0, \alpha} = O(n^{-1}m T) = O((n\Rey)^{\alpha -1}) \label{Eq 7.3.6}.
\end{equation}
Viscosity prevents transient growth if $\alpha > \frac{1}{3}$. Now assume $\boldsymbol{u}$ is an optimal perturbation for $\alpha < \frac{1}{3}$. Then we can evolve this mode \emph{backwards} until times of order $O((n\Rey)^{-\frac{2}{3}})$ before its energy maximum, introduce the result as a new initial condition and thereby attain additional growth. Thus, optimal perturbations must vary on length scales $O(n\Rey)^{-\frac{1}{3}}$ and $t_0 = O((n\Rey)^{-\frac{2}{3}})$ is the natural time scale for transient growth.

Our numerical computations are in perfect agreement with these scaling results. However, we cannot obtain an analytical estimate for the optimal transient growth with this section's zeroth-order approach. Therefore, in the next section we introduce the time scale $t_0$ {into} the evolution equation \eqref{Eq 7.1.7} and analyse it by means of a first-order WKB approximation. Our analysis closely follows the work of \citet[pp. 47-53]{Chapman2002} for oblique modes in channel flows.

\section{WKB analysis of axially independent perturbations}\label{S7}

Following the analysis of the previous section we rescale time as $\bar{t} := \delta ^{-2} t $ with $\delta  := (n\Rey)^{-\frac{1}{3}}$, and rewrite $nB := \delta ^{-3}B_0$, where the factor $B_0$ is independent of $n$ and $\Rey$ (see equation \ref{Eq 2.2.1b}). Substituting these scalings for $t$ and $nB$ in the evolution equation \eqref{Eq 7.1.7}
and multiplying by $\delta ^3$ yields
\begin{multline}
\delta  \partial_{\bar{t}} \left(r^2\partial_r^2 +3r\partial_r -(n^2-1) \right) u_r \\ =\left( \delta ^3 \left( \partial_r^2 -\frac{1}{r} \partial_r - \frac{n^2-1}{r^2} \right)  -\frac{\I B_0}{r^2} \right) \left(r^2\partial_r^2 +3r\partial_r -(n^2-1) \right) u_r, \label{Eq 7.4.1}
\end{multline}
where we have set $A=0$ without loss of generality in accordance with $\S$\ref{SS6.1}. Note that the highest-order spatial derivative in \eqref{Eq 7.4.1} is now multiplied by the factor $\delta^3 \ll 1$, which is small in the limit of high Reynolds numbers $\Rey \rightarrow \infty$.

\subsection{WKB ansatz}\label{SS7.1}

We make a WKB ansatz with amplitude $\tilde{a}$ and rapidly oscillating phase $ \delta^{-1} \phi $
\begin{equation}\label{eq:WKB-Ansatz}
u_r = \tilde{a} \exp\left( \delta^{-1} \phi \right), 
\end{equation}
where both $\tilde{a}$ and $\phi$ depend on $\bar{t}$ and $r$. Together with the divergence condition this yields $u_\varphi = \frac{\I r}{n}\mathcal{D}_+ u_r = O(\partial_r u_r) = O(\delta^{-1}u_r) $. Hence the scaling $\tilde{a}= \delta a$, with $a = O(1)$ and $\phi = O(1)$, is required in order that initial perturbations $\boldsymbol{u} = u_r \boldsymbol{e}_r + u_\varphi \boldsymbol{e}_\varphi$ have unit energy norm ($\|\boldsymbol{u}(0)\|^2 = 1$).

We now substitute the WKB ansatz \eqref{eq:WKB-Ansatz} into the evolution equation \eqref{Eq 7.4.1}. Because of $a,\phi = O(1)$ the evolution equation needs to be independently satisfied at each order in $\delta$. At leading order $O(\delta^{-1})$ the equation reduces to (see $\S$\ref{SSA4})
\begin{equation}
\partial_{\bar{t}} \phi =  -\frac{\I B_0}{r^2}  \,\,\,\,\,\,\Longrightarrow \,\,\,\,\,\, \phi(r,\bar{t}) =  \phi _0(r) -\frac{\I B_0}{r^2} \bar{t} \label{Eq 7.4.2}
\end{equation}
By using this solution to eliminate $\delta^{-1}$ terms in \eqref{Eq 7.4.1}, we obtain
\begin{multline}
r^2 \partial_{\bar{t}} \left((\partial_r\phi)^2 a \right) - r^2 (\partial_r\phi)^4 a = \delta ^1 \left(6r (\partial_r\phi)^3 a + 6r^2 (\partial_r\phi)^2 (\partial_r^2\phi) a + 4 r^2 (\partial_r\phi)^3 (\partial_ra) \right)  \\ - \delta ^1 \partial_{\bar{t}} \left( 2r^2 (\partial_r\phi) (\partial_ra) +r^2 (\partial_r^2\phi) a +3r(\partial_r\phi)a \right)  \label{Eq 7.4.3}
\end{multline}
which at next leading order $O(\delta^0)=O(1)$ reads
\begin{equation}
(\partial_r\phi) \partial_{\bar{t}} a= (\partial_r\phi)^3a - 2 (\partial_{\bar{t}}\partial_r\phi)a. \label{Eq 7.4.35}
\end{equation}
Defining $\tau := \I  (\partial_r\phi)$ and $\partial_{\bar{t}}  = \I  (\partial_{\bar{t}}\partial_r\phi) \partial_\tau = -\frac{2B_0}{r^3} \partial_\tau$ \citep[p. 49]{Chapman2002} yields
\begin{eqnarray}
\frac{\partial_\tau a}{a} = \frac{r^3}{2B_0}\tau^2 -\frac{2 }{\tau}\,\,\,\,\,\, \Longrightarrow& \,\,\,\,\,\, a(r,\tau) &= -\frac{a_0(r) }{\tau^2} \exp\left(\frac{r^3}{6B_0}  \tau^3 \right).\label{Eq 7.4.4}
\end{eqnarray}

\begin{figure}
    \centering
	\includegraphics[width=0.65\textwidth]{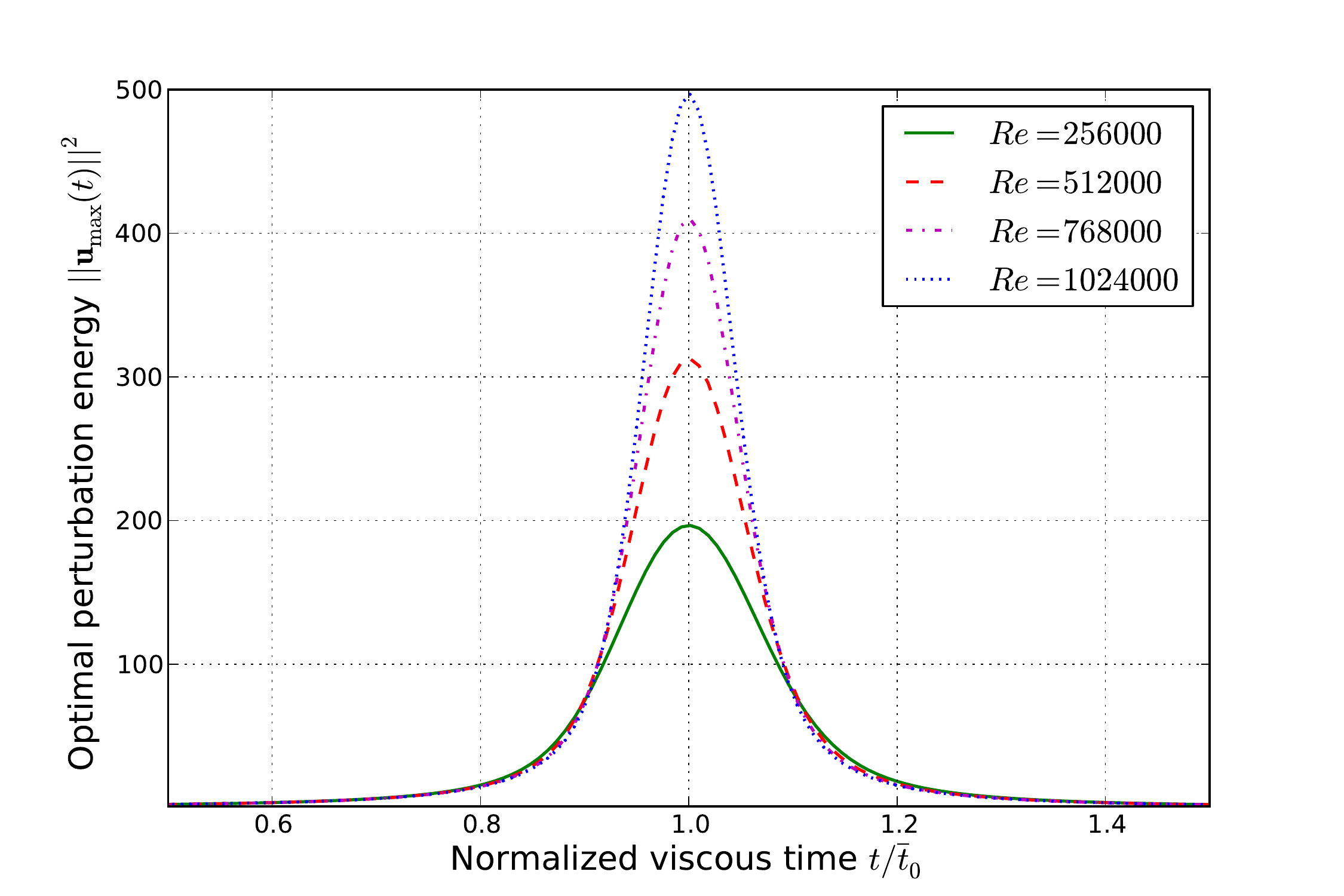}
	\caption{Energy blow-up of numerically computed optimal axially invariant perturbations for $R_\Omega = -2.0$, $\eta = 0.5$ and different shear Reynolds numbers $\Rey$; the time axis is normalized by the respective energy maximum $\bar{t}_0$; the increasingly sharp peaks reflect the singular behaviour of the WKB solution \eqref{Eq 7.4.4} except for $O(\delta)$ neighbourhoods of the maxima} \label{fig 7.1.1}
 \end{figure}

According to this solution, $a$ becomes singular for $\tau \rightarrow 0$, which may raise doubts about its physical correctness. However, in this limit, the underlying separation of orders in the WKB approximation breaks down so that $O(\delta ^1)$ terms in \eqref{Eq 7.4.3} or even in the leading-order equation have to be considered. These bound the blow-up, leading to an overall \emph{nearly} singular amplitude behaviour in the complete linearized dynamics given by \eqref{Eq 7.1.7}. In numerical simulations, this manifests itself in increasingly sharp peaks of the optimal perturbation's
energy for $\Rey \rightarrow \infty$, as visualized in figure \ref{fig 7.1.1}. The larger $\Rey$, the longer the blow-up seems to follow the singular WKB solution \eqref{Eq 7.4.4} before the energy growth is capped near the maximum blow-up time $\bar{t}_0$. Most prominently, for $\Rey = 1024000$ it is only in a neighbourhood $(1 \pm 0.05)\bar{t}_0$ about the maximum that the singularity is smoothed out by additional terms,resulting in the sharpest peak in figure \ref{fig 7.1.1}.

\subsection{Construction of optimal perturbations}\label{sS:WKB-opt}

Assume that the amplitude's growth according to equation \eqref{Eq 7.4.4} is capped as soon as the next-order terms become relevant. Then the optimal \emph{energy growth}
is attained if: 
\begin{enumerate}
 \item  the blow-up occurs at a common time $\bar{t}_0$ over the whole radial domain $r \in (r_i,r_o)$
 \item  the $O(\delta ^1)$ terms in \eqref{Eq 7.4.3} are of the highest attainable order in $\tau$
\end{enumerate}

Condition (\emph{a}) ensures that no averaging effects of the spatial integral evaluated for the computation of $\| \boldsymbol{u}(t) \|^2$ limit the global energy maximum in time. 
It is equivalent to $\partial_r \phi (r, \bar{t}_0) = \partial_r\phi_0(r) + \frac{2\I B_0}{r^3}\bar{t}_0 = 0$, so that $\phi_0 = \frac{\I B_0}{r^2}\bar{t}_0 + c$ and w.l.o.g. $\phi = -\frac{\I B_0}{r^2}(\bar{t}- \bar{t}_0)$.

On the other hand, condition (\emph{b}) implies that the blow-up is capped as late as possible in the evolution in $\tau$. Let us consider the $O(\delta ^1)$ terms in equation \eqref{Eq 7.4.3}
\begin{multline}
- \delta ^1 \partial_{\bar{t}} \left( 2r^2 (\partial_r\phi) (\partial_ra) +r^2 (\partial_r^2\phi) a + 3r(\partial_r\phi)a \right) \\ = \frac{2B_0}{r}\delta ^1 \partial_\tau \left( 2 \tau \partial_r a + (\partial_r \tau) a + \frac{3}{r} \tau a \right). \label{Eq 7.4.5}
 \end{multline}
Recalling that $a = O(\tau^{-2})$ and $\partial_r a = O((\partial_r \tau)\tau^{-3})$ as $\tau \rightarrow 0$, we find that the leading-order terms in \eqref{Eq 7.4.5} are $O(\delta ^1(\partial_r \tau)\tau^{-3})$, whereas the left-hand side of \eqref{Eq 7.4.3} is of order $\tau^{-1}$. Hence, the $O(\delta^1)$ terms become significant as soon as $\tau = O((\delta\partial_r \tau)^{\frac{1}{2}})$. Accordingly, to attain the most sustained blow-up $ \partial_r \tau$ should be as small as possible for $\tau \rightarrow 0$, i.e.
\begin{equation}
0  =  \lim_{\tau \rightarrow 0} (- \I \partial_r \tau) 
= \lim_{\tau \rightarrow 0} \left( \partial_r^2\phi_0-\frac{3}{r} \tau  + \frac{3}{r} \partial_r\phi_0 \right) = \partial_r^2\phi_0 + \frac{3}{r} \partial_r\phi_0. \label{Eq 7.4.6}
 \end{equation}
 Equation \eqref{Eq 7.4.6} is also satisfied for $\phi_0 = \frac{\I B_0}{r^2}\bar{t}_0 + c$. Hence, this is indeed the optimal initial phase giving the optimal perturbation according to WKB theory,
\begin{equation}
 u_r = \delta a \exp\left( \frac{ \phi}{\delta }\right)  \stackrel{\text{Eq. }\eqref{Eq 7.4.4}}{=} \delta a_0(r)  \frac{\exp \left( - \frac{4B_0^2}{3r^6} (\bar{t}-\bar{t}_0)^3\right)}{ \frac{4B_0^2}{r^6} (\bar{t}-\bar{t}_0)^2} \exp \left( - \frac{ \I B_0}{\delta r^2}(\bar{t}-\bar{t}_0) \right). \label{Eq 7.4.7}
\end{equation}
Note that the boundary conditions are satisfied if and only if $a(r_i) =a(r_o) = \partial_r a(r_i) = \partial_r a(r_o) = 0$ so that \eqref{Eq 7.4.7} is indeed an approximate solution to the complete boundary value problem for $\bar{t}-\bar{t}_0 = O(1)$ if $a$ is suitably chosen.

\subsection{Boundedness of the blow-up}\label{sS:uniqe-blow-up}

According to \eqref{Eq 7.4.6} we then have $\partial_r \tau = O(\tau)$ for $\tau \rightarrow 0$ so that the growth is not capped before $\tau = O(\delta)$.  However, it remains to be shown that no further blow-up occurs beyond the domain of the WKB solution \eqref{Eq 7.4.7}. For times $\bar{t}- \bar{t}_0 = O(\delta) $ we obtain $ \partial_r^n a = O(\delta^{-2})$ and $ \partial_r^n \exp \left( - \frac{ \I B_0}{\delta r^2}(\bar{t}-\bar{t}_0) \right) = O(1)$ for all $n \in \mathbb{N}_0$ so that $u_r, \partial_r^n u_r = O(\delta^{-1})$. Therefore the scaling $ \delta \tilde{t}:= \bar{t}- \bar{t}_0$ and $\tilde{u}_r:= \delta^{-1} u_r$ is employed in equation \eqref{Eq 7.4.1} giving
\begin{equation}
\partial_{\tilde{t}} \left(r^2\partial_r^2 +3r\partial_r -(n^2-1) \right) \tilde{u}_r = -\frac{\I B_0}{r^2}  \left(r^2\partial_r^2 +3r\partial_r -(n^2-1) \right) \tilde{u}_r + O(\delta^3). \label{Eq 7.4.8}
\end{equation}

Setting $\tilde{f}_r := \left(r^2\partial_r^2 +3r\partial_r -(n^2-1) \right) \tilde{u}_r$, the leading-order solution of \eqref{Eq 7.4.8} is given by $\tilde{f}_r(r,\tilde{t}) = \tilde{f}_{r,0} (r) \exp\left( -\frac{\I B_0}{r^2} (\tilde{t} -\tilde{t}_0) \right)$. The operator $r^2\partial_r^2 +3r\partial_r -(n^2-1)$ is of Sturm-Liouville type so that a Green's function $G(r,r')$ exists such that
\begin{equation}
\tilde{u}_r (r,\tilde{t}) = \int_{r_i}^{r_o} G(r,r') \tilde{f}_{r,0} (r') \exp\left( -\frac{\I B_0}{r'^2} (\tilde{t} -\tilde{t}_0) \right) r' \D r'. \label{Eq 7.4.9}
\end{equation}
The function $G$ is given in $\S$\ref{SSA5}. Note that with this ansatz only two boundary conditions may be satisfied. However, this affects only a thin boundary layer in the vicinity of the cylinder walls where significant growth is inhibited already for $O(\bar{t}-\bar{t}_0) = O(1)$ due to the {no-slip} condition. Thus, the present focus lies on the \emph{inner} solution in the first place.

By \eqref{Eq 7.4.9} the components $\tilde{u}_r$ and $\tilde{u}_\varphi \sim (1+ r\partial_r)\tilde{u}_r$ are given by $L^2$-kernel integral operators applied to $\tilde{f}_r$. Consequently, they are $L^2$-continuous in $\tilde{f}_r$ so that $\|\boldsymbol{\tilde{u}}\|^2$ depends continuously on $\tilde{t}$. Hence, there is no further blow-up in the time scale $\bar{t}- \bar{t}_0 = O(\delta)$. 

\subsection{A scaling for optimal transient growth}\label{SS7.2}

According to \eqref{Eq 7.4.7} the optimal perturbation's components $u_r$ and $u_\varphi \sim (1+r\partial_r) u_r$ have grown to $O(\delta^{-1})$ by the optimal (blow-up) time. This yields the optimal transient growth
\begin{equation}
G_{\max}^{k=0} = \sup_{\bar{t}\geq 0} G(t) =  \sup_{\bar{t}\geq 0} \|\boldsymbol{u} (\bar{t})\|^2 \stackrel{(a)}{\sim}  \sup_{\bar{t}\geq 0} (|u_{r}(\bar{t})|^2 + |u_{\varphi}(\bar{t})|^2) = O(\delta^{-2}). \label{Eq 7.4.10}
\end{equation}
by condition $(a)$. Since the WKB approximation applies for $\delta \rightarrow 0$ and $\delta = (n\Rey)^{-\frac{1}{3}}$ it has been shown that the optimal transient growth of axially independent perturbations scales like $G_{\max}^{k=0} \sim \Rey^{\frac{2}{3}}$ in the limit of high Reynolds numbers $\Rey \rightarrow \infty$. This result is in perfect agreement with our numerical computations (see $\S$\ref{SS5.5}). 

Notably the scaling exponent $\alpha = \frac{2}{3}$ is independent of $\eta$ and of $R_\Omega$ (see \S\ref{SS6.1}) and equal for all azimuthal wavenumbers. In accordance with this, our numerical results show that as $\Rey \rightarrow \infty$ the optimal azimuthal wavenumber $n_{\max}$ becomes constant; the asymptotic value is selected only by the geometry (specified by $\eta$).

\subsection{Numerical validation}\label{SS7.3}

In order to validate the WKB solution \eqref{Eq 7.4.7} we compute the initial phases $\Imag(\ln u_{r} (r,0))$ of numerically determined optimal perturbations $\boldsymbol{u} = u_{r} \boldsymbol{e}_r+ u_{\varphi} \boldsymbol{e}_\varphi$, as proposed by \citet[p. 51 f.]{Chapman2002}. By equation \eqref{Eq 7.4.7} this should yield
\begin{align}
 \frac{\delta r^2 }{B_0}\Imag(\ln u_{r}(r,0) ) =  \bar{t}_0 + O(\delta).\label{Eq 7.4.11}
\end{align}
Owing to the non-uniqueness of the complex logarithm, relation \eqref{Eq 7.4.11} needs to be assumed to be satisfied for some $r_0 \in (r_i, r_o)$. We choose $r_0 = \frac{1}{2}(r_i + r_o)$.
\begin{figure}
   \centering
   \subfigure[$\Rey = 10^3$, $\delta = 0.069$; computed at $N=30$]
   {\includegraphics[width=0.49\textwidth]{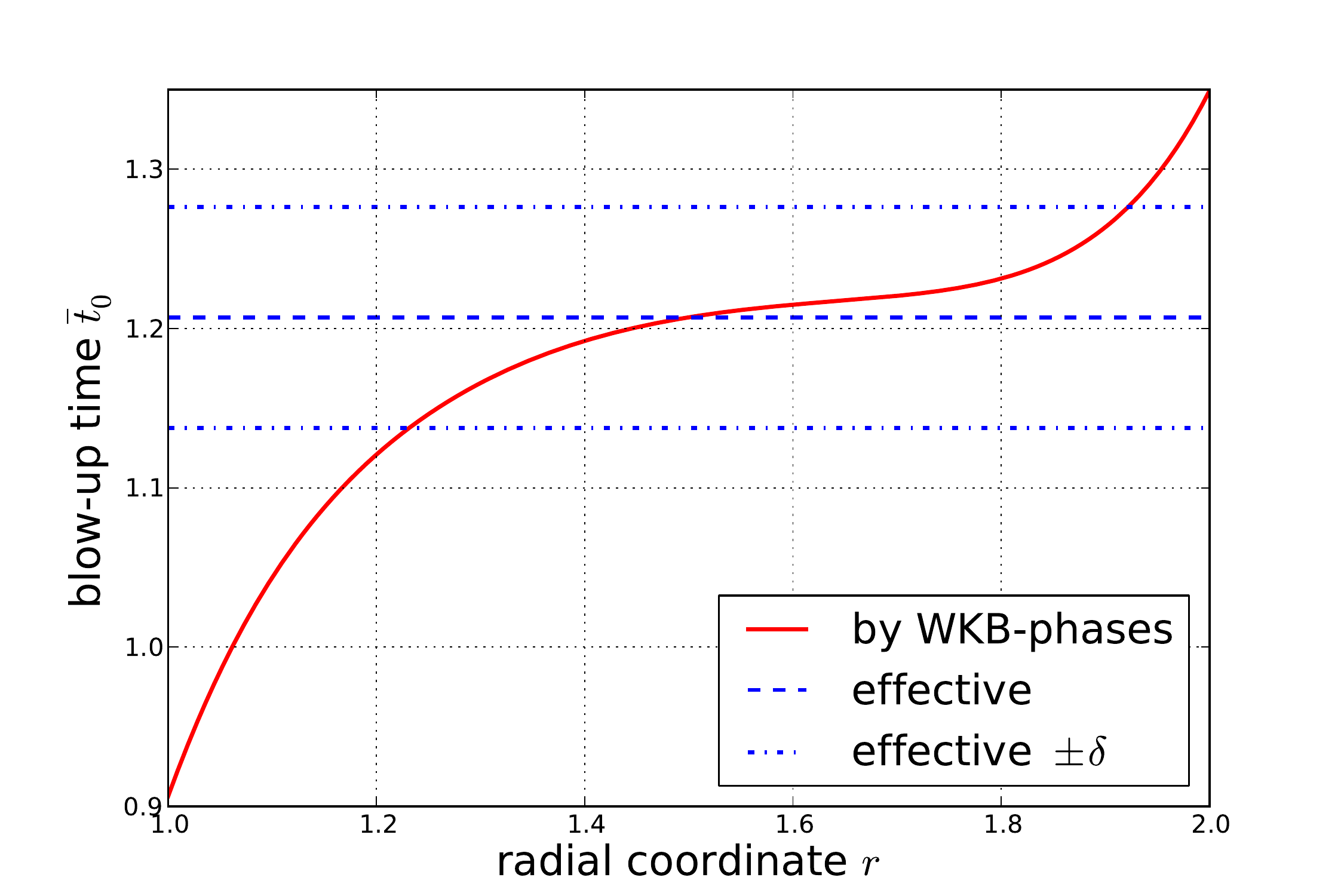}\label{fig 7.3.1a}}
   \hfill
   \subfigure[$\Rey = 10^4$, $\delta = 0.032$; computed at $N=30$]
   {\includegraphics[width=0.49\textwidth]{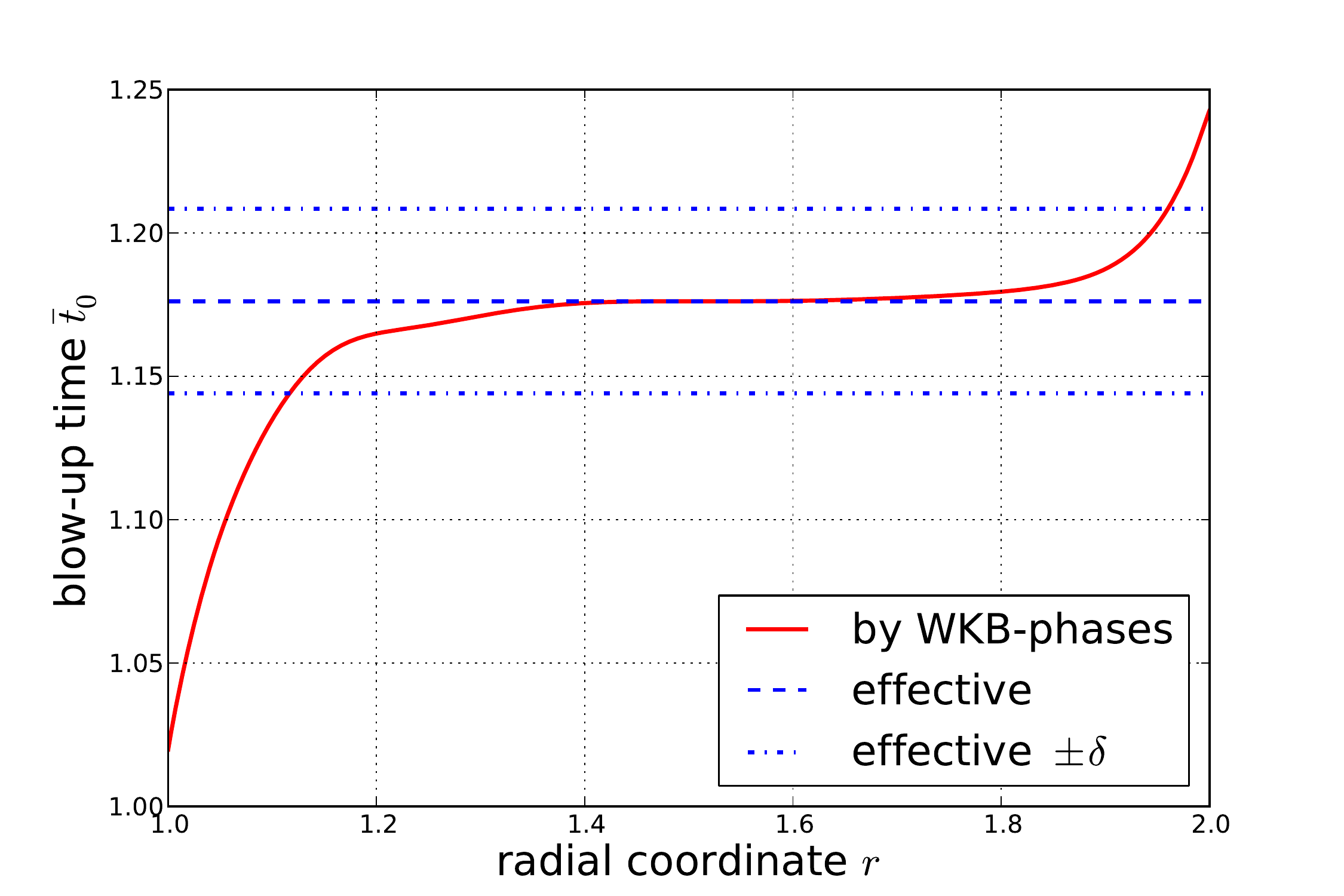}\label{fig 7.3.1b}}
   \\
   \subfigure[$\Rey = 10^5$, $\delta = 0.015$; computed at $N=70$]
   {\includegraphics[width=0.49\textwidth]{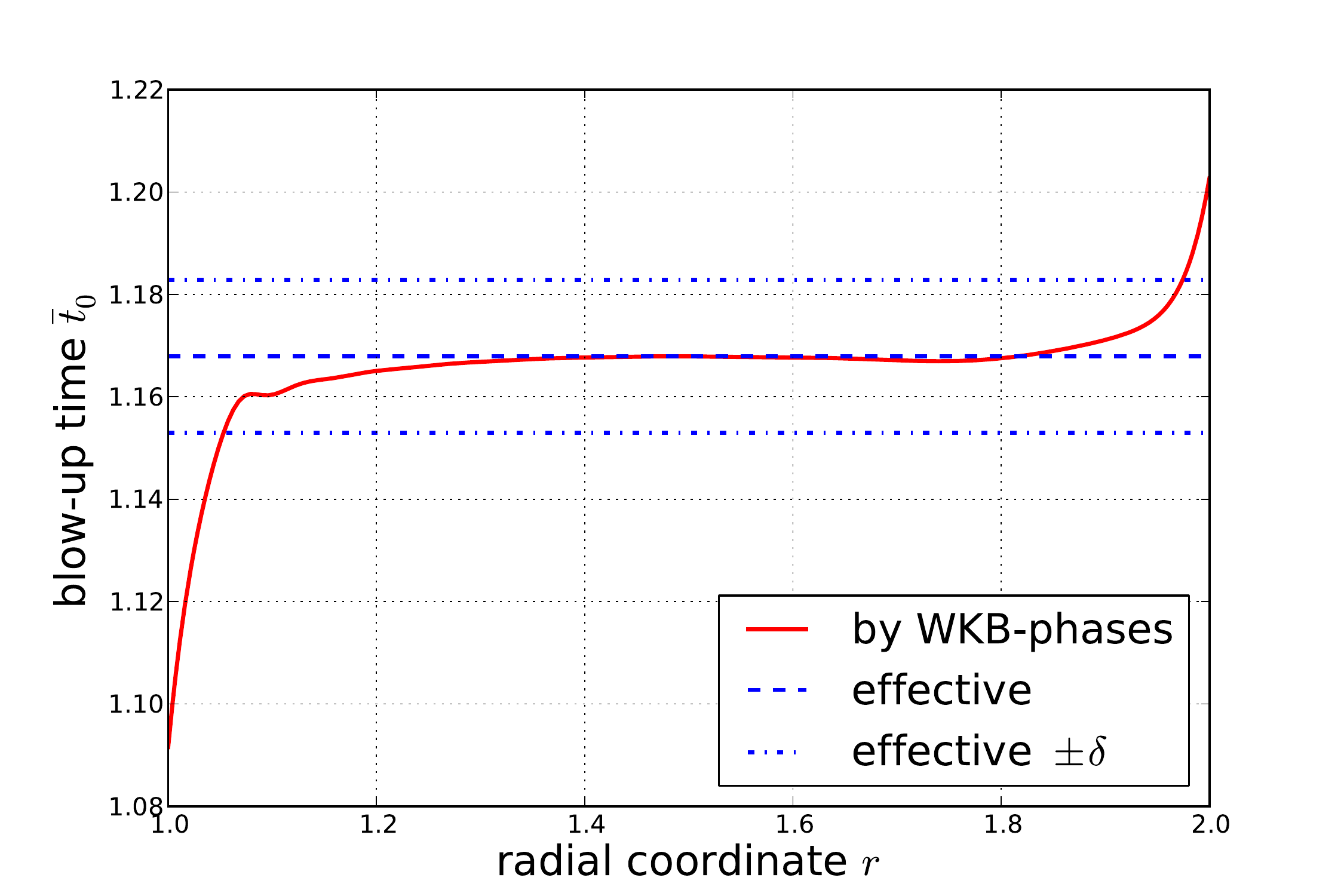}\label{fig 7.3.1c}}
   \hfill
   \subfigure[$\Rey = 10^6$, $\delta = 0.007$; computed at $N=130$]
   {\includegraphics[width=0.49\textwidth]{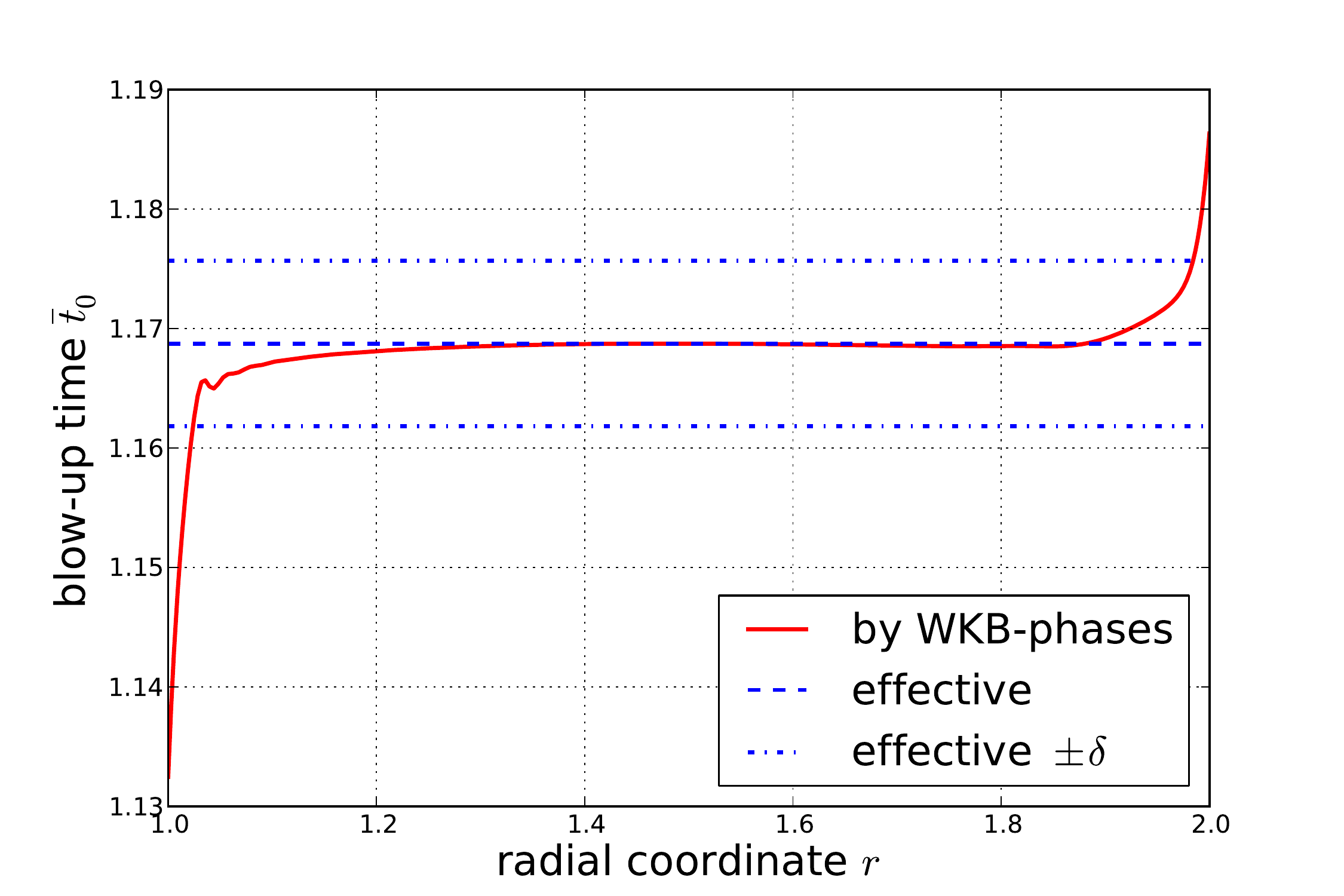}\label{fig 7.3.1d}}
   \caption{Blow-up times $\bar{t}_0$ of numerically determined optimal axially independent perturbations ($k=0$) for $R_\Omega = -2.0$, $\eta = 0.5$, $n=3$ and different shear Reynolds numbers $\Rey$. Results according to the WKB prediction \eqref{Eq 7.4.11} (``by WKB-phases'') are contrasted with the numerically observed transient growth maximum (``effective''). The ``effective $\pm \delta$'' show the expected error range due to finite $\Rey$ effects neglected in the WKB approximation ($\delta = (n \Rey)^{-1/3}$).}
   \label{fig 7.3.1}
\end{figure}

In figure \ref{fig 7.3.1} the blow-up time $\bar{t}_0$ computed from \eqref{Eq 7.4.11} is plotted against the radial coordinate $r$ (solid curves). This WKB prediction is compared for Reynolds numbers $\Rey \in \left\lbrace   10^3, 10^4, 10^5, 10^6 \right\rbrace $, corresponding to $\delta \in \left\lbrace   0.069,  0.032,  0.015,  0.007 \right\rbrace$, to the optimal time determined numerically from the full equations (dashed blue line). The expected error ranges are denoted by $[\bar{t}_0+ \delta; \bar{t}_0 - \delta]$ (dash-dotted blue lines). Excellent agreement between the numerical results and WKB solution within the predicted error of order $\delta$ and convergence for $\Rey \rightarrow \infty$ is found. Significant deviations are confined to a $O(\delta)$ neighbourhood of the cylinder walls in which growth is prevented {\it a priori} by the boundary conditions. Hence, the initial phase's behaviour as a key property of the derived WKB approximation has been numerically verified.

%
%
%
%
%
%
%
%
\section{Discussion}\label{S8}

Rayleigh-stable Taylor--Couette flows with the outer cylinder rotating faster than the inner one tend to become turbulent at moderate Reynolds numbers $\Rey = O(1000)$ \citep{Taylor1936,Borrero2010,Burin2012}. In the case of the quasi-Keplerian regime II, where the inner cylinder rotates faster than the outer one, the existence of turbulence remains debated \citep{JiBurin2006,PaolettiLathrop2011}. At the same time, Rayleigh-unstable but linearly (eigenvalue) stable counter-rotating Taylor--Couette flows are known to undergo subcritical transition \citep{Coles1965}.

In this work, the optimal linear transient growth $G_{\max}$, i.e. the maximum non-normal energy amplification of infinitesimal perturbations, has been investigated. Our analysis covers the whole parameter regime of Taylor--Couette flow, spanned by the shear Reynolds number $\Rey$, the cylinder radius ratio $\eta$ and the rotation number $R_\Omega$. We find that accurate transient growth computations are numerically feasible up to $\Rey = O(10^6)$, even though the characteristic Y-shaped eigenvalue spectrum of the linearized Navier-Stokes operator cannot be resolved for such Reynolds numbers. This is in contrast to previous studies of channel flow  \citep[\eg][]{ReddyHenningson1993}, which suggest that resolving the Y shape of the spctrum is necessary to accurately compute transient growth. For Taylor--Couette flow the transient growth maximum $G_{\max}$ is well converged for resolutions where the approximated spectrum is still far from its natural shape. This allows us to examine the optimal transient 
growth for large $\Rey$. Our numerical computations show an asymptotic scaling $G_{\max} \sim \Rey^{\alpha}$ for $\Rey \geq O(10^4)$ with $\alpha \approx \frac{2}{3}$ for all geometries considered, $\eta \in \lbrace 0.2, 0.5, 0.8 \rbrace$, and all linearly stable flows.

This reveals energy growth of the same order in all regimes and allows for arbitrary transient amplifications if $\Rey$ is sufficiently large. Moreover, the dynamics discussed in $\S$\ref{SS5.4} suggest that the underlying growth mechanisms (interpreted here as a curved analogue of the Orr mechanism) are essentially the same in the studied regimes I, II and IV. In the counter-rotating regime IV there are additional amplifying effects of the Rayleigh instability. {Notably, the observed spiral-shaped structures resemble those of the unstable eigenmodes emerging in the case of an imposed radial inflow at the rotating outer cylinder, reported by \citet{gallet2010}.} A distinction between the regimes is found in the optimal axial wavenumber $k_{\max}$, which reflects the axial dependence of the optimal perturbations attaining maximum energy amplification. Although columnar structures, representing axially invariant modes, dominate within practically the whole quasi-Keplerian regime (II) 
above $\Rey = O(1000)$ corresponding to  $k_{\max} = 0$, weakly three-dimensional optimal perturbations $0 < k_{\max} < 1$ are found in the likewise Rayleigh-stable regime I for $\Rey \rightarrow \infty$. The reason why a weak axial structure enhances transient growth in the latter, but not in the former, remains open. For counter-rotating flows, greater $k_{\max} = \mathcal{O}(1)$ turn out to attain higher energy maxima.

Our numerical results reveal an important role of axially invariant perturbations for transient growth in linearly stable Taylor--Couette flow. Hence, the corresponding linearized Navier-Stokes equations have been studied analytically in $\S$\ref{S6} and $ \S$\ref{S7}. Firstly, the analysis has revealed that transient growth and linear stability are indeed independent of $R_\Omega$ in the case $k = 0$. Then we have shown that optimal perturbations blow up and decay by the Orr mechanism within the time scale $t_0 = O((n\Rey)^{-\frac{2}{3}})$. By introducing this scale in the linearized evolution equations, an optimal transient growth scaling $G_{\max}^{k=0}(\Rey) = a(\eta) \Rey^{\frac{2}{3}}$ for axially independent perturbations has been derived in the limit $\Rey
\rightarrow \infty$, following the channel flow WKB analysis of \citet{Chapman2002}. The results apply for all $R_\Omega$ and thus in all flow regimes. For the coefficient $a(\eta)$ a semi-empirical formula given by \eqref{Eq 6.5.1} and \eqref{Eq 6.5.2} has been obtained by a cubic fit to the numerical data.

The expression $G_{\max}^{k=0}(\Rey) = a(\eta) \Rey^{\frac{2}{3}}$ provides a universal lower bound for the optimal transient growth of general three-dimensional perturbations. This bound attains the optimum in most of regime II according to the numerical results. However, while quasi-Keplerian flows thus indeed have the smallest possible energy amplification potential, the growth is nevertheless of the same order as in the other regimes. Temporary amplifications of disturbances may promote nonlinear instability if growing modes are consistently fed by nonlinear energy redistribution. Hence, by our scaling results, such a transient growth-mediated instability is as likely to exist in quasi-Keplerian flows as in any other regime. However, axially independent perturbations are possibly not equally fit to feed nonlinear instabilities as three-dimensional ones, e.g. because of their sharper growth and decay. {In the future this question could be addressed by studying nonlinear generalizations of transient 
growth, such as applied for instance by \citet{pringle2010}, \citet{monokrousos2011} and \citet{pringle2011}. On the other hand, such investigations are computationally expensive and beyond the present work.}

\citet{Meseguer2002} found a strong correlation between the experimentally observed nonlinear stability boundary \citep{Coles1965} and optimal transient growth $G_{\max}$ in counter-rotating flows. Following these ideas, we estimate the threshold shear Reynolds number $\Rey_T$ for subcritical transition in quasi-Keplerian flows using our universal scaling result. To this end $G_{\max}$ was computed numerically at the subcritical stability boundary of Taylor--Couette flow (results not shown) according to measurements by \citet{Mallock1896}, \citet{Wendt1933}, \citet{Taylor1936}, \citet{Coles1965}, \citet{Borrero2010}, \citet{Burin2012} and \citet{kavila2013}. Not surprisingly, the correlation is not as strong as observed by \citet{Meseguer2002}, who only considered the data of \citet{Coles1965}. Moreover, \citet{Burin2012} have found their experimental results to depend significantly on the applied endcap configurations where the 
sensitivity is stronger for 
wider gaps. Our results indeed range from $G_{\max} \approx 54$ to $G_{\max} \approx 155$. If we translate this to shear Reynolds numbers, the uncertainty roughly agrees with the observed endcap effects. Calculating the mean value of all computed threshold amplifications yields an {\it a priori} estimate for the threshold transient growth in an arbitrary Taylor--Couette flow setting of $G_{\max, T} = 92 \pm 26$.

Applying the estimate formula for $G_{\max}$, we obtain a threshold Reynolds number of $\Rey_T = a(\eta)^{-\frac{3}{2}}(880 \pm 370)$ giving for instance $\Rey_T = 67000 \pm 29000$ if $\eta = 0.7$. For quasi-Keplerian flows, recent experiments have proceeded up to $\Rey = O(10^6)$, yielding contradictory results (see \citet{JiBurin2006, PaolettiLathrop2011}). However, \citet{Avila2012} has shown the state-of-the-art Taylor--Couette apparatus to be possibly unsuited for such measurements because of axial endwall effects. On the other hand, our estimated $\Rey_T$ still lies within the range of direct numerical simulations. Hence, these may be able to resolve the controversy concerning the existence of hydrodynamic turbulence in the quasi-Keplerian regime. If turbulence were found, the value of the threshold $Re_T$ could be used to probe the significance of linear transient growth as a measure for subcritical instability.\\

Support from the Max Planck Society is acknowledged. Simon Maretzke thanks Laurette S. Tuckerman for her enlightening input concerning influence matrix methods.

\appendix
\section{}\label{SA}

\subsection{Calculation of the simplified linearized equations}\label{SSA1}

In this appendix a few supplementary computations for the derivation of the evolution equations in section \ref{S6} are presented.

Firstly, the commutator relation $[r\partial_r, \mathcal{L}_{rr}] =  \mathcal{L}_{rr}^{0}$ is shown. Setting $\alpha:= n^2-1+\I nB$ we obtain
\begin{eqnarray}
 [\mathcal{L}_{rr}, r\partial_r]  &=& \left[\mathcal{D}_+\mathcal{D} -\frac{n^2-1}{r^2} - k^2 - \frac{\I n}{r} v_\varphi^B , r\partial_r\right] \nonumber \\
 &=&\left[\left(\partial_r +\frac{1}{r}\right) \partial_r - \frac{\alpha}{r^2}, r\partial_r \right] = \left[\partial_r^2 , r\partial_r\right] +  \left[\frac{1}{r}\partial_r, r\partial_r\right] - \left[\frac{\alpha}{r^2}, r\partial_r \right] \nonumber \\
&=& 2\partial_r^2 + 2\partial_r - 2 \frac{\alpha}{r^2} = 2\mathcal{L}_{rr} + 2\I nA = 2\mathcal{L}_{rr}^{0}. \label{Eq A.1.1}
\end{eqnarray}
Moreover, the expression $\mathcal{D}_+ \mathcal{L}_{\varphi r} + \mathcal{L}_{r\varphi}\mathcal{D}_+$ can be simplified by
\begin{equation}
\mathcal{D}_+ \mathcal{L}_{\varphi r} + \mathcal{L}_{r\varphi}\mathcal{D}_+ = \mathcal{D}_+ \left(\frac{2\I n}{r^2} - 2A \right) + \left( 2A +\frac{2B}{r^2} -\frac{2\I n}{r}\right)  \mathcal{D}_+  =\frac{2B}{r^2}\mathcal{D}_+ - \frac{4\I n}{r^3}. \label{Eq A.1.2}
\end{equation}
Lastly, the equality $\partial_r\frac{2}{r} \left(r\mathcal{D}_+ r\mathcal{D}_+ -n^2  \right) = 2\mathcal{L}_{rr}^{0}r\mathcal{D}_+ + \I r n \left(   \frac{2B}{r^2}\mathcal{D}_+ - \frac{4\I n}{r^3} \right)$ holds since
\begin{subequations} \label{Eq A.1.3}
\begin{eqnarray}
\partial_r\frac{2}{r} \left(r\mathcal{D}_+ r\mathcal{D}_+ -n^2  \right) &=& \partial_r \frac{2}{r}\left(r^2\partial_r^2 +3r \partial_r -(n^2-1)  \right) \nonumber \\
&=&  2r\partial_r^3 +2 \partial_r^2 + 6 \partial_r^2 - \frac{2(n^2-1)}{r}\partial_r +  \frac{2(n^2-1)}{r^2} \nonumber \\
&=&  2r\partial_r^3 + 8 \partial_r^2 - \frac{2(n^2-1)}{r}\left( \partial_r -\frac{1}{r}\right)  \label{Eq A.1.3a} \\
2\mathcal{L}_{rr}^{0}r\mathcal{D}_+ + \I r n \left(   \frac{2B}{r^2}\mathcal{D}_+ - \frac{4\I n}{r^3} \right) &=& 2 \left(\partial_r^2 + \frac{1}{r} \partial_r -\frac{n^2-1}{r^2} - \frac{\I nB}{r^2}\right)r\mathcal{D}_+ \nonumber \\
&&+\frac{2\I nB}{r^2}r\mathcal{D}_+ + \frac{4n^2}{r^2} \nonumber \\
&=& 2 \left(\partial_r^2 + \frac{1}{r} \partial_r -\frac{n^2-1}{r^2}\right)(r\partial_r +1) + \frac{4n^2}{r^2} \nonumber \\
&=&  2r\partial_r^3 + 8 \partial_r^2 - \frac{2(n^2-1)}{r}\left( \partial_r -\frac{1}{r}\right).  \label{Eq A.1.3b}
\end{eqnarray}
\end{subequations}
\subsection{Analysis of the axial evolution equation}\label{SSA2}

Consider the operator $\mathcal{L}_{zz}$ and the axial component $u_z$ from the evolution equation \eqref{Eq 7.1.5} on the Hilbert space $\mathbb{H}$ introduced in $\S$\ref{SS2.1} and let $u, v \in \mathbb{H} \cap \mathscr{C}^2((r_i; r_o)) $ satisfy homogeneous Dirichlet boundary conditions. Define $\mathcal{A}_1 := \mathcal{D}_+ \mathcal{D}$ and $\mathcal{A}_2:= - \frac{n^2}{r^2}$ and $\mathcal{B} := - \frac{\I n}{r} v_\varphi^B$. $\mathcal{A}_2$ and $\mathcal{B}$ multiply by a real and strictly negative or purely imaginary function, respectively. Hence, $\mathcal{A}_2$ is self-adjoint negative definite and $\mathcal{B}$ is skew hermitian. For $\mathcal{A}_1$ we have by partial integration
\begin{subequations} \label{Eq A.2.1}
\begin{eqnarray}
\left\langle u, \mathcal{A}_1 v \right\rangle &=& \int_{r_i}^{r_o}  r u^\ast (\partial_r^2 + r^{-1} \partial_r) v \D r \stackrel{p.I.}{=} - \int_{r_i}^{r_o}  (\partial_r u^\ast) (\partial_r v) r \D r  \label{Eq A.2.1a} \\ 
&\stackrel{p.I.}{=}& \int_{r_i}^{r_o}  (r\partial_r^2  u^\ast +  \partial_r u^\ast ) v \D r  = \left\langle \mathcal{A}_1 u,  v \right\rangle . \label{Eq A.2.1b} 
\end{eqnarray}
\end{subequations}

Equation \eqref{Eq A.2.1b} reveals $\mathcal{A}_1$ to be self-adjoint and, for $u= v$, \eqref{Eq A.2.1a} shows its negative definiteness. Thus, $\mathcal{L}_{zz}$ is the sum of a self-adjoint strictly negative operator $\mathcal{A}:= \mathcal{A}_1 + \mathcal{A}_2$ and a skew hermitian one, $\mathcal{B}$. For the  commutator $\left[ \cdot , \cdot \right]$ we have
\begin{equation}
\left[\mathcal{A} , \mathcal{B} \right] = (\partial_r^2 + r^{-1} \partial_r) \left(- \frac{\I n}{r} v_\varphi^B \right) \neq 0. \label{Eq A.2.3} 
\end{equation}
Consequently, the adjoint operator $\mathcal{L}_{zz}^{\ast}$ satisfies 
\begin{equation}
 \left[\mathcal{L}_{zz}^\ast , \mathcal{L}_{zz} \right] =  \left[\mathcal{A} - \mathcal{B} , \mathcal{A} + \mathcal{B} \right] = 2 \left[\mathcal{A} , \mathcal{B} \right] \neq 0\label{Eq A.2.4} 
\end{equation}
so that $\mathcal{L}_{zz}$ is a non-normal operator. By definition $\|u_z \|^2 $ is equal to the axial component's portion of the total kinetic energy of $\boldsymbol{u}$. Owing to the evolution $\partial_t u_z = \mathcal{L}_{zz} u_z$ we have
\begin{equation}
\partial_t \| u_z \|^2  = 2 \Real \left\langle u_z, \mathcal{L}_{zz} u_z \right\rangle  = 2 \underbrace{\Real \left\langle u_z, \mathcal{A} u_z \right\rangle }_{<0} + 2 \underbrace{\Real \left\langle u_z, \mathcal{B} u_z \right\rangle}_{=0} < 0 \label{Eq A.2.5} 
\end{equation}
where $2 \Real \left\langle x, \mathcal{T} x \right\rangle = \left\langle x, \mathcal{T} x \right\rangle + \left\langle \mathcal{T} x,  x\right\rangle = \left\langle x, \mathcal{T} x \right\rangle - \left\langle x, \mathcal{T} x \right\rangle =0 $ for $ \mathcal{T}$ skew hermitian has been used. By relation \eqref{Eq A.2.5} there is no transient growth but only monotonic decay in the axial component of $k=0$ perturbations, as claimed in $\S$\ref{SS6.2}.
\newline
\subsection{WKB equations for the radial evolution equation}\label{SSA4}

In what follows, we derive of the WKB equations \eqref{Eq 7.4.2} and \eqref{Eq 7.4.3}.

Application of the operator $\left( r\mathcal{D}_+ r\mathcal{D}_+ -n^2 \right)  =\left(r^2\partial_r^2 +3r\partial_r -(n^2-1) \right)$ to the WKB ansatz $u_r = \delta a  \exp\left(\delta^{-1} \phi\right)$ of $\S$\ref{SS7.1} yields
 \begin{multline*}
  \exp(- \delta^{-1} \phi)\left(r^2\partial_r^2 +3r\partial_r -(n^2-1) \right)u_r \\ = \delta ^{-1} r^2 (\partial_r\phi)^2 a 
 + \delta ^{0} \left( 2r^2 (\partial_r\phi) (\partial_ra) +r^2 (\partial_r^2\phi) a +3r(\partial_r\phi)a \right)  \\
 + \delta ^{1}  \left(r^2\partial_r^2 +3r\partial_r -(n^2-1) \right) a + O(\delta ^2)
 \end{multline*}
The left-hand side of equation \eqref{Eq 7.4.1} thus reads 
 \begin{multline*}
 \delta  \exp(- \delta^{-1} \phi) \partial_{\bar{t}} \left(r^2\partial_r^2 +3r\partial_r -(n^2-1) \right)u_r \\
 = \delta^{-1}\left(\partial_{\bar{t}} \phi \right) \exp(- \delta^{-1} \phi)\left(r^2\partial_r^2 +3r\partial_r -(n^2-1) \right)u_r + \delta ^{0} r^2 \partial_{\bar{t}} \left((\partial_r\phi)^2 A \right) + \\
 \delta ^{1} \partial_{\bar{t}} \left( 2r^2 (\partial_r\phi) (\partial_rA) +r^2 (\partial_r^2\phi) A + +3r(\partial_r\phi)A \right) + O(\delta ^2)
 \end{multline*}
and the right-hand side is
 \begin{multline*}
 \exp(- \delta^{-1} \phi)\left( \delta ^3 \left( \partial_r^2 -\frac{1}{r} \partial_r - \frac{n^2-1}{r^2} \right)  -\frac{\I B_0}{r^2} \right) \left(r^2\partial_r^2 +3r\partial_r -(n^2-1) \right) u_r \\
 = \exp(- \delta^{-1} \phi) \left( -\frac{\I B_0}{r^2} \right) \left(r^2\partial_r^2 +3r\partial_r -(n^2-1) \right)u_r + \delta ^{0} r^2 (\partial_r\phi)^4 a  \\
 + \delta ^1 \left(6r (\partial_r\phi)^3 a + 6r^2 (\partial_r\phi)^2 (\partial_r^2\phi) a + 4 r^2 (\partial_r\phi)^3 (\partial_ra) \right)  + O(\delta ^2)
 \end{multline*}
Hence, for the leading-order terms $= O(\delta^{-1})$, equation \eqref{Eq 7.4.2} is obtained:
 \begin{equation}
(\partial_{\bar{t}} \phi) (r^2 (\partial_r\phi)^2 a) = -\frac{\I B_0}{r^2} (r^2 (\partial_r\phi)^2 a) \,\,\,\,\,\, \Longleftrightarrow \,\,\,\,\,\, \partial_{\bar{t}} \phi  = -\frac{\I B_0}{r^2} \label{Eq A.4.3}
 \end{equation}
The next-order $ O(\delta^0)$ equation reads
 \begin{multline}
r^2 \partial_{\bar{t}} \left((\partial_r\phi)^2 a \right) - r^2 (\partial_r\phi)^4 a = - \left(\partial_{\bar{t}} \phi + \frac{\I B_0}{r^2} \right) \exp(- \delta^{-1} \phi)\left(r^2\partial_r^2 +3r\partial_r -(n^2-1) \right)u_r\\
+\delta ^1 \left(6r (\partial_r\phi)^3 a + 6r^2 (\partial_r\phi)^2 (\partial_r^2\phi) a + 4 r^2 (\partial_r\phi)^3 (\partial_ra) \right)  \\ - \delta ^1 \partial_{\bar{t}} \left( 2r^2 (\partial_r\phi) (\partial_ra) +r^2 (\partial_r^2\phi) a +3r(\partial_r\phi)a \right) \label{Eq A.4.4}
 \end{multline}
Consequently, applying $\partial_{\bar{t}} \phi + \frac{\I B_0}{r^2} = 0$ from expression \eqref{Eq A.4.3} to equation \eqref{Eq A.4.4}, the next-to-leading-order WKB equation \eqref{Eq 7.4.3} follows.
\newline
\subsection{Green's function for the radial evolution equation}\label{SSA5}

In this appendix the Green's function $G$ used in $\S$\ref{sS:uniqe-blow-up} is derived and thereby the regularity of the approximate solution $\tilde{u}_r$ defined by equation \eqref{Eq 7.4.9} is proven.

Consider the eigenvalue problem $\left(r^2\partial_r^2 +3r\partial_r -(n^2-1) \right) \psi_\lambda (r) = -\lambda \psi_\lambda(r)$ in the interval $r \in (r_i; r_o)$. With $p:=r^3$, $q:=(n^2-1)r$, $w:=r$ and boundary conditions $\psi_\lambda (r_i) = \psi_\lambda(r_o) = 0$. this is a Sturm-Liouville problem of the form
 \begin{equation}
 -\partial_r(p\cdot (\partial_r \psi_\lambda)) + q = \lambda w \psi_\lambda  \label{Eq A.5.1}
 \end{equation}
The eigenvalues $\left\lbrace \lambda_m\right\rbrace_{m\in \mathbb{N}}$ are thus discrete and the corresponding normalized eigenfunctions form a complete orthonormal set $\left\lbrace \psi_m\right\rbrace_{m\in \mathbb{N}}$ with respect to the inner product $\left\langle \psi_l, \psi_m \right\rangle  = \int_{r_i}^{r_o} \psi_l^\ast \psi_m w \D r$ of the Hilbert space $\mathbb{H}$ introduced in $\S$\ref{SS2.1}.

A solution to the inhomogeneous problem $\left(r^2\partial_r^2 +3r\partial_r -(n^2-1) \right) \psi = g$ and $\psi (r_i) = \psi (r_o) = 0$ is consequently given by
 \begin{equation}
\psi(r)   =  \int_{r_i}^{r_o} G(r,r')g(r') r' \D r'  \,\,\,\,\,\, \text{with} \,\,\,\,\,\, G(r,r') := - \sum_{m\in \mathbb{N}}  \frac{\psi_m (r')^\ast \psi_m (r)}{\lambda_m} \label{Eq A.5.2}
 \end{equation}
where $G$ is the Green's function. By definition $G$ is continuous and thus bounded on $[r_i; r_o]^2$. For the given problem the normalized solution to the eigenvalue problem reads
 \begin{equation}
\lambda_m = n^2 -  \frac{\pi^2 m^2}{\ln\eta}\,\,\,\,\,\, \text{and} \,\,\,\,\,\, \psi_m (r) = \frac{\sqrt{-2\ln\eta}}{r} \sin \left( - \frac{\pi m}{\ln \eta} \ln \frac{r}{r_i}\right),\,\,\,\,\,\, m \in \mathbb{N} \label{Eq A.5.3}
 \end{equation}

\bibliographystyle{jfm}
\bibliography{Literature}

\end{document}